\documentclass[fleqn,usenatbib]{mnras}

\usepackage{newtxtext,newtxmath}

\usepackage[T1]{fontenc}

\DeclareRobustCommand{\VAN}[3]{#2}
\let\VANthebibliography\thebibliography
\def\thebibliography{\DeclareRobustCommand{\VAN}[3]{##3}\VANthebibliography}

\usepackage{graphicx}	
\usepackage{amsmath}	

\title[SNe Ia with Late-time Spectra]{Implications for the Explosion Mechanism of Type Ia Supernovae from their Late-time Spectra}

\author[J. Liu et al.]{
Jialian Liu,$^{1}$\thanks{E-mail: liu-jl22@mails.tsinghua.edu.cn}
Xiaofeng Wang,$^{1,2}$\thanks{E-mail: wang\_xf@mail.tsinghua.edu.cn}
Alexei V. Filippenko,$^{3}$
Thomas G. Brink,$^{3,4}$
Yi Yang,$^{3,5}$
Weikang Zheng,$^{3,6}$
\newauthor
Hanna Sai,$^{1}$
Gaobo Xi,$^{1}$
Shengyu Yan,$^{1}$
Nancy Elias-Rosa,$^{7,8}$
Wenxiong Li,$^{9}$
Xiangyun Zeng$^{10,11}$
\newauthor
and Abdusamatjan Iskandar$^{10,11}$\\
$^{1}$Physics Department and Tsinghua Center for Astrophysics, Tsinghua University, Beijing 100084, China\\
$^{2}$Beijing Planetarium, Beijing Academy of Science and Technology, Beijing 100044, China\\
$^{3}$Department of Astronomy, University of California, Berkeley, CA 94720-3411, USA\\
$^{4}$Wood Specialist in Astronomy\\
$^{5}$Bengier-Winslow-Robertson Postdoctoral Fellow\\
$^{6}$Eustace Specialist in Astronomy\\
$^{7}$INAF Osservatorio Astronomico di Padova, Vicolo dell’Osservatorio 5, 35122 Padova, Italy\\
$^{8}$Institute of Space Sciences (ICE, CSIC), Campus UAB, Carrer de Can Magrans s/n, 08193 Barcelona, Spain\\
$^{9}$The School of Physics and Astronomy, Tel Aviv University, Tel Aviv 69978, Israel\\
$^{10}$Xinjiang Astronomical Observatory, Chinese Academy of Sciences, Urumqi, Xinjiang 830011, China\\
$^{11}$School of Astronomy and Space Science, University of Chinese Academy of Sciences, Beijing 100049, China\\
}

\date{Accepted XXX. Received YYY; in original form ZZZ}

\pubyear{2023}

\begin{document}
\label{firstpage}
\pagerange{\pageref{firstpage}--\pageref{lastpage}}
\maketitle

\begin{abstract}
Late-time spectra of Type Ia supernovae (SNe~Ia) are important in clarifying the physics of their explosions, as they provide key clues to the inner structure of the exploding white dwarfs. We examined late-time optical spectra of 36 SNe~Ia, including \textcolor{black}{five from our own project (SNe 2019np, 2019ein, 2021hpr, 2021wuf, and 2022hrs),} with phase coverage of $\sim 200$ to $\sim 400$ days after maximum light. At this late phase, the outer ejecta have become transparent and the features of inner iron-group elements emerge in the spectra. Based on multicomponent Gaussian fits and reasonable choices for the pseudocontinuum around Ni and Fe emission features, we get reliable estimates of the Ni to Fe ratio, which is sensitive to the explosion models of SNe~Ia. Our results show that the majority (about \textcolor{black}{67\%}) of our SNe~Ia are more consistent with the sub-Chandrasekhar-mass (i.e., double-detonation) model, although they could be affected by evolutionary or ionisation effects. Moreover, we find that the Si~II $\lambda$6355 velocity measured around the time of maximum light tends to increase with the Ni to Fe ratio \textcolor{black}{for the subsample with either redshifted or blueshifted nebular velocities}, suggesting that progenitor metallicity might play an important role in accounting for the observed velocity diversity of SNe~Ia. 
\end{abstract}




\section{Introduction} \label{sec:intro}

It is widely accepted that thermonuclear explosions of carbon-oxygen (CO) white dwarfs (WDs) with masses close to the Chandrasekhar limit \citep{1997Sci...276.1378N,2000ARA&A..38..191H,2014ARA&A..52..107M} produce Type Ia supernovae (SNe~Ia; see, e.g., \citealt{1997ARA&A..35..309F} for a review of supernova classification). However, the mechanism that triggers the explosion and drives the propagation of the burning front, together with the nature of the donor, still remain unclear. Different models (that probably yield multiple valid channels of the explosion) invoke for the mass donor a non-WD companion such as a red giant or a helium star (the ``single-degenerate'' channel; \citealp{1973ApJ...186.1007W}), or another WD  (the ``double-degenerate'' channel; \citealp{1984ApJS...54..335I,1984ApJ...277..355W}). 
Also, increasing attention is being paid to sub-Chandrasekhar-mass (sub-$M_{\rm Ch}$) models \citep{2007ApJ...662L..95B,2009ApJ...699.1365S}.
In addition to the vast parameter space for the progenitor systems, since the spectral features of SNe~Ia exhibit both intermediate-mass elements (IMEs, i.e., from Si to Ca) and iron-group elements (IGEs), a transition of the nuclear burning front from subsonic to supersonic phases is expected to take place. Various models are mainly differentiated by the mechanism that triggers the detonation, as follows.

(i) In WDs near $M_{\rm Ch}$, owing to the compressional heating, an initial burning starts as a subsonic deflagration near its mass centre. As the burning front propagates outward, some parts of the front would be accelerated due to the Rayleigh-Taylor (RT) instabilities. Subsequent transitions from the subsonic deflagration to the supersonic detonation would take place at the plume structures that develop \citep{2005ApJ...623..337G, 2013MNRAS.429.1156S}. Such a later-ignited ``delayed-detonation'' (DDT) front will synthesise the remaining WD into IGE-dominated products.

(ii) In WDs below $M_{\rm Ch}$ (i.e., $\lesssim 1.2$\,M$_{\odot}$), the detonation of the CO WD can be triggered by an initial instability-induced detonation of an accreted helium shell on top of the WD (the ``double-detonation''; \citealp{2010ApJ...714L..52S, 2018ApJ...854...52S, 2019ApJ...878L..38T}).

(iii) Based on the result of hydrodynamic simulations, dynamical procedures such as the merger or the head-on collision of a double WD binary would also be able to meet the criteria that detonate the merged WD \citep{2009MNRAS.399L.156R, 2013ApJ...778L..37K, 2013ApJ...770L...8P, 2017hsn..book.1257P}.

Because WDs are mostly electron-degenerate matter consisting of carbon and oxygen, all models that blow up the WD are anchored to the same network of nuclear reactions, which dominate the electromagnetic signatures of SNe~Ia around their peak luminosity. Such a commonality thus lead to universal chemical compositions and energetics regardless of the model details. This also roughly explains why the photometric and spectroscopic evolution of SNe~Ia around the time of maximum light can be well reproduced by a broad range of models.

When WDs explode as SNe~Ia, the timescale of their photometric evolution is well correlated with its peak luminosity. Both quantities are determined by the content of the heavy elements synthesised during the explosion.
The peak luminosities of SNe~Ia, including the slowly-declining, hot, luminous SN~1991T-like or SN~1999aa-like objects, the rapidly-declining, cool, subluminous SN~1991bg-like objects, and the ``Branch-normal" objects \citep{1993AJ....106.2383B}, show a prominent correlation with the post-peak decline rate ${\Delta}m_{15}(B)$, with brighter objects having smaller ${\Delta}m_{15}(B)$ \citep{1993ApJ...413L.105P}. This correlation has been dubbed the ``Phillips relation." It serves as the basic recipe for the cosmological use of SNe~Ia, and it is believed to be governed mainly by the amount of $^{56}$Ni produced in the explosion \citep{1996ApJ...457..500H}. Moreover, there is increasing evidence showing that SN~Ia peak luminosity is not parameterised only by the decline rate. The inclusion of a color parameter helps tighten the dispersion of normalised peak luminosity \citep{1998A&A...331..815T, 2005ApJ...620L..87W} and improves distance estimates from SNe~Ia (e.g., \citealt{2007A&A...466...11G, 2014A&A...568A..22B}). 

However, despite the first-order simplicity and commonality of SNe~Ia, diversity among a range of subtypes has also been explored by various studies.
For example, \citet{2005ApJ...623.1011B} found that SNe~Ia exhibit large scatter in the velocity evolution of the ejecta, and the velocity gradient measured from Si~II $\lambda$6355 absorption is not correlated with $\Delta m_{15}(B)$ for Branch-normal SNe~Ia. \citet{2009ApJ...699L.139W} divided Branch-normal SNe~Ia into two groups based on Si~II $\lambda$6355 velocities measured around their time of $B$-band maximum light, 
with the high-velocity (HV) group having Si velocities $\gtrsim 12,000$ $\rm km\ s^{-1}$ and the normal-velocity (NV) group having Si velocities $\lesssim 12,000$ $\rm km\ s^{-1}$. Such velocity diversity is independent of ${\Delta}m_{15}(B)$, and the origin of the spectral differences was interpreted as a geometric viewing-angle effect \citep{2010Natur.466...82M,2013MNRAS.430.1030S}. However, \citet{2013Sci...340..170W} studied the birthplace environments of SNe~Ia and found that HV SNe~Ia tend to occur in inner, brighter regions of more-massive galaxies compared with the NV counterparts, suggesting that these two subclasses may have different progenitor properties. For example, HV SNe~Ia are likely associated with metal-rich progenitors \citep{2015MNRAS.446..354P, 2020ApJ...895L...5P}, and they have more circumstellar matter (CSM) than NV SNe~Ia \citep{2019ApJ...882..120W}. Thus, the idea that all SNe~Ia originate from one family or one explosion mechanism is challenged even if we do {\it not} consider peculiar subclasses of SNe~Ia such as those defined by subluminous SN~1991bg-like \citep{1992AJ....104.1543F}, overluminous SN~1991T-like \citep{1992AJ....103.1632P}, SN~2002es-like \citep{2012ApJ...751..142G},  and SN~2009dc-like \citep{2011MNRAS.410..585S} SNe~Ia.

To further explore the explosion mechanism and progenitor physics of SNe~Ia, we need to inspect the inner regions of their ejecta. At early times, the outer ejecta are opaque, and the deeper regions are hidden. At a phase of over 200 days after peak brightness, the ejecta have expanded substantially and become transparent to the radiation from the inner core, which is dominated by emission from the Fe-group elements. At such nebular phases, the shape of the spectral profile over the wavelength range of 6800--7800~\AA, dominated by [Fe~II] and [Ni~II] features \citep{2018MNRAS.477.3567M,2020MNRAS.491.2902F}, delivers critical constraints on the structure and abundance ratio of the Fe-group elements. For instance, the nonzero velocity shifts of [Fe~II] and [Ni~II] features indicate an asymmetric explosion \citep{2010ApJ...708.1703M}. In addition, the iron is mainly contributed by the end product of the radioactive decay of $^{56}$Ni, and the stable nickel was synthesised by the explosion. Thus, the Ni/Fe ratio in the nebular phase reflects the ratio of stable to radioactive isotopes of Fe-group elements produced in the explosion, which is sensitive to the central density of the exploding WD. \citet{2018MNRAS.477.3567M} used a multicomponent Gaussian fit to measure the Ni/Fe ratio in the 7300~{\AA} region in optical spectra to distinguish $M_{\rm Ch}$ and sub-$M_{\rm Ch}$ models since the central densities of WDs are quite different for these two explosion models. \citet{2020MNRAS.491.2902F} and \citet{2022MNRAS.511.3682G} also measured the Ni/Fe ratio using different methods, but their results were generally lower than those given by \citet{2018MNRAS.477.3567M}. \citet{2020MNRAS.491.2902F} mentioned that the differences are mainly due to the placement of the pseudocontinuum across the 7300~{\AA} region. In this work, we generally follow the fitting method of  \citet{2018MNRAS.477.3567M} but adopt the pseudocontinuum \textcolor{black}{in a different way.}  

Using nebular-phase spectra of SNe~Ia published in the literature as well as data collected through our own program, we attempt to provide more-accurate measurements of the Ni/Fe ratio to put tighter constraints on the explosion mechanism of SNe~Ia. Since the Ni/Fe ratio can be affected by the progenitor metallicity \citep{2003ApJ...590L..83T, 2018ApJ...854...52S, 2020MNRAS.492.2029S}, we also examine correlations of this ratio with SN~Ia observables measured at early times, such as the Si~II velocity and $\Delta m_{15}(B)$. 

The paper is structured as follows. We outline the sample used in our analysis in Section~\ref{sec:source}, and Section~\ref{sec:method} shows the fitting methods. Our results are presented in Section~\ref{sec:results} and discussed in Section~\ref{sec:discussion}. 
Section~\ref{sec:conclusion} summarises our conclusions.

\section{Data Sources} \label{sec:source} 
In order to examine the distribution of the Ni/Fe ratio, we collect a sample of 58 late-time spectra taken at $t \approx +200$--400 days after maximum light for 36 SNe~Ia. Those showing peculiar properties such as SN~1991T-like and SN~1991bg-like SNe~Ia are also included except when their spectra exhibit strong calcium features or have a flat profile in the 7300~{\AA} region. The publicly available data were retrieved using the Open Supernova Catalog (OSC; \citealt{2017ApJ...835...64G}), the Weizmann Interactive Supernova data REPository (WISeREP; \citealt{2012PASP..124..668Y}), and the Supernovae Database (SNDB; \citealt{2012MNRAS.425.1789S,2019MNRAS.482.1545S}). The $t \approx +384$~day spectrum of SN~2017fgc comes from \citet{2021ApJ...919...49Z}. Information on the spectra collected by our own project is presented in Section~\ref{sec:observations}. An overview of the observations for all the SNe~Ia in this work is listed in Table~\ref{tab:base_para}. References for all the late-time spectra can be found in Table~\ref{tab:Multi}.

To study the connections between early-time and late-time properties, \textcolor{black}{we used the spectrum around the time of maximum light (i.e., within about 3 days from the $B$-band peak, except for SN~2012hr at +5 days and ASASSN-14jg at +6 days) to measure the Si~II $\lambda$6355 velocity from the absorption minimum for each object of our sample. The Si velocities and the phase of the spectra used in the measurements are presented in Table~\ref{tab:base_para}. We collected the post-peak decline rate ${\Delta}m_{15}(B)$ and the date of $B$-band maximum from the literature when available. For SNe~2003kf, 2012hr, and 2013cs, these two parameters are estimated by applying the SALT2 \citep{2007A&A...466...11G} fit to their light curves.} All spectra used in the analysis have been corrected for host-galaxy redshift from the NASA/IPAC Extragalactic Database (NED) and extinction due to the Milky Way \textcolor{black}{\citep{2011ApJ...737..103S}} and host galaxy whenever possible (for details, see Table~\ref{tab:base_para}). As we only focus on the 7300~{\AA} region in the spectra, the results should suffer little from uncertainties in extinction corrections. 

\subsection{Late-Time Spectra from our Project \label{sec:observations}}
\begin{figure}
    \centering
    \includegraphics[width=0.95\columnwidth]{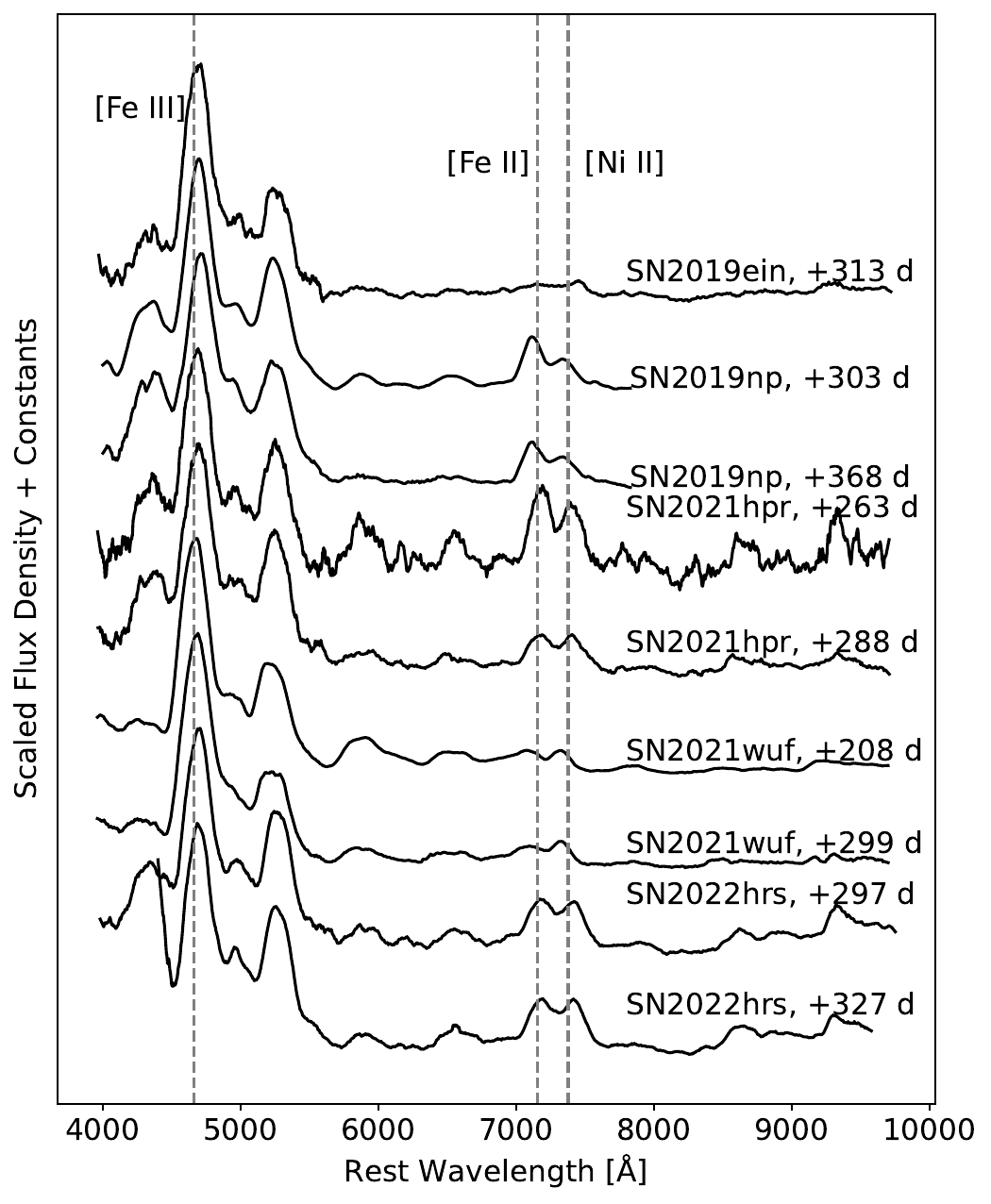}
    \caption{Late-time spectra of SNe 2019ein, 2019np, 2021hpr, 2021wuf, and 2022hrs. Flux densities ($f_\lambda$) are all normalised to the [Fe~III] 4700~\AA\ feature \textcolor{black}{ and smoothed with a Savitsky-Golay filter of width $\sim 50$~\AA\ and order 1.} The rest-frame wavelengths of [Fe~III] 4659~\AA, [Fe~II] 7155~\AA, and [Ni~II] 7378~\AA\ are marked by dashed lines. }
    \label{fig:new_spectra}
\end{figure}

Our program aims to collect some late-time spectra for SNe~Ia that were well observed at early phases. Thus far, we have obtained late-time spectra of SN~2019np \citep{2022MNRAS.514.3541S}, SN~2019ein \citep{2022MNRAS.517.4098X}, SN~2021hpr (Iskandar et al., in prep.), SN~2021wuf (Zeng et al., in prep.), and SN~2022hrs (Liu et al., in prep.). The nebular spectra of these last four SNe~Ia were taken with LRIS \citep{1995PASP..107..375O} mounted on the 10~m Keck-I telescope and the DEIMOS spectrograph \citep{2003SPIE.4841.1657F} on the 10~m Keck-II telescope on Maunakea, and the Kast double spectrograph \citep{miller1993lick} mounted on the Shane 3~m telescope at Lick Observatory. The late-time spectra of SN~2019np were obtained with OSIRIS mounted on the 10.4~m Gran Telescopio CANARIAS (GTC) at the Roque de Los Muchachos Observatory (Spain). \textcolor{black}{To minimise slit losses caused by atmospheric dispersion \citep{1982PASP...94..715F}, the slit was oriented at or near the parallactic angle. The Keck~I/LRIS spectra were reduced using the \texttt{LPipe} pipeline \citep{2019PASP..131h4503P}. The GTC/OSIRIS, Keck II/DEIMOS, and Shane/Kast observations were reduced using standard \texttt{IRAF}\footnote{{IRAF} is distributed by the National Optical Astronomy Observatories, which are operated by the Association of Universities for Research in Astronomy, Inc., under cooperative agreement with the National Science Foundation (NSF).} routines for CCD processing (e.g., \citealt{2012MNRAS.425.1789S}) and optimal spectrum extraction \citep{1986PASP...98..609H}. The spectra were flux calibrated using observations of appropriate spectrophotometric standard stars observed on the same night, at similar airmasses, and with an identical instrument configuration. We corrected for the atmospheric extinction using the extinction curves of local observatories.} A journal of observations is given in Table~\ref{tab:new} and the corresponding spectra are shown in Fig.~\ref{fig:new_spectra}.  

\begin{table*}
\centering
\caption{Overview of the Observations.}
\begin{tabular}{ccccccccccccc}
\hline
\hline
 {Name} &  {Observation} &  {Observation} &  {Phase} &  {Range} &
 {Instrument} &  {Telescope} &  {Exposure}  \\
 {} &  {MJD} &  {Date (UTC)} &  {(d)} &
 {[{\AA}]} &  {} &  {} &  {[s]} \\
\hline
SN~2019ein   & 58932 & 2020 Mar. 24 & +313 & 3147--10,278 & LRIS & Keck I & 1077.5\\
SN~2019np    & 58813 & 2019 Nov. 26  & +303 & 3812--7830 & OSIRIS & GTC & 1380\\
SN~2019np    & 58878 & 2020 Jan. 30  & +368 & 3832--7831 & OSIRIS & GTC & 1380\\
SN~2021hpr	 & 59585 & 2022 Jan. 6  & +263 & 3622--10,380 & Kast & Shane & 3600\\
SN~2021hpr   & 59610 & 2022 Jan. 31 & +288 & 3166--10,275 & LRIS & Keck I & 1200\\ 
SN~2021wuf	 & 59670 & 2022 Apr. 1  & +208 & 3142--10,272 & LRIS & Keck I & 1200\\
SN~2021wuf   & 59761 & 2022 July 1  & +299 & 3138--10,259 & LRIS & Keck I & 2400\\
SN~2022hrs   & 59995 & 2023 Feb. 20 & +297 & 3634--10,752 & Kast & Shane & 3660\\
SN~2022hrs   & 60025 & 2023 Mar. 22 & +327 & 4420--9626  & DEIMOS & Keck II & 1500\\

\hline
\end{tabular}%
  \label{tab:new}%
\end{table*}

\section{Fitting Method} \label{sec:method} 

\subsection{Multicomponent Gaussian Fit \label{subsec:fitting}}
To derive the Ni/Fe ratio in an SN~Ia explosion, we focus on the 7300~{\AA} region in the nebular-phase spectra, where the emission features are dominated by [Fe~II] (7155, 7172, 7388, 7453~{\AA}) and [Ni~II] (7378, 7412~{\AA}) \citep{2018MNRAS.477.3567M}. The [Fe~II] features mainly come from $^{56}$Fe which results from the decay of $^{56}$Co, while the [Ni~II] features are mostly contributed by stable $^{58}$Ni, since radioactive $^{56}$Ni has a half-life of only 6.1~days and its contribution is negligible at late times. 

To measure accurately the velocity shift, width, and strength of the above features, a multicomponent Gaussian function is used to fit the line profiles following \citet{2018MNRAS.477.3567M},

\begin{equation}
f(\lambda)=\sum_{i=1}^6 A_i {\rm exp}(-\frac{(\lambda-\lambda_i^{\rm centre})^2}{2\sigma_i^2})\, ,
\label{eq:Gaussian}
\end{equation}
where $A_i$ is the strength of the emission feature, $\lambda_i^{\rm centre}$ is the central wavelength, and $\sigma_i$ represents the width. The subscript $i$ denotes different components. With these parameters, we can calculate the velocity shift ($v_i$), full width at half-maximum intensity ($\rm FWHM_i$), and flux ($F_i$) for different emission components:
\begin{equation}
v_i = (\frac{\lambda_i^{\rm centre})^2-(\lambda_i^{\rm rest})^2}{\lambda_i^{\rm centre})^2+(\lambda_i^{\rm rest})^2} c \, ,
\label{eq:velocity}
\end{equation}

\begin{equation}
{\rm FWHM_i} = \frac{(\sqrt{2\,{\rm ln}\, 2}\sigma_i+\lambda_i^{\rm rest})^2-(\lambda_i^{\rm rest})^2}{(\sqrt{2\,{\rm ln}\, 2}\sigma_i+\lambda_i^{\rm rest})^2+(\lambda_i^{\rm rest})^2} 2c \, ,
\label{eq:FWHM}
\end{equation}

\begin{equation}
{\rm and}~ F_i = \sqrt{2\pi}A_i\sigma_i \, ,
\label{eq:Flux}
\end{equation}
where $\lambda_i^{\rm rest}$ is the rest-frame wavelength of the corresponding emission component and $c$ is the speed of light, \textcolor{black}{and we use the relativistic Doppler formula to convert wavelengths to velocities.} The Ni/Fe ratio is inferred from the flux ratio of [Ni~II] 7378~{\AA} and [Fe~II] 7155~{\AA}, $F_{7378}/F_{7155}$; more details are presented in Section~\ref{subsec:ratio}. 

Before fitting the Ni and Fe emission features, each spectrum \textcolor{black}{is smoothed with a Savitsky-Golay filter of width $\sim 20$--50~{\AA} and order 1 using the {\tt scipy} package’s $\rm signal.savgol\_filter$ function.} The pseudocontinuum is defined as a straight line connecting the interactively chosen points to the red and blue of the features in the 7300~{\AA} region. We notice that the red side of the features near 7600~{\AA} were often affected by telluric absorption lines, so we choose a region longward of this (i.e., at $\sim 7700$~{\AA}) as the red continuum point when such absorption appears in the spectrum, and we mask the absorption before fitting. \textcolor{black}{After subtracting the pseudocontinuum, we apply a multicomponent Gaussian function to fit the observed features.} Following \citet{2018MNRAS.477.3567M}, line widths and velocity shifts of the same species are set to the same values, and the relative strengths of the same species are fixed as well. Following \citet{2015A&A...573A..12J}, the strengths of the [Fe~II] 7172, 7388, and 7453~{\AA} features relative to [Fe~II] 7155~{\AA} are set to 0.24, 0.19, and 0.31, respectively, while the [Ni~II] 7412~{\AA} feature has a relative strength of 0.31 compared to [Ni~II] 7378~{\AA}. Thus, there are only six free parameters for the multicomponent Gaussian function: width, velocity shift, and strength of each of [Fe~II] 7155~{\AA} and [Ni~II] 7378~{\AA}. Similar to \citet{2022MNRAS.511.3682G}, we also place the boundary of the nickel width to be $\lesssim 13,000$~km~s$^{-1}$. 

\textcolor{black}{It should be noted that possible [Ca~II] emission at 7291.5 and 7323.9~{\AA} could contaminate the late-time spectra. We try to add two additional Gaussian components to represent the weak [Ca~II] emission lines in the fit, where the velocity shifts of the [Fe~II] and [Ni~II] are set to the same values; otherwise the [Ca~II] could dominate the blue or red peak. We find that the influence of the [Ca~II] emission on our fitting is very limited for most objects except the SN~1991bg-like SNe~Ia such as SN~1986G and SN~2003gs, \textcolor{black}{which is consistent with previous work (e.g., \citealt{2017MNRAS.472.3437G, 2018MNRAS.477.3567M, 2020MNRAS.491.2902F, 2022ApJ...926L..25T}).} In fact, the spectra of SN~1986G and SN~2003gs cannot be fit well by only [Fe~II] and [Ni~II], so we fit them by adding the [Ca~II] emission lines.} 

To estimate the uncertainties, \textcolor{black}{the endpoints of the continuum vary within 10--50~{\AA} (smaller variation ranges for lower-quality spectra to avoid improperly determining of the continuum;} if some regions are masked, their edges are varied within 10~{\AA}) and the emission features at 7100--7400~\AA\ are refit 1000 times through the Monte Carlo
method. A fit is rejected when the FWHM of nickel emission features is $> 12,500$~km~s$^{-1}$. The standard deviation in the measurement is taken as one part of the uncertainty of the Gaussian parameters. Additional uncertainty comes from smoothing the spectrum, which is taken as the difference between the results given by the smoothed spectrum and the observed one. Following \citet{2018MNRAS.477.3567M}, an uncertainty of 200 $\rm km\ s^{-1}$ is added to the velocity-shift measurement to account for the peculiar velocities of host galaxies.

\citet{2022MNRAS.511.3682G} also adopted the analysis method of \citet{2018MNRAS.477.3567M} to fit the 7300~{\AA} region, finding that the nickel line profile can be unusually broad. Thus, they proposed a two-stage {\it minimum-Ni} fit to further suppress the nickel contribution, 
which led to a lower Ni/Fe ratio. \citet{2020MNRAS.491.2902F} used a non-local-thermodynamic-equilibrium (NLTE) level population model of the first and second ionisation stages of iron, nickel, and cobalt to fit the full late-time spectra for their SN~Ia sample. The Ni/Fe ratios derived by them are lower than those obtained by \citet{2018MNRAS.477.3567M} for the same objects and the deviations were attributed to the placement of the pseudocontinuum. Although in this work we mainly follow the fitting method of \citet{2018MNRAS.477.3567M}, we try to give a more reasonable placement of the pseudocontinuum to make the Ni/Fe ratio more accurate; see Section~\ref{subsec:continuum}.

\subsection{Selection of Pseudocontinuum \label{subsec:continuum}}
\begin{figure*}
    \centering
    \includegraphics[width=1\columnwidth]{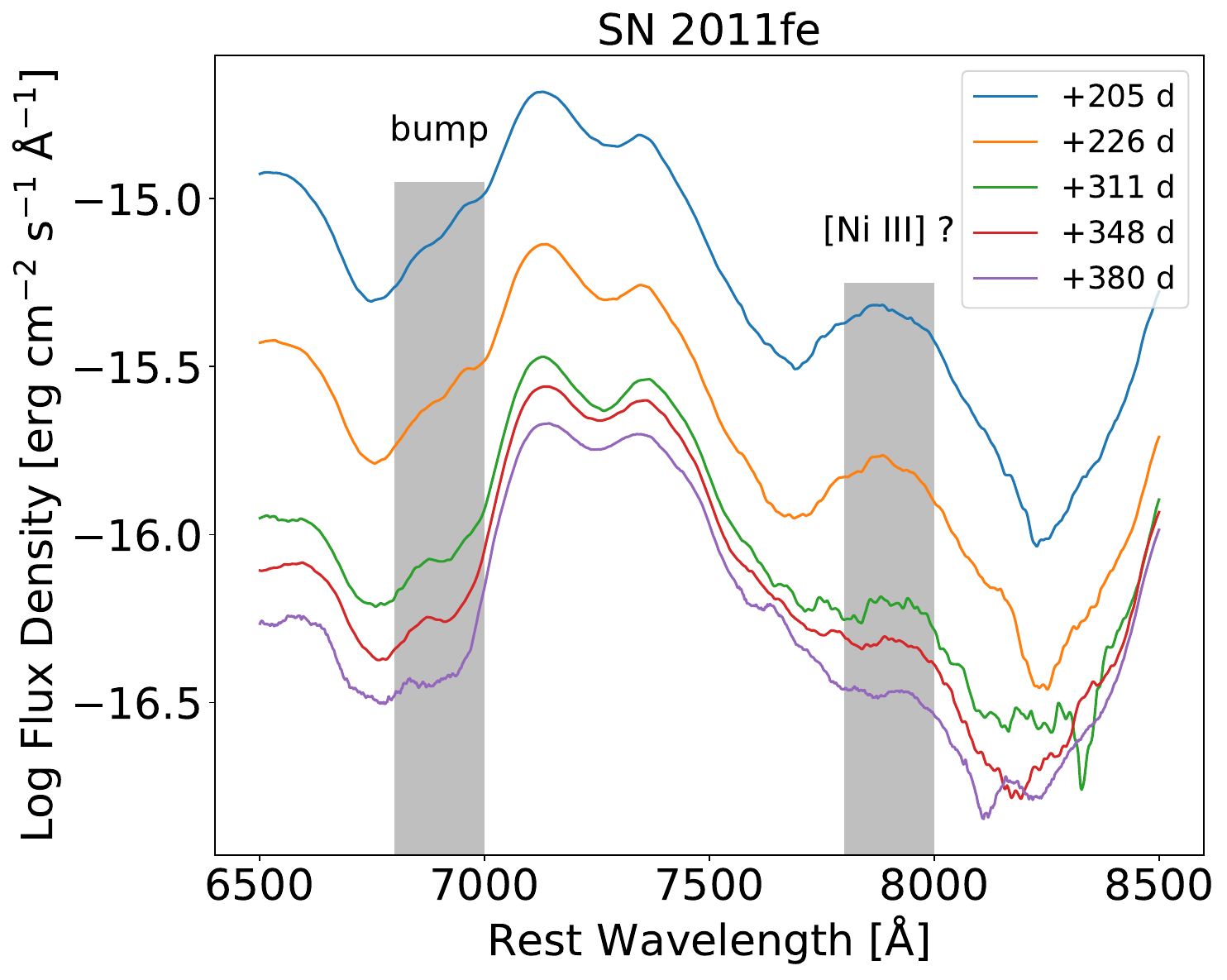}
    \includegraphics[width=1\columnwidth]{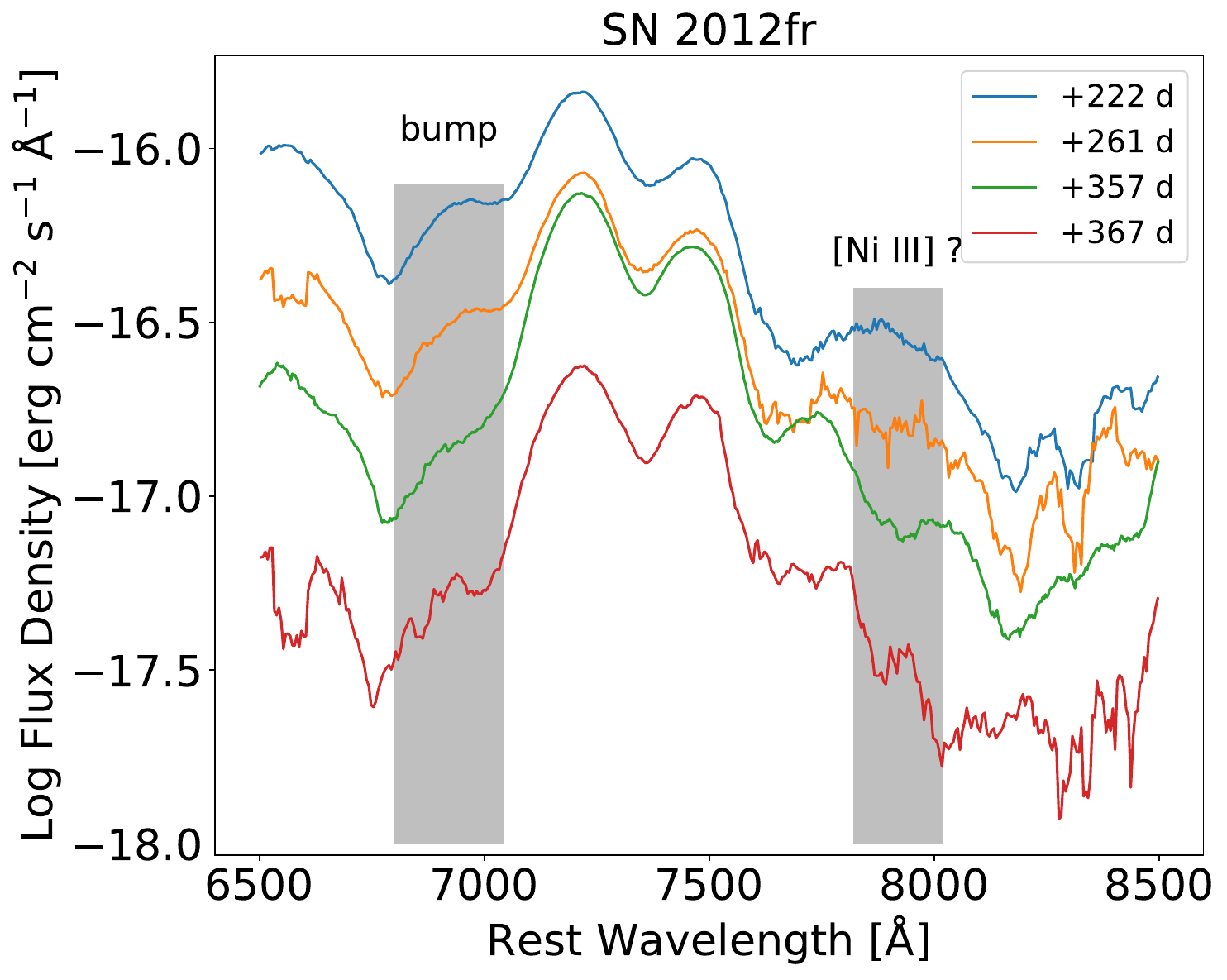}
    \includegraphics[width=1\columnwidth]{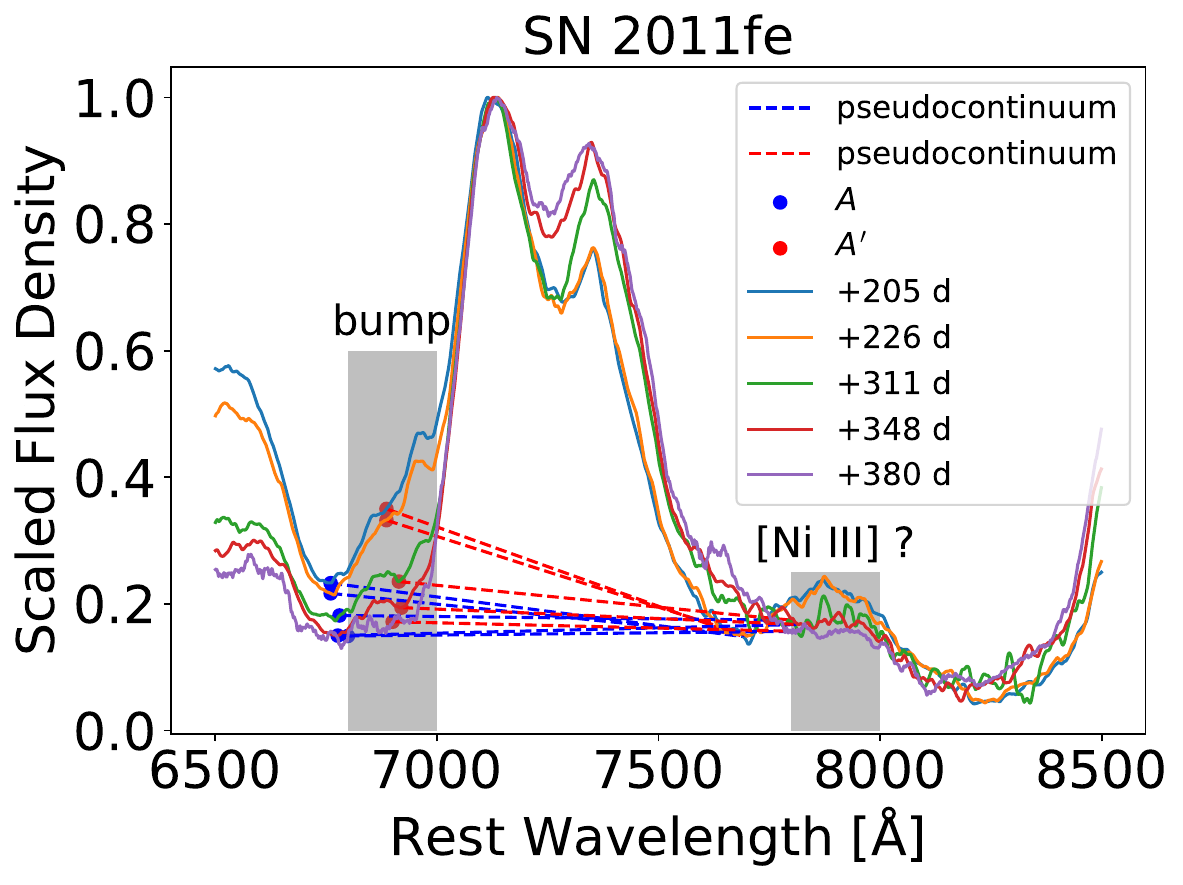}
    \includegraphics[width=1\columnwidth]{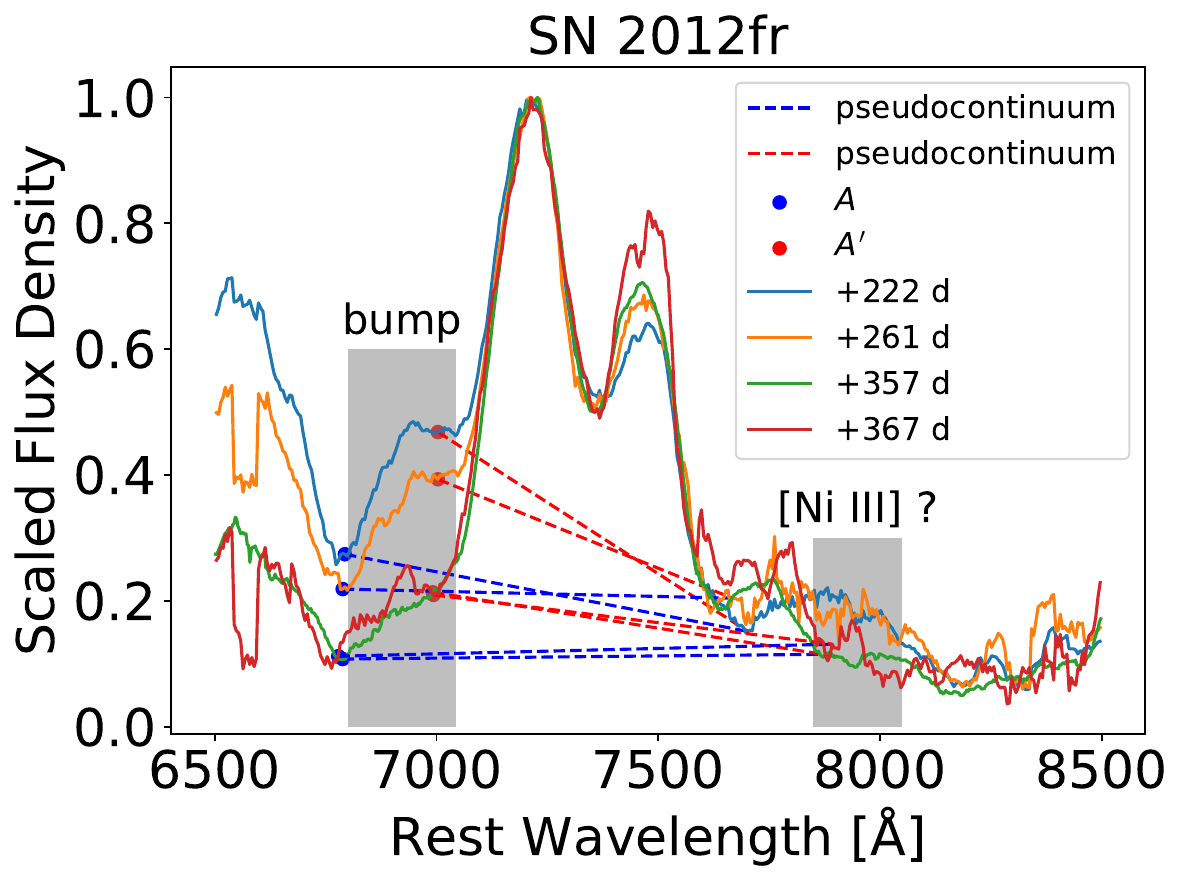}
    \includegraphics[width=1\columnwidth]{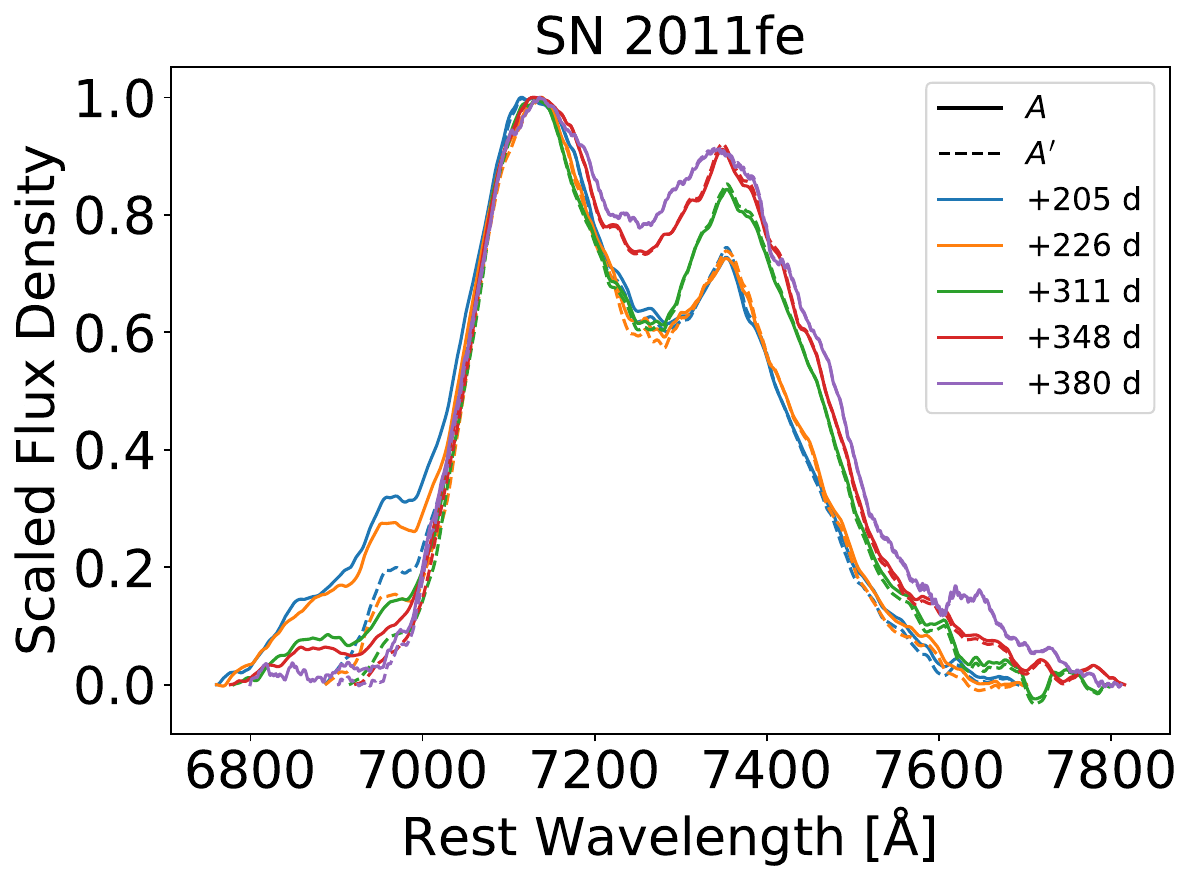}
    \includegraphics[width=1\columnwidth]{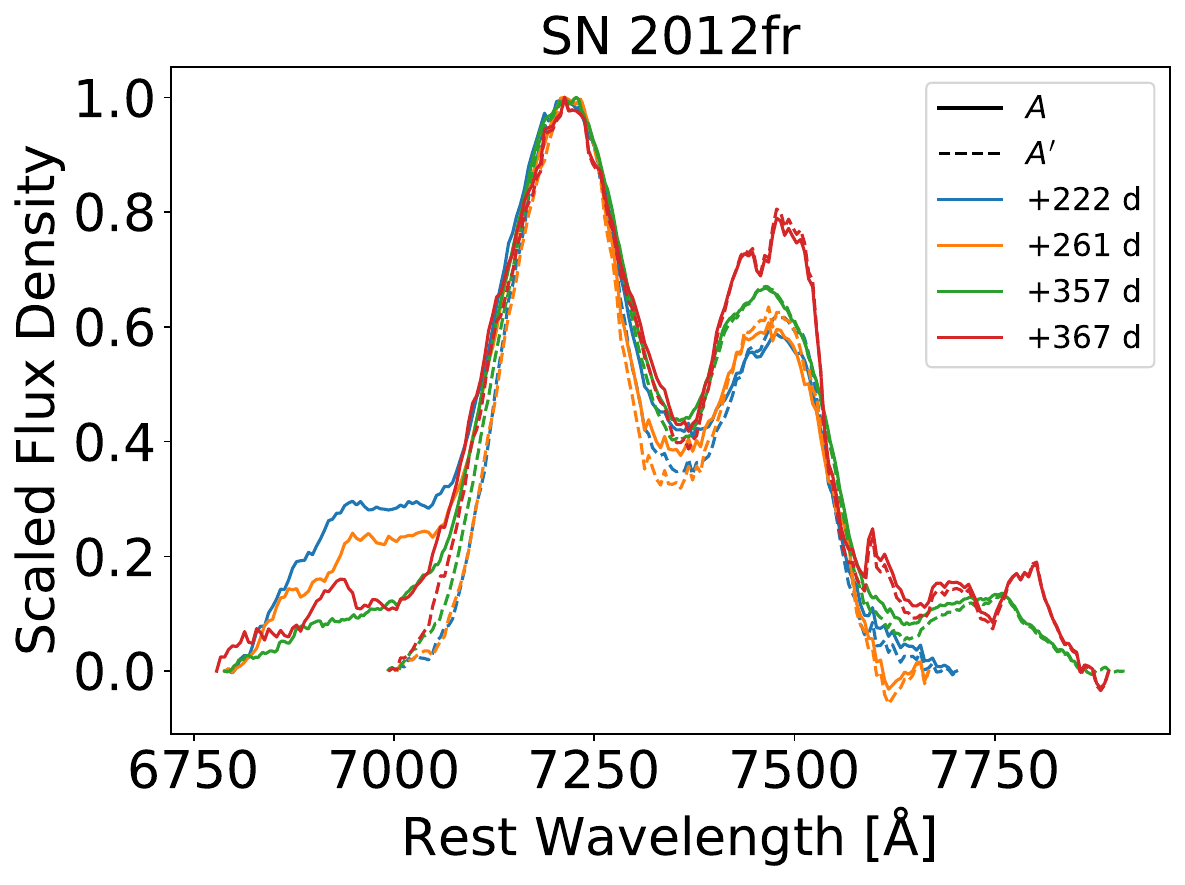}
    \caption{Evolution of the 7300~{\AA} region shown for late-time spectra of SN~2011fe and SN~2012fr. \textcolor{black}{All spectra have been smoothed with a Savitsky-Golay filter of width $\sim 50$~\AA\ and order 1. {\it Upper panel:} Evolution of unscaled spectra. {\it Middle panel:} Evolution of spectra scaled to the 7155~\AA\ peak.} The pseudocontinuum is defined as a straight line connecting the red and blue endpoints of the fitting region. The pseudocontinuum with point $A'$ or point $A$ as the blue endpoint are shown with red and blue dashed lines, respectively. The bump near 6900~{\AA} affecting the definition of the pseudocontinuum \textcolor{black}{and the region near 7900~{\AA} which might be contaminated by [Ni~III] are} labeled by the shaded region. \textcolor{black}{{\it Bottom panel:} Evolution of spectra from which the pseudocontinuum defined in the middle panel has been subtracted. The spectra obtained by subtracting the pseudocontinuum defined by $A$ or $A'$ are shown as solid and dashed, respectively.}}
    \label{fig:excess}
\end{figure*}

We notice that there exists a small bump at 6800--7000~{\AA} in most spectra of our sample, which tends to disappear at later phases and might be contributed by Co lines such as [Co~III] $\lambda$6855 \citep{2020MNRAS.491.2902F}. In the analysis by \citet{2018MNRAS.477.3567M}, the blue end of the continuum for the emission features near 7100--7400~{\AA} was chosen as the point around 6900~{\AA} (called point $A'$), which would induce a steeper pseudocontinuum. In our analysis, we prefer to choose the minimum in the blue wing of the bump as the blue end of the continuum (called point $A$) to place a flatter pseudocontinuum. Fig.~\ref{fig:excess} shows a comparison of these two placements for spectra of SN~2011fe and SN~2012fr, two well-observed nearby SNe~Ia (e.g., \citealt{2016ApJ...820...67Z,2015MNRAS.454.3816C}). The pseudocontinuum defined with point $A'$ is steeper than that defined with point $A$, significantly affecting the fitting results. In the following analysis, we call these two cases of choosing the blue side of the continuum as case $A$ and case $A'$, respectively.

\textcolor{black}{We mask} the bump feature at 6800--7000~{\AA} before fitting. However, this bump may be blended with iron emission [Fe~II] 7155~{\AA}. To roughly estimate the influence on the fitting, we \textcolor{black}{try to} use another Gaussian component to fit the bump feature near 6900~{\AA} instead of masking it. We find that the $F_{7378}/F_{7155}$ ratio derived by these two methods can differ by about 15\% when using the $t \approx +222$~day spectrum of SN~2012fr. Such a discrepancy becomes less significant with time, and it nearly disappears in the $t \approx +357$~day spectrum of SN~2012fr. Assuming that the bump feature near 6900~{\AA}\ is indeed mainly contributed by [Co~III] 6855~{\AA}, it has a limited effect on the iron feature unless the [Co~III] emission has an unreasonably broad width.

\begin{figure}
    \centering
    \includegraphics[width=0.95\columnwidth]{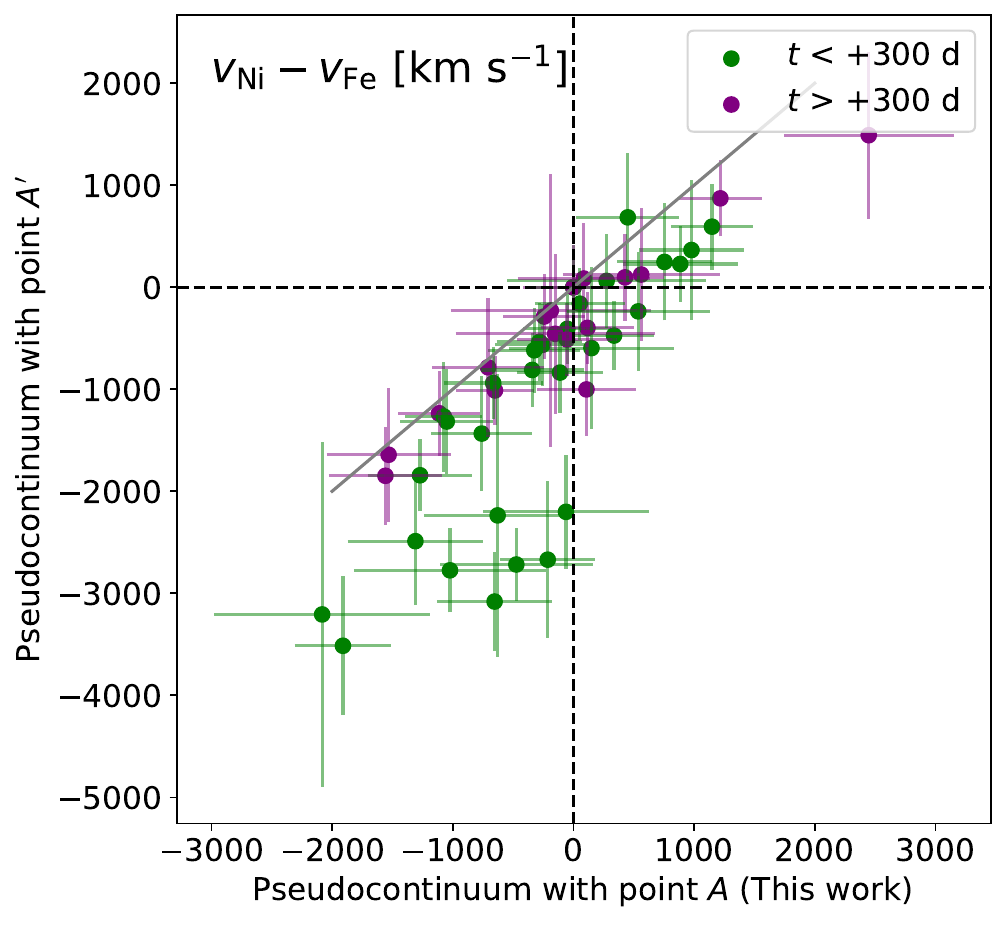}
    \caption{Comparisons of the velocity difference of [Ni~II] 7378~{\AA} and [Fe~II] 7155~{\AA} between case $A$ and case $A'$. The velocity difference inferred from the spectra taken at $t < +300$~days are shown with green dots while those taken at $t > +300$~days are shown with purple dots.}
\label{fig:cpr_fit_v}
\end{figure}

An important reason for us to choose case $A$ is that it gives a smaller systematic error for the estimation of feature velocities. Assuming that velocities of [Fe~II] 7155~{\AA} and [Ni~II] 7378~{\AA} are similar \citep{2010ApJ...708.1703M}, the velocity difference of these two features should be close to zero or at least has a symmetric distribution with respect to zero for our sample. Fig.~\ref{fig:cpr_fit_v} compares the velocity fitting results between case $A$ and case $A'$. We find that the velocity difference of case $A'$ with a mean value of $\sim -900$~km~s$^{-1}$ tends to be bluer --- \textcolor{black}{that is,} the nickel velocity tends to be bluer than the iron velocity. This bias is more serious for the spectra taken at $t < +300$~days. Meanwhile, the velocity difference distribution of case $A$ is consistent with our expectations, with a mean value of $\sim -200$~km~s$^{-1}$,  close to zero.

Another problem of case $A'$ is the overly broad nickel component in some fitting cases, which is also mentioned by \citet{2022MNRAS.511.3682G}. We find that, in most cases, the unreasonably broad nickel component can be avoided by using point $A$ to set the pseudocontinuum and adjusting the masking regions. When the adjustment doesn't work (only for the spectrum of SN~2017cbv at +317 days in our sample), we assume that the velocity shifts of [Fe~II] and [Ni~II] are the same, which can also avoid the overly broad nickel.

\begin{figure*}
    \centering
    \includegraphics[width=2.05\columnwidth]{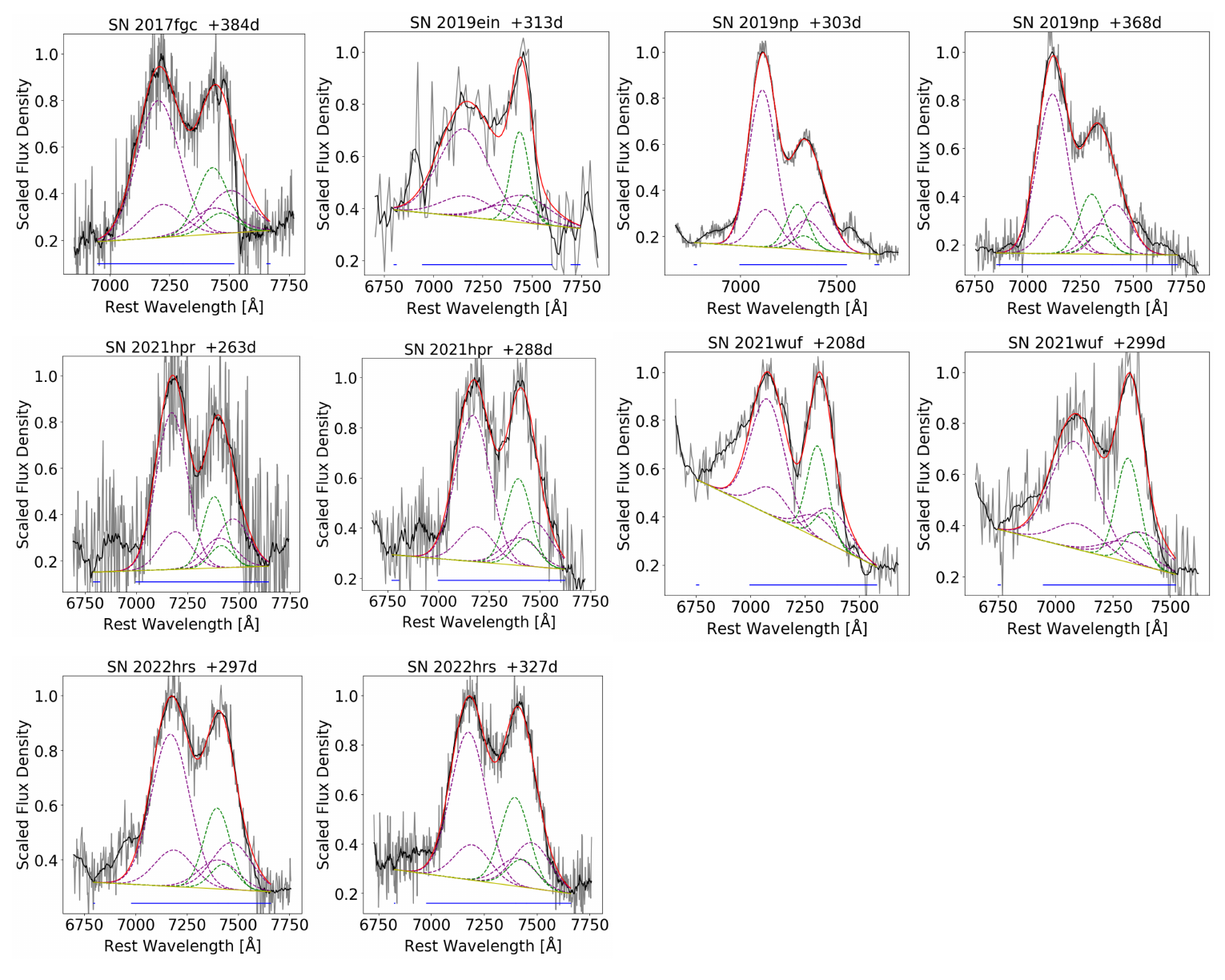} 
    \caption{Best fits to the 7300~{\AA} region dominated by [Fe~II] and [Ni~II] features. The reddening-corrected spectra are shown in grey while the smoothed spectra are in black. The overall fits are shown with a red line, while the fitting regions are indicated by blue lines. The [Fe~II] and [Ni~II] features are represented by purple and green dashed lines, respectively.}
    \label{fig:new_spectra_fits}
\end{figure*}

\textcolor{black}{Fig.~\ref{fig:new_spectra_fits} shows the best fit to the 7300~{\AA} region for the late-time spectra of SNe 2017fgc, 2019ein, 2019np, 2021hpr, 2021wuf, and 2022hrs.}    

\section{Results} \label{sec:results}

The velocity shifts and line widths (i.e., FWHM) of the [Fe~II] 7155~{\AA} and [Ni~II] 7378~{\AA} features as well as their flux ratios are listed in Table~\ref{tab:Multi}. \textcolor{black}{Significant differences are found to exist in the velocity shifts and FWHM between the [Fe~II] 7155~{\AA} and [Ni~II] 7378~{\AA} features for some objects, as discussed in \ref{subsec:N_v}}. We use the velocity shifts of [Fe~II] 7155~{\AA} and [Ni~II] 7378~{\AA} to identify whether the nebular velocities are redshifted or blueshifted, which might have a connection with asymmetric explosion models \citep{2010Natur.466...82M}. We then use the flux ratio of [Fe~II] 7155~{\AA} and [Ni~II] 7378~{\AA} to roughly estimate the Ni/Fe ratio and constrain the explosion models. The fits for all the spectra used in this work are shown in Fig~\ref{fig:all_fits}. The inferred nebular velocities and Ni/Fe ratios are also listed in Table~\ref{tab:Multi}. Some potential correlations can be found between nebular velocities, Ni/Fe ratios, Si~II velocities at maximum light, and post-maximum decline rates $\Delta m_{15}(B)$. 

\subsection{\textcolor{black}{Velocities and Widths of [Fe~II] and [Ni~II] features} \label{subsec:N_v}}

\textcolor{black}{Nebular velocities were often estimated from the mean velocity shift inferred from the [Fe~II] 7155~{\AA} and [Ni~II] 7378~{\AA} emission features (e.g., \citealt{2010Natur.466...82M, 2013MNRAS.430.1030S}). However, we note that the velocities inferred from [Fe~II] and [Ni~II] lines tend to show noticeable differences (e.g., over 1000\ km\ s$^{-1}$) when their emission peaks are not of comparable strength. The velocity inferred from the weaker feature usually has a larger uncertainty relative to the stronger one. To quantify the difference between these two velocities, we apply a simple linear fit and find that $v_{\rm Ni} = 1.24\,v_{\rm Fe} - 170$ km s$^{-1}$. For an absolute mean velocity of $v_{\rm Fe}=1260$ km s$^{-1}$, the difference in $v_{\rm Ni}$ and $v_{\rm Fe}$ is within about 500 km s$^{-1}$, which is not significant. As the [Fe~II] 7155~\AA\ feature is usually stronger and has smaller measurement uncertainties in our sample, we adopt the [Fe~II] velocity to represent the nebular velocity except when the [Ni~II] 7378~\AA\ feature is stronger. In these cases, we use the [Ni~II] velocity to represent the nebular velocity (e.g., for SN~2003hv).}

\begin{figure*} 
    \centering
    \includegraphics[width=2.05\columnwidth]{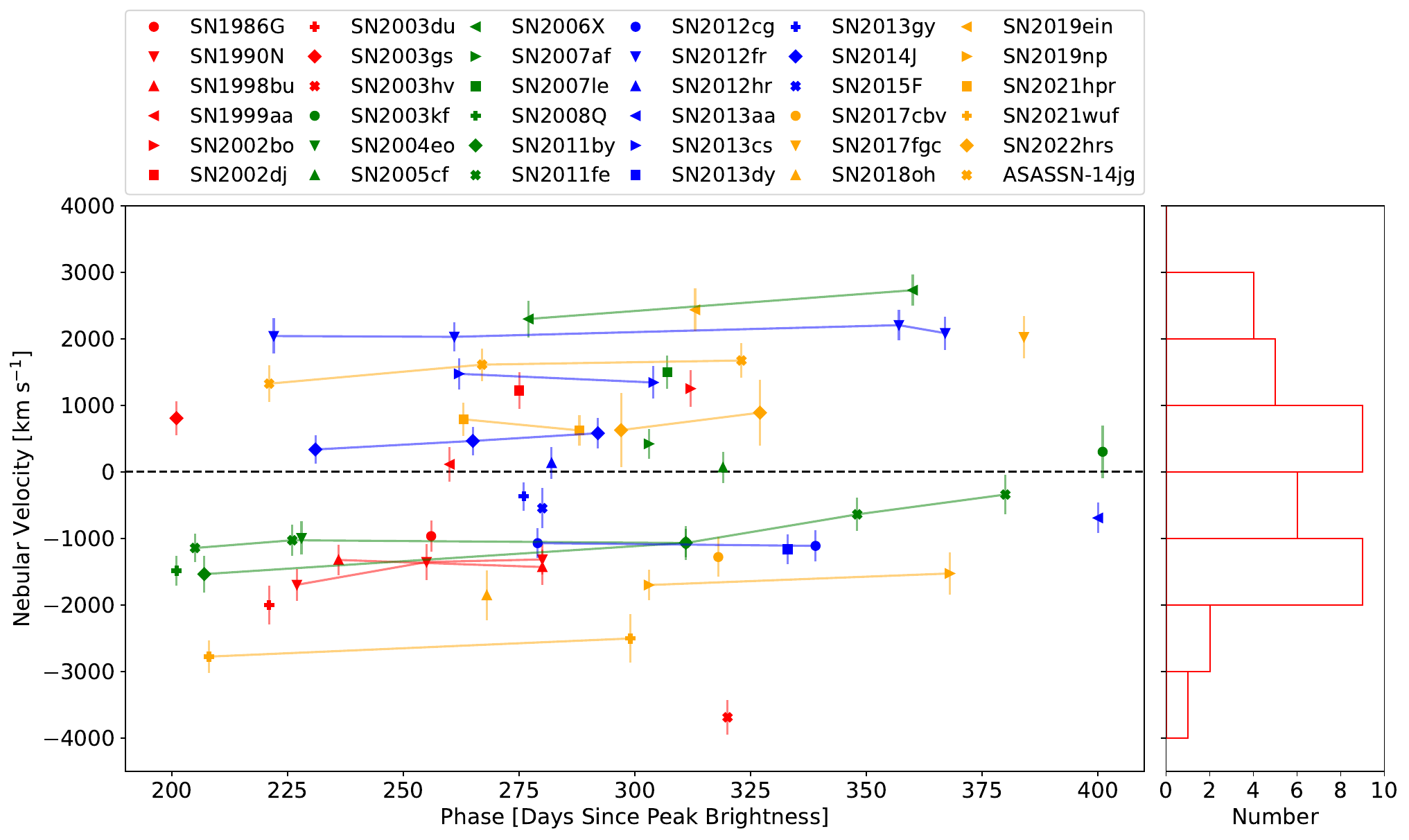}
    \caption{\textcolor{black}{{\it Left panel:} Nebular velocities inferred from the [Fe~II] 7155~{\AA} or [Ni~II] 7378~{\AA} features as a function of time. {\it Right panel:} The corresponding histogram for average nebular velocity measured for our sample. 
    }} 
    \label{fig:n_p}{}
\end{figure*}

Fig.~\ref{fig:n_p} shows the nebular velocities inferred from the [Fe~II] 7155~{\AA} \textcolor{black}{or} [Ni~II] 7378~{\AA} features in spectra taken at different phases. As can be seen, the nebular velocity does not show significant evolution in a given object. When the nebular velocity is close to zero within the quoted uncertainty, we consider its nebular velocity as zero instead of identifying it as a redshifted or blueshifted subclass. Among the 36 SNe~Ia collected in our study, we find that 14 can be classified as having redshifted nebular velocities and 18 as having blueshifted ones. The numbers of redshifted and blueshifted objects are comparable, and the nebular velocities have a roughly symmetric distribution in the approximate range $-3000$ to 3000~km~s$^{-1}$.

\textcolor{black}{For the line widths, there is considerable scatter between the FWHM of [Fe~II] and [Ni~II] features. The [Fe~II] FWHM is usually larger than the [Ni~II] FWHM, with mean values of $8429 \pm 961$ km s$^{-1}$ and $6078 \pm 1445$ km s$^{-1}$ (respectively) --- approximately consistent with the DDT models of \citet{2013MNRAS.429.1156S} where $^{56}$Ni usually has a slightly wider distribution than $^{58}$Ni (see their Fig. 8). But in some extreme cases, the [Fe~II] FWHM is much larger than the [Ni~II] FWHM (e.g., more than a factor of two), not expected from the models. Note that the difference between [Fe~II] and [Ni~II] FWHM evolves with phase, with a tendency of being larger at earlier nebular phases (e.g., ASASSN-14jg), which can partially explain the large FWHM differences as revealed in Table A2. Nevertheless, inspecting the measurements for those spectra taken at similar phases (i.e., $t \approx 250$~days), we still find large FWHM scatter between [Ni~II] and  [Fe~II] lines, suggesting that this scatter might be genuine and its origin should be further explored.}

\subsection{Estimation of Ni/Fe Ratio \label{subsec:ratio}}
The Ni/Fe ratio is estimated following the method of \citet{2018MNRAS.477.3567M},

\begin{equation}
\frac{n_{\rm Ni~II}}{n_{\rm Fe~II}\ }=\frac{L_{7378}}{L_{7155}}\exp{\left(-\frac{0.28}{kT}\right)}\frac{d_{C_{\rm Fe~II}}}{d_{C_{\rm Ni~II}}}/4.9 \, ,
\label{eq:Ni_Fe_ratio}
\end{equation}
where $d_{C_{\rm Fe~II}}/d_{C_{\rm Ni~II}}$ denotes the ratio of departure coefficients, $k$ is the Boltzmann constant \textcolor{black}{in units of eV per K}, and $T$ is the temperature. $L_{7378}/L_{7155}$ represents the luminosity ratio and is estimated as the measured flux ratio of [Fe~II] 7155~{\AA} and [Ni~II] 7378~{\AA}. Note that \cite{2018MNRAS.477.3567M} used the ratio of pseudoequivalent width of these two lines. 

To obtain a rough estimate of the Ni/Fe ratio, \citet{2018MNRAS.477.3567M} adopted the value of the ratio of departure coefficients, temperature, and ionisation balance from a day 330 W7 model of \citet{2015ApJ...814L...2F}. The ratio $d_{C_{\rm Fe~II}}/d_{C_{\rm Ni~II}}$ has a range of 1.2--2.4 over the phase of interest, with smaller coefficients at earlier times when the conditions are slightly closer to LTE. The temperature is assumed to range from 3000~K to 8000~K, while the Ni/Fe ratio $n_{\rm Ni}/n_{\rm Fe}$ is considered to be equal to $n_{\rm Ni~II}/n_{\rm Fe~II}$. We also use these estimates to infer the Ni/Fe ratio. Then the mass ratio of Ni and Fe is estimated using $^{56}$Fe and $^{58}$Ni. The relative uncertainty of the Ni/Fe ratio obtained in this way is about 40\% \citep{2018MNRAS.477.3567M}, and it is combined with that of the measured flux ratio to give a total uncertainty. 

Considering the decay of $^{56}$Co, the abundance of $^{56}$Fe will increase with time while the resulting Ni/Fe ratio will decrease accordingly. To avoid such an effect of temporal evolution, we scale the Ni/Fe ratio to $t \rightarrow \infty$ under the assumption that all of the $^{56}$Fe comes from the final decay product of $^{56}$Ni; the contribution from other stable irons is negligible.

\begin{figure*} 
    \centering
    \includegraphics[width=2.05\columnwidth]{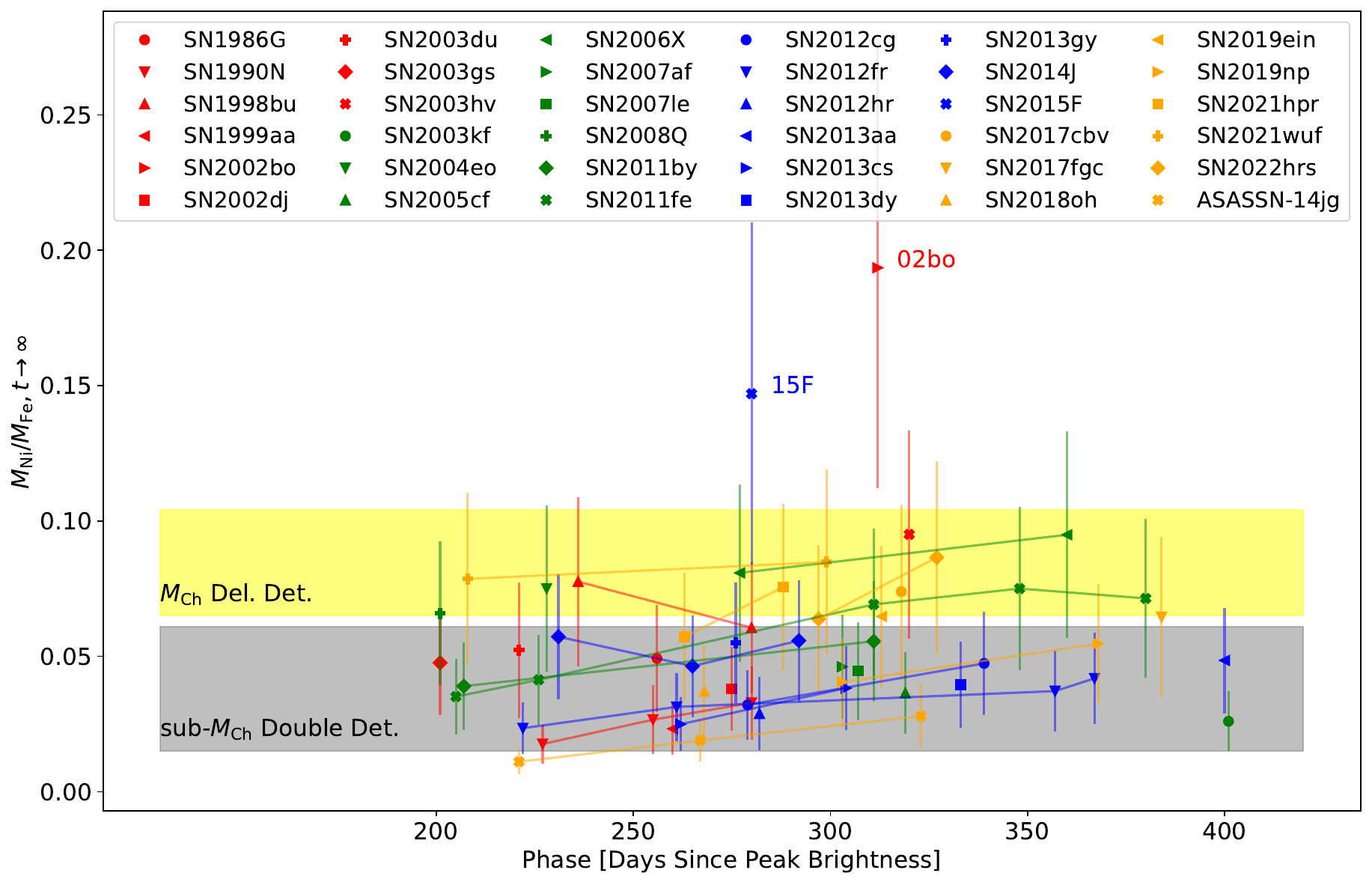}
    \caption{The inferred mass ratio of Ni and Fe as a function of phase after maximum light for SNe~Ia collected in this work. The large error bars are mainly due to the $\sim 40$\% relative uncertainty of the estimate of \citet{2018MNRAS.477.3567M} that we adopt. Following \citet{2020MNRAS.491.2902F},  the Ni/Fe ratio range predicted for the DDT models \citep{2013MNRAS.429.1156S} is shown by the yellow band and that of the sub-$M_{\rm Ch}$ models \citep{2018ApJ...854...52S} is represented by the grey band. Both the Ni/Fe ratios measured from the observed spectra and those predicted by the explosion models are scaled to $t \rightarrow \infty$ by assuming a rise time of $\sim 18$~d \citep{2011MNRAS.416.2607G}.} 
    \label{fig:r_p}{}
\end{figure*}

Fig.~\ref{fig:r_p} shows the estimated mass ratio of Ni and Fe as a function of phase since maximum light. We notice that the Ni/Fe ratio inferred from nebular-phase spectra might increase with time, as seen in SN~2011fe, which is also discussed in Section~\ref{subsec:evolution}. For those SNe~Ia with multiple late-time spectra, we usually adopt the Ni/Fe ratios derived from the latest ones. The final results shown in Fig.~\ref{fig:r_p} indicate that the inferred Ni/Fe ratios for about $70 \pm 6\%$ of our SN~Ia sample favour the double-detonation sub-$M_{\rm Ch}$ model. Here the proportion uncertainty is simply estimated by using a Monte Carlo method to vary the Ni/Fe ratios according to their uncertainty. This proportion is higher than that derived by \citet{2018MNRAS.477.3567M}, mainly because of the different pseudocontinuum adoption in the fitting. 
The remaining objects are in better agreement with delayed-detonation models \citep{2013MNRAS.429.1156S} or have a critical Ni/Fe ratio lying between the above two models, except for SN~2015F and SN~2002bo which have too high Ni/Fe ratios. \citet{2022MNRAS.tmp.1645T} use late-time light curves to constrain the mass of SN~Ia progenitors and find that SN~2015F is only consistent with the sub-$M_{\rm Ch}$ model, which suggests that its Ni/Fe ratio should be small. In addition, we notice that the shapes of the 7300~{\AA} feature are peculiar and cannot be well fit, so we exclude SN~2015F from the following analysis. For SN~2002bo, the red half of the 7300~{\AA} feature seems much stronger than the blue half, compared with other objects, which leads to a Ni/Fe ratio that is much higher than the upper limit given by the models involved in our analysis. Thus, we also exclude SN~2002bo from our analysis.

\subsection{Connection between Ni/Fe Ratios and Si Velocities \label{subsec:ratio_v}}
To find some potential correlations between Ni/Fe ratios and early-time observed properties, we use the $t \approx 300$~days spectra whenever possible, since the parameters that we adopt to infer the Ni/Fe ratio come from the W7 model at $t \approx 330$~days \citep{2015ApJ...814L...2F}. Note that $t \approx +300$~days is the median of the phase range ($\sim +200$--400~days) of our sample.  

\begin{figure*} 
    \centering
    \includegraphics[width=2.05\columnwidth]{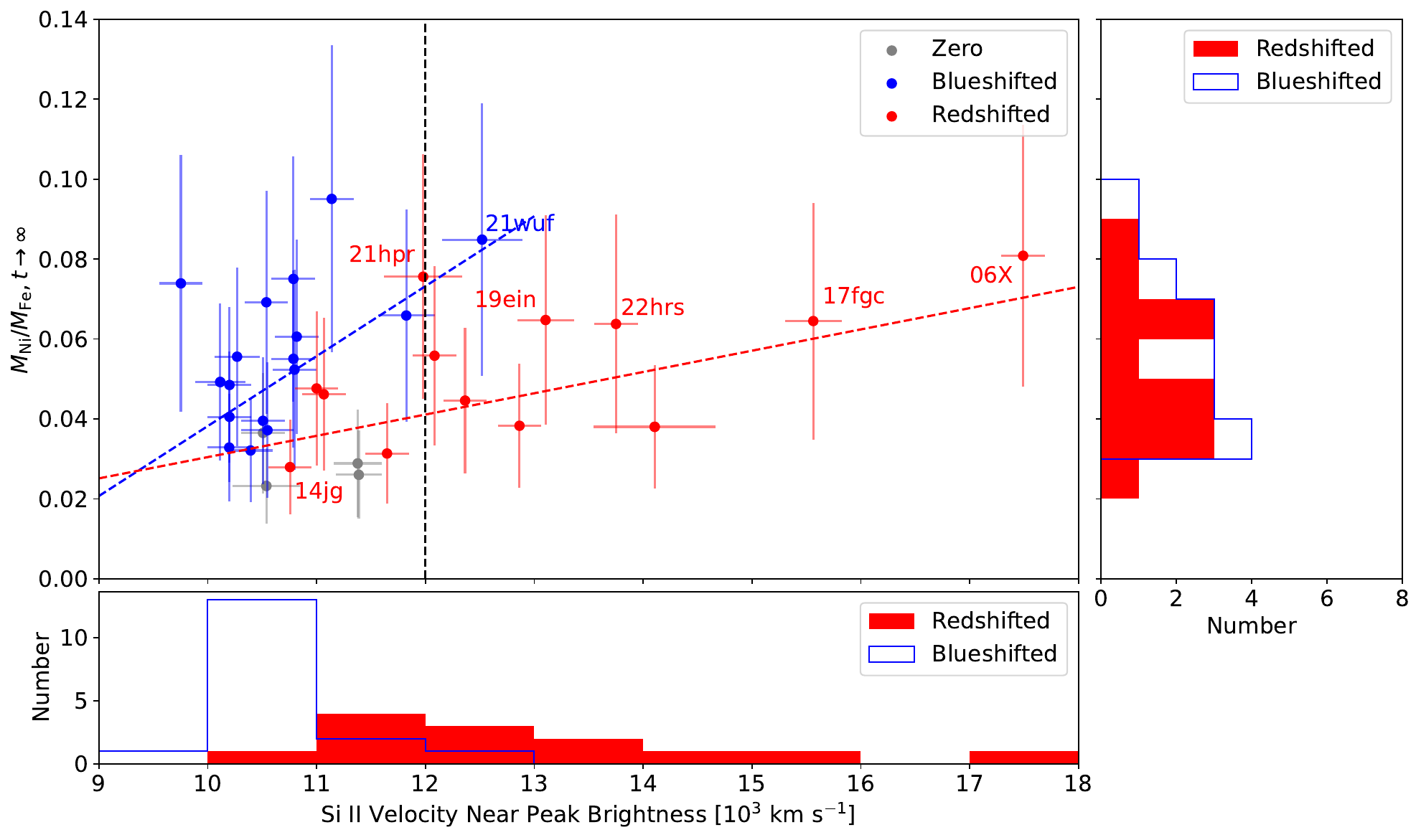}
    \caption{Inferred Ni/Fe mass ratio versus the Si~II 6355~{\AA} velocity measured around the time of maximum light. SNe~Ia that have redshifted or blueshifted nebular velocities are shown by red and blue circles, respectively, while those with nebular velocities close to zero are represented by grey dots. The red and blue dashed lines represent the best linear fits to the data points with redshifted and blueshifted nebular velocities, respectively. The horizontal and vertical histograms represent the distributions of Si velocities and Ni/Fe ratios.}
    \label{fig:R_Si}
\end{figure*}

The distribution of the inferred Ni/Fe mass ratios and the Si~II $\lambda$6355 velocities measured around the time of maximum light is shown in Fig.~\ref{fig:R_Si}, where one can see that the $M_{\rm Ni}/M_{\rm Fe}$ ratio tends to increase with Si~II velocity. Moreover, the subsamples with redshifted or blueshifted nebular velocities seem to follow distinct $M_{\rm Ni}/M_{\rm Fe}$--$v_{\rm Si~II}$ relations. To test the robustness of the above correlation, we further use the {\tt scipy} package’s stats.pearsonr function to calculate the Pearson coefficient. The correlations are found to be prominent for both the blueshifted and redshifted subsamples when not considering the error bars, with the $r$-values being \textcolor{black}{0.53 and 0.61}, and the $p$-values being \textcolor{black}{0.03 and 0.03}, respectively. \textcolor{black}{When combining the whole sample, the correlation seems to be less prominent, with the $r$-value being only 0.32, and the $p$-value being 0.07.} For the case including error bars, we assume a Gaussian distribution for the quoted uncertainties and repeat the calculation 20,001 times with a Monte Carlo method. If selecting the middle number of the Monte Carlo results, the significance of the correlations becomes more ambiguous, with the $r$-values being \textcolor{black}{0.32 and 0.38}, and the $p$-values being \textcolor{black}{0.21 and 0.19}, for the blueshifted and redshifted subsamples, respectively. Nevertheless, the connection is in agreement with some previous work and will be further discussed in Section~\ref{subsec:implications}.

From the histograms shown in Fig.~\ref{fig:R_Si}, we can see that all SNe~Ia with blueshifted Ni/Fe velocities in the nebular phase have Si velocities $< 12,000$~km~s$^{-1}$ except for SN~2021wuf, while those with redshifted nebular velocities have Si velocities ranging from 10,000~km~s$^{-1}$ to \textcolor{black}{$\sim 17,000$~km~s$^{-1}$}. The range of the Ni/Fe ratio is similar for these two subgroups. We also notice that when the Si velocity is near 10,000~km~s$^{-1}$, the two groups with blueshifted and redshifted Ni/Fe velocities tend to mix.   

\subsection{Connection between the Ni/Fe Ratio and $\Delta m_{15}(B)$ \label{subsec:ratio_15}}

\begin{figure} 
    \centering
    \includegraphics[width=0.95\columnwidth]{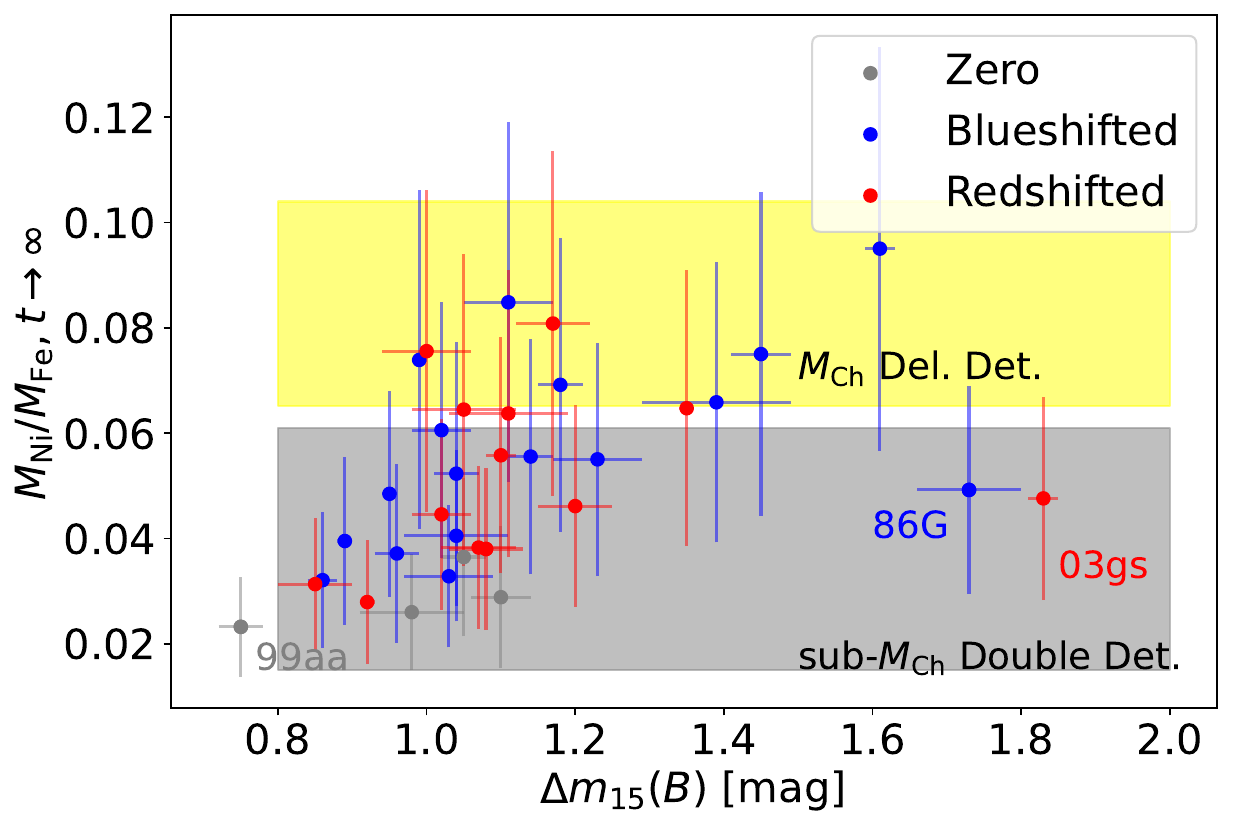}
    \caption{Inferred mass ratio of Ni and Fe versus the post-peak decline rate $\Delta m_{15}(B)$. The symbols are the same as in Fig.~\ref{fig:R_Si}. Yellow and grey regions represent the explosion models as indicated in Fig.~\ref{fig:r_p}.}
    \label{fig:R_15}
\end{figure}

Fig.~\ref{fig:R_15} shows the relation between the Ni/Fe mass ratio and $\Delta m_{15}(B)$. Note that SNe~Ia with blueshifted and redshifted nebular velocities tend to mix in the plot, in contrast to the distinct distribution seen in Fig.~\ref{fig:R_Si}. The Ni/Fe mass ratio seems to increase with $\Delta m_{15}(B)$, with fast decliners having larger Ni/Fe mass ratio. However, this correlation reverses for SNe~Ia with larger post-peak decline rates --- that is, the SNe~Ia with \textcolor{black}{$\Delta m_{15}(B) > 1.7$}~mag have small Ni/Fe mass \textcolor{black}{ratio.} \textcolor{black}{This }discrepancy is likely related to the intrinsic difference existing between the SN~1991bg-like (SN~1986G and SN~2003gs in Fig.~\ref{fig:R_15}) and other SNe~Ia. In addition, we caution about the low Ni/Fe ratio inferred for the overluminous SN~1991T-like SN~1999aa, which favours a sub-$M_{\rm Ch}$ explosion. Note that the sub-$M_{\rm Ch}$ detonation model of \citet{2018ApJ...854...52S} of a relatively massive WD (1.1\,M$_\odot$) shows lower luminosity than the overluminous SN~Ia; thus, it seems to be hard for the sub-$M_{\rm Ch}$ model to produce the high luminosity.

We calculate the Pearson's $r$-values for all of the sample shown in Fig.~\ref{fig:R_15}, except SN~1986G and SN~2003gs. We obtain an $r$-value of \textcolor{black}{0.68} and a $p$-value of \textcolor{black}{$2.18 \times 10^{-5}$} without considering the error bars, and an $r$-value of \textcolor{black}{0.43} and a $p$-value of \textcolor{black}{0.01} when including the error bars with the Monte Carlo method adopted in Section~\ref{subsec:ratio_v}. These results strongly suggest that the inferred Ni/Fe mass ratio has a positive correlation with $\Delta m_{15}(B)$.   

\section{Discussion} \label{sec:discussion}
In this paper, we applied multicomponent Gaussian fits to the spectral features near 7300~{\AA} for 36 SNe~Ia. Nebular velocities and Ni/Fe mass ratios are inferred from the velocity shifts and flux ratios, respectively. In this section, we further discuss the evolution of the inferred Ni/Fe ratio, and we explore possible connections between inferred late-time parameters and early-time parameters and their implications for explosion mechanisms.   

\subsection{\textcolor{black}{Evolution of the 7300~\AA\ region}} \label{subsec:evolution}

For the well-observed object SN~2011fe, we notice that the inferred Ni/Fe ratio shows temporal evolution, moving from the regions favouring the sub-$M_{\rm Ch}$ double-detonation model to that favouring the $M_{\rm Ch}$ delayed detonation model.  \textcolor{black}{The increasing trend is also found for other events having at least three spectra except for SN~2014J. Considering the uncertainties using a Monte Carlo method, the mean Pearson $r$-value derived for SNe 1990N, 2011fe, 2012fr, and ASASSN-14jg is $0.73\pm0.11$. Note that the variation inferred for the Ni/Fe ratio with phase only reflects the evolution of the flux ratio of [Ni~II] 7378~\AA\ and [Fe~II] 7155~\AA, since we use the same temperature and departure coefficients to infer the Ni/Fe ratio in all phases as \citet{2018MNRAS.477.3567M} did. It should be noted that the errors associated with the Ni/Fe ratio shown in Fig. 6 are dominated by uncertainties in temperature and departure coefficients ratio, which might be overestimated when analysing the evolution for the same object. This is due to the fact that the temperature usually decreases with time while the departure coefficients increase with time for the same object, and the net evolution effect of these two parameters on the inferred Ni/Fe ratio is thus limited.}

\textcolor{black}{Note that we use the same temperature and ratio of departure coefficients to infer the Ni/Fe ratio, hence the evolution of the Ni/Fe ratio is due to the evolution of the flux ratio of [Ni~II] 7378~{\AA} and [Fe~II] 7155~{\AA}.} This evolution is clearly indicated in Fig.~\ref{fig:excess}, where the [Ni~II] 7378~{\AA} feature tends to become increasingly prominent compared with the [Fe~II] 7155~{\AA} feature, which might be due to the suppression of the [Ni~II] lines as a result of higher ionisation of nickel in the inner ejecta \citep{2022A&A...660A..96B}. As the inner ejecta cool gradually, the flux ratio of [Ni~II] to [Ni~III] increases with time, leading to the increase of the inferred Ni/Fe ratio. \textcolor{black}{In fact, we also notice that the scaled flux of the 7900~{\AA} region decreases with time and this region could have a contribution from [Ni~III] 7890~{\AA}. Thus, the spectral evolution might indicate the recombination of [Ni~III], since the flux-density ratio of [Ni~II] 7378~{\AA} and the 7900~{\AA} region increases with time. Moreover, the high-quality spectra of SN~2011fe show that there exists a weak feature in the 7900~{\AA} region, which nearly disappears at later phases.}

\textcolor{black}{Another possible reason for the increase of the flux ratio of [Ni~II] 7378~{\AA} and [Fe~II] 7155~{\AA} is that the [Fe~II] 7155~{\AA} feature is contributed by the bump that gets weaker with time. Fig.~\ref{fig:excess} also shows the evolution of the spectra subtracting the pseudocontinuums defined by point $A'$, where the bump (likely related to some additional features that only contribute at early times) is subtracted. In this case, the trend that the flux ratio of [Ni~II] 7378~\AA\ and [Fe~II] 7155~\AA\ increase with time is still visible. But if the bump contributes little to the [Ni~II] 7378~{\AA} feature, the strength of the [Ni~II] 7378~{\AA} feature will be underestimated when subtracting the pseudocontinuums defined by point $A'$. A more careful fitting that includes some new emission lines related to the bump is needed to confirm the evolution, which is more complicated and beyond the scope of this paper.} 

\begin{figure}
    \centering
    \includegraphics[width=0.95\columnwidth]{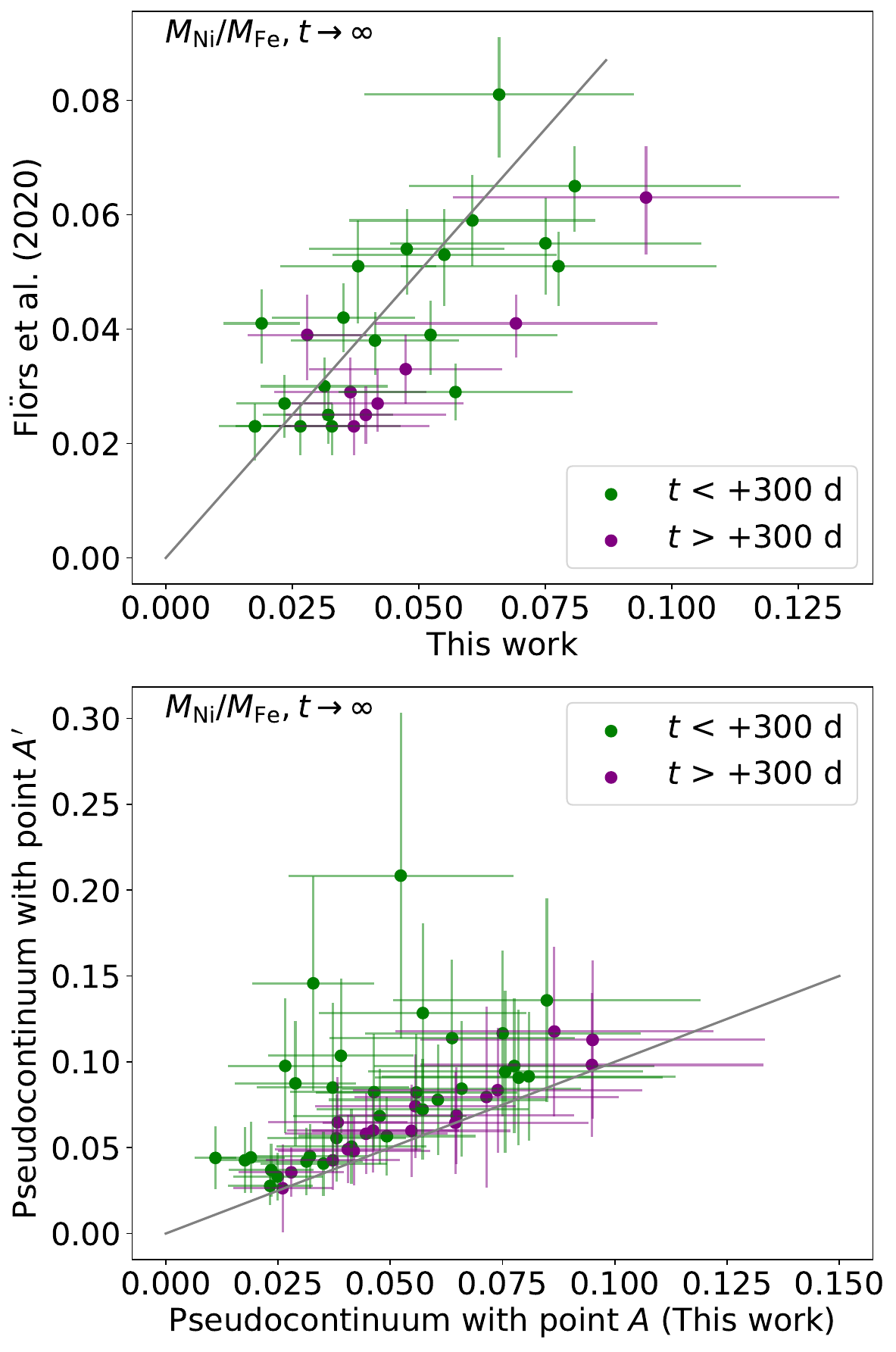}
    \caption{Comparisons of the Ni/Fe ratios given by different methods. The top panel compares our fits with those from the NLTE model of \citet{2020MNRAS.491.2902F} for the same spectra. The labeled objects are those whose Ni/Fe ratios deviate by more than 1$\sigma$. The bottom panel compares the fits using point $A$, which defines a similar pseudocontinuum to that of \citet{2018MNRAS.477.3567M}, with the fits using point $A'$, which is adopted in this work. The Ni/Fe ratios inferred from the spectra taken at $t < +300$~days are shown with green dots while those taken at $t > +300$~days are shown with purple dots.}
    \label{fig:cpr_ratio}
\end{figure}

\citet{2020MNRAS.491.2902F} used an NLTE level population model to fit the full late-time spectra for their SN~Ia sample and found no evident evolution for Ni/Fe ratios for a given object. In Fig.~\ref{fig:cpr_ratio}, we compare our inferred Ni/Fe ratios with those predicted by \citet{2020MNRAS.491.2902F} for the same spectra. We also show the comparison of the Ni/Fe ratios between case $A$ and case $A'$. \textcolor{black}{The results from \citet{2020MNRAS.491.2902F} and ours are in good agreement at phases $t \lesssim +300$~days. When approaching $t  \gtrsim +300$~days, our fits tend to give systematically larger Ni/Fe ratios. A possible interpretation is that the evolution of the Ni/Fe ratio disappears when more accurate temperatures, departure coefficients, and flux ratios are obtained through the NLTE calculations and full spectral fitting. But we notice that the 7900~\AA\ region is not fit well by \citet{2020MNRAS.491.2902F}, which might also lead to the different results.} For case $A$ and case $A'$, the Ni/Fe mass ratios inferred from the spectra taken at $t > +300$~days are more similar, since the small bump at 6800--7000~{\AA} that leads to the different pseudocontinua becomes weak at such phases.

It should be noted that the evolution of the inferred Ni/Fe ratio makes the distinction between $M_{\rm Ch}$ models and sub-$M_{\rm Ch}$ models difficult. If the evolution is caused by suppression of [Ni~II], more-accurate estimates of the Ni/Fe ratio could be inferred from very late-phase spectra, at $t \gtrsim +350$~days when the minor emission on the left side of the $\sim 7300$~{\AA} feature disappears in the spectra. Considering such an evolutionary trend, a relatively lower Ni/Fe ratio inferred from earlier nebular-phase spectra (i.e., at $t \approx 250$~days) should be used carefully to confine the explosion models. For our sample, \textcolor{black}{7 and 12 out of 19} SNe~Ia are found to be more in favour of the $M_{\rm Ch}$ and sub-$M_{\rm Ch}$ models, respectively, if restricting the SN~Ia sample to those with at least one spectrum at $t > +300$~days. The relative fractions (i.e., about \textcolor{black}{$37\%$} for $M_{\rm Ch}$ and \textcolor{black}{$63\%$} for sub-$M_{\rm Ch}$) are consistent with the results derived for the whole sample. However, the proportion favouring an origin in sub-$M_{\rm Ch}$ explosions still could be overestimated because the inferred low Ni/Fe ratio might be affected by spectral evolution.      

\subsection{Implications for the Variation of Si Velocities \label{subsec:implications}} 

\citet{2010Natur.466...82M} found that Si velocity gradients were correlated with nebular velocities, and they proposed an asymmetric and off-centre explosion toy model to interpret this correlation as a geometric effect. In particular, one will find low-velocity-gradient (LVG) SNe~Ia when viewing from the direction of off-centre ignition and high-velocity-gradient (HVG) SNe~Ia from the opposite direction. A similar correlation can be found between Si velocities at maximum light and nebular velocities \citep{2013MNRAS.430.1030S}. Fig.~\ref{fig:R_Si} shows that almost all HV SNe~Ia display a redshifted nebular velocity, in agreement with \citet{2013MNRAS.430.1030S}. Note that the HV subclass is not bound exactly to the HVG subclass; for example, SN~2021wuf can be put into the HV group while its Si velocity gradient is relatively low (LVG). Nevertheless, to do the analysis we use the Si velocity around the time of maximum light instead of the Si velocity gradient, since the former parameter is more convenient to measure. 

As indicated in Fig.~\ref{fig:R_Si}, the two branches of SNe~Ia characterised by redshifted or blueshifted nebular velocities show distinct distributions in the Ni/Fe ratio and Si~II velocity plane, especially at the larger velocity end where no blueshifted SNe~Ia are found except for SN~2021wuf. The histogram distribution of the Ni/Fe ratio and Si~II velocity indicates that the separation between redshifted and blueshifted objects is caused mainly by Si velocity instead of the Ni/Fe ratio. 

To examine whether the blueshifted and redshifted objects at lower velocities also come from different groups, we applied a two-dimensional K-S test \citep{1987MNRAS.225..155F} and derived the $p$-value as \textcolor{black}{0.05}. Such a small $p$-value suggests that the blueshifted and redshifted branches still show a significant difference. Nevertheless, we caution that this result may suffer from small-number statistics (especially for redshifted NV objects), and a larger sample is needed for further analysis. \textcolor{black}{Note that the correlation between the Ni/Fe ratio and Si velocity is weak for the whole sample, with a Pearson $r$-value being 0.31 even when not considering the uncertainties. Thus, we focus on the subsamples with redshifted or blueshifted nebular velocities in the following analysis in this subsection.}

In the study by \citet{2010Natur.466...82M}, the LVG SNe~Ia displaying redshifted nebular velocities were not regarded as exceptions, and they suggested that SNe~Ia of the LVG and HVG subclasses do not have intrinsic differences as long as the redshifted nebular velocity is not too large (i.e., $\lesssim 1000$~km~s$^{-1}$). However, the \textcolor{black}{possible} LVG SN~Ia ASASSN-14jg \citep{2017MNRAS.472.3437G}, with Si velocity \textcolor{black}{$\sim 10,760$~km~s$^{-1}$ at +6 days}, has an unusually large redshifted nebular velocity \textcolor{black}{($\sim 1600$~km~s$^{-1}$)} relative to the expected upper limit of $\sim 1000$~km~s$^{-1}$. This suggests that ASASSN-14jg could still be observed in the NV subclass when viewing from the opposite direction of off-centre ignition owing to its overall low Si velocity. Moreover, among the redshifted objects, \textcolor{black}{$\sim 40$\%} are found to have normal Si velocity (i.e., $\lesssim 12,000$~km~s$^{-1}$), which means that it is not rare that NV SNe~Ia display redshifted nebular velocities. Thus, other factors are needed to interpret the variation of Si velocities. 

On the other hand, \citet{2013Sci...340..170W} found that HV SNe~Ia tend to occur in inner regions of their host galaxies, arguing that the diversity of Si velocities in SNe~Ia cannot be completely attributed to geometric effects and HV SNe~Ia may be related to metal-rich stellar populations \citep{2015MNRAS.446..354P, 2020ApJ...895L...5P}. This is consistent with the recent result that HV SNe~Ia tend to have abundant circumstellar dust \citep{2019ApJ...882..120W}. The analysis of late-time spectral features presented in this paper also supports that HV SNe~Ia may have metal-rich progenitors. 
\citet{2003ApJ...590L..83T} explored how the ratio of stable-to-radioactive nucleosynthetic products increases with the metallicity of SN~progenitors. A higher ratio of stable-to-radioactive nucleosynthetic products means a higher late-time Ni/Fe ratio, since all of the radioactive $^{56}$Ni would decay to $^{56}$Co and finally $^{56}$Fe. This has also been discussed by \citet{2022MNRAS.511.3682G}. If we only focus on the SNe~Ia that have redshifted or blueshifted nebular velocities, the Ni/Fe ratio tends to increase with Si velocity, so the Si velocity has a positive correlation with metallicity. This correlation is also in agreement with the simulated result that the blueshift of Si~II absorption increases with the metallicity of the C+O layer of SN~Ia progenitors \citep{2000ApJ...530..966L}. Thus, the low Si velocity of ASASSN-14jg may be due to its low Ni/Fe ratio (low metallicity). However, we caution that whether the variation of metallicity can completely account for the observed variation of Si velocity is not clear, and more-detailed explosion simulations are needed in future work.

\subsection{Sub-$M_{\rm Ch}$ Models Prefer Smaller $\Delta m_{15}(B)$?}

As shown in Fig.~\ref{fig:R_15}, a positive correlation between the Ni/Fe ratio and $\Delta m_{15}(B)$ seems to exist for SNe~Ia with \textcolor{black}{$\Delta m_{15} < 1.7$}~mag. A higher Ni/Fe ratio means a higher ratio of stable-to-radioactive nucleosynthetic products and hence a lower mass of synthetic radioactive $^{56}$Ni. The mass of radioactive $^{56}$Ni is a dominant parameter describing the explosion energy of SNe~Ia, with larger $^{56}$Ni corresponding to a more energetic explosion and hence smaller $\Delta m_{15}(B)$. Thus, given a similar total mass of ejecta, a higher ratio of stable-to-radioactive nucleosynthetic products would naturally result in a fainter SN~Ia with larger $\Delta m_{15}(B)$. 

However, the positive correlation between the Ni/Fe ratio and $\Delta m_{15}(B)$ also implies that the sub-$M_{\rm Ch}$ models would tend to produce more-luminous SNe~Ia than the $M_{\rm Ch}$ models, contradicting the previous understanding of explosion models. Considering the [Ni~II] suppression that we have discussed in Section~\ref{subsec:evolution}, an alternative interpretation is that SNe~Ia with smaller $\Delta m_{15}(B)$ might have more radioactive materials, which would ionise more nickel to Ni~III at late phases. \textcolor{black}{This interpretation is supported by the finding of \citet{2022A&A...660A..96B} that the strong [Ni~II] lines predicted by their $M_{\rm Ch}$ models can be completely suppressed when $^{56}$Ni is sufficiently mixed with the stable iron-group elements.} Thus, the Ni/Fe ratio inferred from our methods would be underestimated for more-luminous SNe~Ia. In other words, the positive correlation between the inferred Ni/Fe ratio and $\Delta m_{15}(B)$ is caused by different degrees of suppression of [Ni~II] features.   

\section{Conclusions} \label{sec:conclusion}
We have performed multicomponent Gaussian fits to the 7300~{\AA} emission features (due to [Fe~II] and [Ni~II]) in the nebular spectra of 36 SNe~Ia. This has allowed us to measure the velocity shifts and flux ratios of [Ni~II] and [Fe~II] features in this region, which can be used to infer nebular velocities and the late-time Ni/Fe ratio, respectively. The nebular velocity might be connected to the geometry of the explosion, while the Ni/Fe ratio can be used to distinguish sub-$M_{\rm Ch}$ and $M_{\rm Ch}$ models. Connecting these inferred late-time parameters with the early-time observations (Si~II $\lambda$6355 velocity at maximum light and $\Delta m_{15}(B)$), we find some interesting correlations.
Our main results are as follows.

(i) The majority (about \textcolor{black}{67\%}) of SNe~Ia in this work favour sub-$M_{\rm Ch}$ models. However, this fraction could be overestimated owing to a potentially increasing trend of the inferred Ni/Fe ratio with SN age. Moreover, the increasing trend can be interpreted as the suppression of [Ni~II] lines as a result of higher ionisation of nickel in the inner ejecta.   

(ii) \textcolor{black}{Although the whole sample do not show a prominent correlation between the inferred Ni/Fe ratio and Si velocity, a positive correlation tends to exist for subsamples with redshifted or blueshifted nebular velocities.} This indicates that the progenitor metallicity should be at least partially responsible for the variation of Si velocity observed in SNe~Ia.  

(iii) The Ni/Fe ratio has a positive correlation with $\Delta m_{15}(B)$, except for the SN~1991bg-like SNe~Ia. This correlation seems to suggest that sub-$M_{\rm Ch}$ explosions tend to produce more-luminous SNe~Ia than $M_{\rm Ch}$ explosions, which violates common sense. However, this correlation could be caused by serious suppression of [Ni~II] (ionised to [Ni~III]) for luminous SNe~Ia with smaller $\Delta m_{15}(B)$. 

Late-time spectra reveal the inner region of SNe~Ia and are important in studying the explosion properties. Connections between the inner region and the outer ejecta also have many implications for the study of explosion mechanisms. In future work, more high-quality late-time spectra are needed to confirm these connections and explore new correlations. Additional well-observed SNe~Ia at late phases are also needed to study the evolution of late-time spectra and give more-accurate estimates of the Ni/Fe ratio to confine explosion models.

\section*{Acknowledgements}

We thank the anonymous referee for his/her constructive comments which help improve the manuscript.

This work is supported by the National Science Foundation of China (NSFC grants 12288102, 12033003, 11633002, and 12090044), the Ma Huateng Foundation, the Scholar Program of Beijing Academy of Science and Technology (DZ:BS202002), and the Tencent Xplorer Prize.  A.V.F.'s team received support from the Christopher R. Redlich Fund, Gary and Cynthia Bengier, Clark and Sharon Winslow, Sanford Robertson, Alan Eustace, Briggs and Kathleen Wood, and many other donors.
N.E.R. acknowledges partial support from MIUR, PRIN 2017 (grant 20179ZF5KS), PRIN-INAF 2022, the Spanish MICINN grant PID2019-108709GB-I00 and FEDER funds, and the program Unidad de Excelencia María de Maeztu CEX2020-001058-M. 

We thank the staffs at the observatories where data were obtained.
Some of the data presented herein were obtained at the W. M. Keck Observatory, which is operated as a scientific partnership among the California Institute of Technology, the University of California, and NASA; the observatory was made possible by the generous financial support of the W. M. Keck Foundation.
A major upgrade of the Kast spectrograph on the Shane 3~m telescope at Lick Observatory, led by Brad Holden, was made possible through generous gifts from the Heising-Simons Foundation, William and Marina Kast, and the University of California Observatories. Research at Lick Observatory is partially supported by a generous gift from Google.
Based in part on observations made with the Gran Telescopio Canarias (GTC), installed in the Spanish Observatorio del Roque de los Muchachos of the Instituto de Astrofísica de Canarias, in the island of La Palma.
 
This work made use of the OCS (https://sne.space), SNDB (http://heracles.astro.berkeley.edu), and WISeREP (https://wiserep.weizmann.ac.il) databases.
We have used the NASA/IPAC Extragalactic Database (NED), which is funded by NASA and operated by the California Institute of Technology.
This research utilised Scipy (https://www.scipy.org), Extinction (https://pypi.org/project/extinction), Matplotlib \citep{2007CSE.....9...90H}, and Numpy (https://numpy.org).

\section*{Data Availability}

The data underlying this article are available in the article. Information on the spectra collected by our own project is presented in Table~\ref{tab:new}. Some basic information about the SNe~Ia in this work is presented in Table~\ref{tab:base_para}. The fits for all spectra in this work are shown in Fig.~\ref{fig:all_fits}, and the corresponding parameters are presented in Table~\ref{tab:Multi}.



\bibliographystyle{mnras}
\bibliography{paper} 

\begin{thebibliography}{}
\makeatletter
\relax
\def\mn@urlcharsother{\let\do\@makeother \do\$\do\&\do\#\do\^\do\_\do\%\do\~}
\def\mn@doi{\begingroup\mn@urlcharsother \@ifnextchar [ {\mn@doi@}
  {\mn@doi@[]}}
\def\mn@doi@[#1]#2{\def\@tempa{#1}\ifx\@tempa\@empty \href
  {http://dx.doi.org/#2} {doi:#2}\else \href {http://dx.doi.org/#2} {#1}\fi
  \endgroup}
\def\mn@eprint#1#2{\mn@eprint@#1:#2::\@nil}
\def\mn@eprint@arXiv#1{\href {http://arxiv.org/abs/#1} {{\tt arXiv:#1}}}
\def\mn@eprint@dblp#1{\href {http://dblp.uni-trier.de/rec/bibtex/#1.xml}
  {dblp:#1}}
\def\mn@eprint@#1:#2:#3:#4\@nil{\def\@tempa {#1}\def\@tempb {#2}\def\@tempc
  {#3}\ifx \@tempc \@empty \let \@tempc \@tempb \let \@tempb \@tempa \fi \ifx
  \@tempb \@empty \def\@tempb {arXiv}\fi \@ifundefined
  {mn@eprint@\@tempb}{\@tempb:\@tempc}{\expandafter \expandafter \csname
  mn@eprint@\@tempb\endcsname \expandafter{\@tempc}}}

\bibitem[\protect\citeauthoryear{{Amanullah} et~al.,}{{Amanullah}
  et~al.}{2015}]{2015MNRAS.453.3300A}
{Amanullah} R.,  et~al., 2015, \mn@doi [\mnras] {10.1093/mnras/stv1505}, \href
  {https://ui.adsabs.harvard.edu/abs/2015MNRAS.453.3300A} {453, 3300}

\bibitem[\protect\citeauthoryear{{Anupama}, {Sahu}  \& {Jose}}{{Anupama}
  et~al.}{2005}]{2005A&A...429..667A}
{Anupama} G.~C.,  {Sahu} D.~K.,   {Jose} J.,  2005, \mn@doi [\aap]
  {10.1051/0004-6361:20041687}, \href
  {https://ui.adsabs.harvard.edu/abs/2005A&A...429..667A} {429, 667}

\bibitem[\protect\citeauthoryear{{Benetti} et~al.,}{{Benetti}
  et~al.}{2004}]{2004MNRAS.348..261B}
{Benetti} S.,  et~al., 2004, \mn@doi [\mnras]
  {10.1111/j.1365-2966.2004.07357.x}, \href
  {https://ui.adsabs.harvard.edu/abs/2004MNRAS.348..261B} {348, 261}

\bibitem[\protect\citeauthoryear{{Benetti} et~al.,}{{Benetti}
  et~al.}{2005}]{2005ApJ...623.1011B}
{Benetti} S.,  et~al., 2005, \mn@doi [\apj] {10.1086/428608}, \href
  {https://ui.adsabs.harvard.edu/abs/2005ApJ...623.1011B} {623, 1011}

\bibitem[\protect\citeauthoryear{{Betoule} et~al.,}{{Betoule}
  et~al.}{2014}]{2014A&A...568A..22B}
{Betoule} M.,  et~al., 2014, \mn@doi [\aap] {10.1051/0004-6361/201423413},
  \href {https://ui.adsabs.harvard.edu/abs/2014A&A...568A..22B} {568, A22}

\bibitem[\protect\citeauthoryear{{Bildsten}, {Shen}, {Weinberg}  \&
  {Nelemans}}{{Bildsten} et~al.}{2007}]{2007ApJ...662L..95B}
{Bildsten} L.,  {Shen} K.~J.,  {Weinberg} N.~N.,   {Nelemans} G.,  2007,
  \mn@doi [\apjl] {10.1086/519489}, \href
  {https://ui.adsabs.harvard.edu/abs/2007ApJ...662L..95B} {662, L95}

\bibitem[\protect\citeauthoryear{{Blondin} \& {Tonry}}{{Blondin} \&
  {Tonry}}{2007}]{2007ApJ...666.1024B}
{Blondin} S.,  {Tonry} J.~L.,  2007, \mn@doi [\apj] {10.1086/520494}, \href
  {https://ui.adsabs.harvard.edu/abs/2007ApJ...666.1024B} {666, 1024}

\bibitem[\protect\citeauthoryear{{Blondin} et~al.,}{{Blondin}
  et~al.}{2012}]{2012AJ....143..126B}
{Blondin} S.,  et~al., 2012, \mn@doi [\aj] {10.1088/0004-6256/143/5/126}, \href
  {https://ui.adsabs.harvard.edu/abs/2012AJ....143..126B} {143, 126}

\bibitem[\protect\citeauthoryear{{Blondin}, {Bravo}, {Timmes}, {Dessart}  \&
  {Hillier}}{{Blondin} et~al.}{2022}]{2022A&A...660A..96B}
{Blondin} S.,  {Bravo} E.,  {Timmes} F.~X.,  {Dessart} L.,   {Hillier} D.~J.,
  2022, \mn@doi [\aap] {10.1051/0004-6361/202142323}, \href
  {https://ui.adsabs.harvard.edu/abs/2022A&A...660A..96B} {660, A96}

\bibitem[\protect\citeauthoryear{{Branch}, {Fisher}  \& {Nugent}}{{Branch}
  et~al.}{1993}]{1993AJ....106.2383B}
{Branch} D.,  {Fisher} A.,   {Nugent} P.,  1993, \mn@doi [\aj]
  {10.1086/116810}, \href
  {https://ui.adsabs.harvard.edu/abs/1993AJ....106.2383B} {106, 2383}

\bibitem[\protect\citeauthoryear{{Brown}, {Breeveld}, {Holland}, {Kuin}  \&
  {Pritchard}}{{Brown} et~al.}{2014}]{2014Ap&SS.354...89B}
{Brown} P.~J.,  {Breeveld} A.~A.,  {Holland} S.,  {Kuin} P.,   {Pritchard} T.,
  2014, \mn@doi [\apss] {10.1007/s10509-014-2059-8}, \href
  {https://ui.adsabs.harvard.edu/abs/2014Ap&SS.354...89B} {354, 89}

\bibitem[\protect\citeauthoryear{{Burke}, {Arcavi}, {Howell}, {Hiramatsu},
  {McCully}  \& {Valenti}}{{Burke} et~al.}{2019}]{2019TNSCR.701....1B}
{Burke} J.,  {Arcavi} I.,  {Howell} D.~A.,  {Hiramatsu} D.,  {McCully} C.,
  {Valenti} S.,  2019, Transient Name Server Classification Report, \href
  {https://ui.adsabs.harvard.edu/abs/2019TNSCR.701....1B} {2019-701, 1}

\bibitem[\protect\citeauthoryear{{Burns} et~al.,}{{Burns}
  et~al.}{2020}]{2020ApJ...895..118B}
{Burns} C.~R.,  et~al., 2020, \mn@doi [\apj] {10.3847/1538-4357/ab8e3e}, \href
  {https://ui.adsabs.harvard.edu/abs/2020ApJ...895..118B} {895, 118}

\bibitem[\protect\citeauthoryear{{Cartier} et~al.,}{{Cartier}
  et~al.}{2017}]{2017MNRAS.464.4476C}
{Cartier} R.,  et~al., 2017, \mn@doi [\mnras] {10.1093/mnras/stw2678}, \href
  {https://ui.adsabs.harvard.edu/abs/2017MNRAS.464.4476C} {464, 4476}

\bibitem[\protect\citeauthoryear{{Childress} et~al.,}{{Childress}
  et~al.}{2013}]{2013ApJ...770...29C}
{Childress} M.~J.,  et~al., 2013, \mn@doi [\apj] {10.1088/0004-637X/770/1/29},
  \href {https://ui.adsabs.harvard.edu/abs/2013ApJ...770...29C} {770, 29}

\bibitem[\protect\citeauthoryear{{Childress} et~al.,}{{Childress}
  et~al.}{2015}]{2015MNRAS.454.3816C}
{Childress} M.~J.,  et~al., 2015, \mn@doi [\mnras] {10.1093/mnras/stv2173},
  \href {https://ui.adsabs.harvard.edu/abs/2015MNRAS.454.3816C} {454, 3816}

\bibitem[\protect\citeauthoryear{{Childress} et~al.,}{{Childress}
  et~al.}{2016}]{2016PASA...33...55C}
{Childress} M.~J.,  et~al., 2016, \mn@doi [\pasa] {10.1017/pasa.2016.47}, \href
  {https://ui.adsabs.harvard.edu/abs/2016PASA...33...55C} {33, e055}

\bibitem[\protect\citeauthoryear{{Cristiani} et~al.,}{{Cristiani}
  et~al.}{1992}]{1992A&A...259...63C}
{Cristiani} S.,  et~al., 1992, \aap, \href
  {https://ui.adsabs.harvard.edu/abs/1992A&A...259...63C} {259, 63}

\bibitem[\protect\citeauthoryear{{Evans} et~al.,}{{Evans}
  et~al.}{2003}]{2003IAUC.8171....1E}
{Evans} R.,  et~al., 2003, \iaucirc, \href
  {https://ui.adsabs.harvard.edu/abs/2003IAUC.8171....1E} {8171, 1}

\bibitem[\protect\citeauthoryear{{Faber} et~al.,}{{Faber}
  et~al.}{2003}]{2003SPIE.4841.1657F}
{Faber} S.~M.,  et~al., 2003, in {Iye} M.,  {Moorwood} A. F.~M.,  eds,  Society
  of Photo-Optical Instrumentation Engineers (SPIE) Conference Series Vol.
  4841, Instrument Design and Performance for Optical/Infrared Ground-based
  Telescopes. pp 1657--1669, \mn@doi{10.1117/12.460346}

\bibitem[\protect\citeauthoryear{{Fasano} \& {Franceschini}}{{Fasano} \&
  {Franceschini}}{1987}]{1987MNRAS.225..155F}
{Fasano} G.,  {Franceschini} A.,  1987, \mn@doi [\mnras]
  {10.1093/mnras/225.1.155}, \href
  {https://ui.adsabs.harvard.edu/abs/1987MNRAS.225..155F} {225, 155}

\bibitem[\protect\citeauthoryear{{Filippenko}}{{Filippenko}}{1982}]{1982PASP...94..715F}
{Filippenko} A.~V.,  1982, \mn@doi [\pasp] {10.1086/131052}, \href
  {https://ui.adsabs.harvard.edu/abs/1982PASP...94..715F} {94, 715}

\bibitem[\protect\citeauthoryear{{Filippenko}}{{Filippenko}}{1997}]{1997ARA&A..35..309F}
{Filippenko} A.~V.,  1997, \mn@doi [\araa] {10.1146/annurev.astro.35.1.309},
  \href {https://ui.adsabs.harvard.edu/abs/1997ARA&A..35..309F} {35, 309}

\bibitem[\protect\citeauthoryear{{Filippenko} et~al.,}{{Filippenko}
  et~al.}{1992}]{1992AJ....104.1543F}
{Filippenko} A.~V.,  et~al., 1992, \mn@doi [\aj] {10.1086/116339}, \href
  {https://ui.adsabs.harvard.edu/abs/1992AJ....104.1543F} {104, 1543}

\bibitem[\protect\citeauthoryear{{Fl{\"o}rs} et~al.,}{{Fl{\"o}rs}
  et~al.}{2020}]{2020MNRAS.491.2902F}
{Fl{\"o}rs} A.,  et~al., 2020, \mn@doi [\mnras] {10.1093/mnras/stz3013}, \href
  {https://ui.adsabs.harvard.edu/abs/2020MNRAS.491.2902F} {491, 2902}

\bibitem[\protect\citeauthoryear{{Folatelli} et~al.,}{{Folatelli}
  et~al.}{2013}]{2013ApJ...773...53F}
{Folatelli} G.,  et~al., 2013, \mn@doi [\apj] {10.1088/0004-637X/773/1/53},
  \href {https://ui.adsabs.harvard.edu/abs/2013ApJ...773...53F} {773, 53}

\bibitem[\protect\citeauthoryear{{Foley} et~al.,}{{Foley}
  et~al.}{2014}]{2014MNRAS.443.2887F}
{Foley} R.~J.,  et~al., 2014, \mn@doi [\mnras] {10.1093/mnras/stu1378}, \href
  {https://ui.adsabs.harvard.edu/abs/2014MNRAS.443.2887F} {443, 2887}

\bibitem[\protect\citeauthoryear{{Foley} et~al.,}{{Foley}
  et~al.}{2016}]{2016MNRAS.461.1308F}
{Foley} R.~J.,  et~al., 2016, \mn@doi [\mnras] {10.1093/mnras/stw1440}, \href
  {https://ui.adsabs.harvard.edu/abs/2016MNRAS.461.1308F} {461, 1308}

\bibitem[\protect\citeauthoryear{{Fransson} \& {Jerkstrand}}{{Fransson} \&
  {Jerkstrand}}{2015}]{2015ApJ...814L...2F}
{Fransson} C.,  {Jerkstrand} A.,  2015, \mn@doi [\apjl]
  {10.1088/2041-8205/814/1/L2}, \href
  {https://ui.adsabs.harvard.edu/abs/2015ApJ...814L...2F} {814, L2}

\bibitem[\protect\citeauthoryear{{Gamezo}, {Khokhlov}  \& {Oran}}{{Gamezo}
  et~al.}{2005}]{2005ApJ...623..337G}
{Gamezo} V.~N.,  {Khokhlov} A.~M.,   {Oran} E.~S.,  2005, \mn@doi [\apj]
  {10.1086/428767}, \href
  {https://ui.adsabs.harvard.edu/abs/2005ApJ...623..337G} {623, 337}

\bibitem[\protect\citeauthoryear{{Ganeshalingam}, {Li}  \&
  {Filippenko}}{{Ganeshalingam} et~al.}{2011}]{2011MNRAS.416.2607G}
{Ganeshalingam} M.,  {Li} W.,   {Filippenko} A.~V.,  2011, \mn@doi [\mnras]
  {10.1111/j.1365-2966.2011.19213.x}, \href
  {https://ui.adsabs.harvard.edu/abs/2011MNRAS.416.2607G} {416, 2607}

\bibitem[\protect\citeauthoryear{{Ganeshalingam} et~al.,}{{Ganeshalingam}
  et~al.}{2012}]{2012ApJ...751..142G}
{Ganeshalingam} M.,  et~al., 2012, \mn@doi [\apj]
  {10.1088/0004-637X/751/2/142}, \href
  {https://ui.adsabs.harvard.edu/abs/2012ApJ...751..142G} {751, 142}

\bibitem[\protect\citeauthoryear{{Garavini} et~al.,}{{Garavini}
  et~al.}{2007}]{2007A&A...471..527G}
{Garavini} G.,  et~al., 2007, \mn@doi [\aap] {10.1051/0004-6361:20066971},
  \href {https://ui.adsabs.harvard.edu/abs/2007A&A...471..527G} {471, 527}

\bibitem[\protect\citeauthoryear{{Gerardy}}{{Gerardy}}{2005}]{2005ASPC..342..250G}
{Gerardy} C.~L.,  2005, in {Turatto} M.,  {Benetti} S.,  {Zampieri} L.,
  {Shea} W.,  eds,  Astronomical Society of the Pacific Conference Series Vol.
  342, 1604-2004: Supernovae as Cosmological Lighthouses. p.~250

\bibitem[\protect\citeauthoryear{{G{\'o}mez} \& {L{\'o}pez}}{{G{\'o}mez} \&
  {L{\'o}pez}}{1998}]{1998AJ....115.1096G}
{G{\'o}mez} G.,  {L{\'o}pez} R.,  1998, \mn@doi [\aj] {10.1086/300248}, \href
  {https://ui.adsabs.harvard.edu/abs/1998AJ....115.1096G} {115, 1096}

\bibitem[\protect\citeauthoryear{{Graham}, {Nugent}, {Sullivan}, {Filippenko},
  {Cenko}, {Silverman}, {Clubb}  \& {Zheng}}{{Graham}
  et~al.}{2015}]{2015MNRAS.454.1948G}
{Graham} M.~L.,  {Nugent} P.~E.,  {Sullivan} M.,  {Filippenko} A.~V.,  {Cenko}
  S.~B.,  {Silverman} J.~M.,  {Clubb} K.~I.,   {Zheng} W.,  2015, \mn@doi
  [\mnras] {10.1093/mnras/stv1888}, \href
  {https://ui.adsabs.harvard.edu/abs/2015MNRAS.454.1948G} {454, 1948}

\bibitem[\protect\citeauthoryear{{Graham} et~al.,}{{Graham}
  et~al.}{2017}]{2017MNRAS.472.3437G}
{Graham} M.~L.,  et~al., 2017, \mn@doi [\mnras] {10.1093/mnras/stx2224}, \href
  {https://ui.adsabs.harvard.edu/abs/2017MNRAS.472.3437G} {472, 3437}

\bibitem[\protect\citeauthoryear{{Graham} et~al.,}{{Graham}
  et~al.}{2022}]{2022MNRAS.511.3682G}
{Graham} M.~L.,  et~al., 2022, \mn@doi [\mnras] {10.1093/mnras/stac192}, \href
  {https://ui.adsabs.harvard.edu/abs/2022MNRAS.511.3682G} {511, 3682}

\bibitem[\protect\citeauthoryear{{Guillochon}, {Parrent}, {Kelley}  \&
  {Margutti}}{{Guillochon} et~al.}{2017}]{2017ApJ...835...64G}
{Guillochon} J.,  {Parrent} J.,  {Kelley} L.~Z.,   {Margutti} R.,  2017,
  \mn@doi [\apj] {10.3847/1538-4357/835/1/64}, \href
  {https://ui.adsabs.harvard.edu/abs/2017ApJ...835...64G} {835, 64}

\bibitem[\protect\citeauthoryear{{Guy} et~al.,}{{Guy}
  et~al.}{2007}]{2007A&A...466...11G}
{Guy} J.,  et~al., 2007, \mn@doi [\aap] {10.1051/0004-6361:20066930}, \href
  {https://ui.adsabs.harvard.edu/abs/2007A&A...466...11G} {466, 11}

\bibitem[\protect\citeauthoryear{{Hicken} et~al.,}{{Hicken}
  et~al.}{2009}]{2009ApJ...700..331H}
{Hicken} M.,  et~al., 2009, \mn@doi [\apj] {10.1088/0004-637X/700/1/331}, \href
  {https://ui.adsabs.harvard.edu/abs/2009ApJ...700..331H} {700, 331}

\bibitem[\protect\citeauthoryear{{Hillebrandt} \& {Niemeyer}}{{Hillebrandt} \&
  {Niemeyer}}{2000}]{2000ARA&A..38..191H}
{Hillebrandt} W.,  {Niemeyer} J.~C.,  2000, \mn@doi [\araa]
  {10.1146/annurev.astro.38.1.191}, \href
  {https://ui.adsabs.harvard.edu/abs/2000ARA&A..38..191H} {38, 191}

\bibitem[\protect\citeauthoryear{{Hoeflich} \& {Khokhlov}}{{Hoeflich} \&
  {Khokhlov}}{1996}]{1996ApJ...457..500H}
{Hoeflich} P.,  {Khokhlov} A.,  1996, \mn@doi [\apj] {10.1086/176748}, \href
  {https://ui.adsabs.harvard.edu/abs/1996ApJ...457..500H} {457, 500}

\bibitem[\protect\citeauthoryear{{Holmbo} et~al.,}{{Holmbo}
  et~al.}{2019}]{2019A&A...627A.174H}
{Holmbo} S.,  et~al., 2019, \mn@doi [\aap] {10.1051/0004-6361/201834389}, \href
  {https://ui.adsabs.harvard.edu/abs/2019A&A...627A.174H} {627, A174}

\bibitem[\protect\citeauthoryear{{Horne}}{{Horne}}{1986}]{1986PASP...98..609H}
{Horne} K.,  1986, \mn@doi [\pasp] {10.1086/131801}, \href
  {https://ui.adsabs.harvard.edu/abs/1986PASP...98..609H} {98, 609}

\bibitem[\protect\citeauthoryear{{Hosseinzadeh} et~al.,}{{Hosseinzadeh}
  et~al.}{2017}]{2017ApJ...845L..11H}
{Hosseinzadeh} G.,  et~al., 2017, \mn@doi [\apjl] {10.3847/2041-8213/aa8402},
  \href {https://ui.adsabs.harvard.edu/abs/2017ApJ...845L..11H} {845, L11}

\bibitem[\protect\citeauthoryear{{Hunter}}{{Hunter}}{2007}]{2007CSE.....9...90H}
{Hunter} J.~D.,  2007, \mn@doi [Computing in Science and Engineering]
  {10.1109/MCSE.2007.55}, \href
  {https://ui.adsabs.harvard.edu/abs/2007CSE.....9...90H} {9, 90}

\bibitem[\protect\citeauthoryear{{Iben} \& {Tutukov}}{{Iben} \&
  {Tutukov}}{1984}]{1984ApJS...54..335I}
{Iben} I. J.,  {Tutukov} A.~V.,  1984, \mn@doi [\apjs] {10.1086/190932}, \href
  {https://ui.adsabs.harvard.edu/abs/1984ApJS...54..335I} {54, 335}

\bibitem[\protect\citeauthoryear{{Jerkstrand}, {Ergon}, {Smartt}, {Fransson},
  {Sollerman}, {Taubenberger}, {Bersten}  \& {Spyromilio}}{{Jerkstrand}
  et~al.}{2015}]{2015A&A...573A..12J}
{Jerkstrand} A.,  {Ergon} M.,  {Smartt} S.~J.,  {Fransson} C.,  {Sollerman} J.,
   {Taubenberger} S.,  {Bersten} M.,   {Spyromilio} J.,  2015, \mn@doi [\aap]
  {10.1051/0004-6361/201423983}, \href
  {https://ui.adsabs.harvard.edu/abs/2015A&A...573A..12J} {573, A12}

\bibitem[\protect\citeauthoryear{{Jha} et~al.,}{{Jha}
  et~al.}{1999}]{1999ApJS..125...73J}
{Jha} S.,  et~al., 1999, \mn@doi [\apjs] {10.1086/313275}, \href
  {https://ui.adsabs.harvard.edu/abs/1999ApJS..125...73J} {125, 73}

\bibitem[\protect\citeauthoryear{{Krisciunas}, {Hastings}, {Loomis},
  {McMillan}, {Rest}, {Riess}  \& {Stubbs}}{{Krisciunas}
  et~al.}{2000}]{2000ApJ...539..658K}
{Krisciunas} K.,  {Hastings} N.~C.,  {Loomis} K.,  {McMillan} R.,  {Rest} A.,
  {Riess} A.~G.,   {Stubbs} C.,  2000, \mn@doi [\apj] {10.1086/309263}, \href
  {https://ui.adsabs.harvard.edu/abs/2000ApJ...539..658K} {539, 658}

\bibitem[\protect\citeauthoryear{{Krisciunas} et~al.,}{{Krisciunas}
  et~al.}{2009}]{2009AJ....138.1584K}
{Krisciunas} K.,  et~al., 2009, \mn@doi [\aj] {10.1088/0004-6256/138/6/1584},
  \href {https://ui.adsabs.harvard.edu/abs/2009AJ....138.1584K} {138, 1584}

\bibitem[\protect\citeauthoryear{{Kushnir}, {Katz}, {Dong}, {Livne}  \&
  {Fern{\'a}ndez}}{{Kushnir} et~al.}{2013}]{2013ApJ...778L..37K}
{Kushnir} D.,  {Katz} B.,  {Dong} S.,  {Livne} E.,   {Fern{\'a}ndez} R.,  2013,
  \mn@doi [\apjl] {10.1088/2041-8205/778/2/L37}, \href
  {https://ui.adsabs.harvard.edu/abs/2013ApJ...778L..37K} {778, L37}

\bibitem[\protect\citeauthoryear{{Leloudas} et~al.,}{{Leloudas}
  et~al.}{2009}]{2009A&A...505..265L}
{Leloudas} G.,  et~al., 2009, \mn@doi [\aap] {10.1051/0004-6361/200912364},
  \href {https://ui.adsabs.harvard.edu/abs/2009A&A...505..265L} {505, 265}

\bibitem[\protect\citeauthoryear{{Lentz}, {Baron}, {Branch}, {Hauschildt}  \&
  {Nugent}}{{Lentz} et~al.}{2000}]{2000ApJ...530..966L}
{Lentz} E.~J.,  {Baron} E.,  {Branch} D.,  {Hauschildt} P.~H.,   {Nugent}
  P.~E.,  2000, \mn@doi [\apj] {10.1086/308400}, \href
  {https://ui.adsabs.harvard.edu/abs/2000ApJ...530..966L} {530, 966}

\bibitem[\protect\citeauthoryear{{Li} et~al.,}{{Li}
  et~al.}{2019a}]{2019ApJ...870...12L}
{Li} W.,  et~al., 2019a, \mn@doi [\apj] {10.3847/1538-4357/aaec74}, \href
  {https://ui.adsabs.harvard.edu/abs/2019ApJ...870...12L} {870, 12}

\bibitem[\protect\citeauthoryear{{Li} et~al.,}{{Li}
  et~al.}{2019b}]{2019ApJ...882...30L}
{Li} W.,  et~al., 2019b, \mn@doi [\apj] {10.3847/1538-4357/ab2b49}, \href
  {https://ui.adsabs.harvard.edu/abs/2019ApJ...882...30L} {882, 30}

\bibitem[\protect\citeauthoryear{{Lira} et~al.,}{{Lira}
  et~al.}{1998}]{1998AJ....115..234L}
{Lira} P.,  et~al., 1998, \mn@doi [\aj] {10.1086/300175}, \href
  {https://ui.adsabs.harvard.edu/abs/1998AJ....115..234L} {115, 234}

\bibitem[\protect\citeauthoryear{{Maeda} et~al.,}{{Maeda}
  et~al.}{2010a}]{2010Natur.466...82M}
{Maeda} K.,  et~al., 2010a, \mn@doi [\nat] {10.1038/nature09122}, \href
  {https://ui.adsabs.harvard.edu/abs/2010Natur.466...82M} {466, 82}

\bibitem[\protect\citeauthoryear{{Maeda}, {Taubenberger}, {Sollerman},
  {Mazzali}, {Leloudas}, {Nomoto}  \& {Motohara}}{{Maeda}
  et~al.}{2010b}]{2010ApJ...708.1703M}
{Maeda} K.,  {Taubenberger} S.,  {Sollerman} J.,  {Mazzali} P.~A.,  {Leloudas}
  G.,  {Nomoto} K.,   {Motohara} K.,  2010b, \mn@doi [\apj]
  {10.1088/0004-637X/708/2/1703}, \href
  {https://ui.adsabs.harvard.edu/abs/2010ApJ...708.1703M} {708, 1703}

\bibitem[\protect\citeauthoryear{{Maguire} et~al.,}{{Maguire}
  et~al.}{2018}]{2018MNRAS.477.3567M}
{Maguire} K.,  et~al., 2018, \mn@doi [\mnras] {10.1093/mnras/sty820}, \href
  {https://ui.adsabs.harvard.edu/abs/2018MNRAS.477.3567M} {477, 3567}

\bibitem[\protect\citeauthoryear{{Maoz}, {Mannucci}  \& {Nelemans}}{{Maoz}
  et~al.}{2014}]{2014ARA&A..52..107M}
{Maoz} D.,  {Mannucci} F.,   {Nelemans} G.,  2014, \mn@doi [\araa]
  {10.1146/annurev-astro-082812-141031}, \href
  {https://ui.adsabs.harvard.edu/abs/2014ARA&A..52..107M} {52, 107}

\bibitem[\protect\citeauthoryear{{Marion} et~al.,}{{Marion}
  et~al.}{2016}]{2016ApJ...820...92M}
{Marion} G.~H.,  et~al., 2016, \mn@doi [\apj] {10.3847/0004-637X/820/2/92},
  \href {https://ui.adsabs.harvard.edu/abs/2016ApJ...820...92M} {820, 92}

\bibitem[\protect\citeauthoryear{{Matheson} et~al.,}{{Matheson}
  et~al.}{2008}]{2008AJ....135.1598M}
{Matheson} T.,  et~al., 2008, \mn@doi [\aj] {10.1088/0004-6256/135/4/1598},
  \href {https://ui.adsabs.harvard.edu/abs/2008AJ....135.1598M} {135, 1598}

\bibitem[\protect\citeauthoryear{{Mazzali}, {Lucy}, {Danziger}, {Gouiffes},
  {Cappellaro}  \& {Turatto}}{{Mazzali} et~al.}{1993}]{1993A&A...269..423M}
{Mazzali} P.~A.,  {Lucy} L.~B.,  {Danziger} I.~J.,  {Gouiffes} C.,
  {Cappellaro} E.,   {Turatto} M.,  1993, \aap, \href
  {https://ui.adsabs.harvard.edu/abs/1993A&A...269..423M} {269, 423}

\bibitem[\protect\citeauthoryear{{Mazzali} et~al.,}{{Mazzali}
  et~al.}{2015}]{2015MNRAS.450.2631M}
{Mazzali} P.~A.,  et~al., 2015, \mn@doi [\mnras] {10.1093/mnras/stv761}, \href
  {https://ui.adsabs.harvard.edu/abs/2015MNRAS.450.2631M} {450, 2631}

\bibitem[\protect\citeauthoryear{Miller \& Stone}{Miller \&
  Stone}{1993}]{miller1993lick}
Miller J.,  Stone R.,  1993, Technical report, Lick Obs.
Tech. Rep. 66. Lick Obs., Santa Cruz

\bibitem[\protect\citeauthoryear{{Milne} et~al.,}{{Milne}
  et~al.}{2010}]{2010ApJ...721.1627M}
{Milne} P.~A.,  et~al., 2010, \mn@doi [\apj] {10.1088/0004-637X/721/2/1627},
  \href {https://ui.adsabs.harvard.edu/abs/2010ApJ...721.1627M} {721, 1627}

\bibitem[\protect\citeauthoryear{{Nomoto}, {Iwamoto}  \& {Kishimoto}}{{Nomoto}
  et~al.}{1997}]{1997Sci...276.1378N}
{Nomoto} K.,  {Iwamoto} K.,   {Kishimoto} N.,  1997, \mn@doi [Science]
  {10.1126/science.276.5317.1378}, \href
  {https://ui.adsabs.harvard.edu/abs/1997Sci...276.1378N} {276, 1378}

\bibitem[\protect\citeauthoryear{{Oke} et~al.,}{{Oke}
  et~al.}{1995}]{1995PASP..107..375O}
{Oke} J.~B.,  et~al., 1995, \mn@doi [\pasp] {10.1086/133562}, \href
  {https://ui.adsabs.harvard.edu/abs/1995PASP..107..375O} {107, 375}

\bibitem[\protect\citeauthoryear{{Pakmor}}{{Pakmor}}{2017}]{2017hsn..book.1257P}
{Pakmor} R.,  2017, in {Alsabti} A.~W.,  {Murdin} P.,  eds, , Handbook of
  Supernovae.
p.~1257, \mn@doi{10.1007/978-3-319-21846-5_61}

\bibitem[\protect\citeauthoryear{{Pakmor}, {Kromer}, {Taubenberger}  \&
  {Springel}}{{Pakmor} et~al.}{2013}]{2013ApJ...770L...8P}
{Pakmor} R.,  {Kromer} M.,  {Taubenberger} S.,   {Springel} V.,  2013, \mn@doi
  [\apjl] {10.1088/2041-8205/770/1/L8}, \href
  {https://ui.adsabs.harvard.edu/abs/2013ApJ...770L...8P} {770, L8}

\bibitem[\protect\citeauthoryear{{Pan}}{{Pan}}{2020}]{2020ApJ...895L...5P}
{Pan} Y.-C.,  2020, \mn@doi [\apjl] {10.3847/2041-8213/ab8e47}, \href
  {https://ui.adsabs.harvard.edu/abs/2020ApJ...895L...5P} {895, L5}

\bibitem[\protect\citeauthoryear{{Pan}, {Sullivan}, {Maguire}, {Gal-Yam},
  {Hook}, {Howell}, {Nugent}  \& {Mazzali}}{{Pan}
  et~al.}{2015a}]{2015MNRAS.446..354P}
{Pan} Y.~C.,  {Sullivan} M.,  {Maguire} K.,  {Gal-Yam} A.,  {Hook} I.~M.,
  {Howell} D.~A.,  {Nugent} P.~E.,   {Mazzali} P.~A.,  2015a, \mn@doi [\mnras]
  {10.1093/mnras/stu2121}, \href
  {https://ui.adsabs.harvard.edu/abs/2015MNRAS.446..354P} {446, 354}

\bibitem[\protect\citeauthoryear{{Pan} et~al.,}{{Pan}
  et~al.}{2015b}]{2015MNRAS.452.4307P}
{Pan} Y.~C.,  et~al., 2015b, \mn@doi [\mnras] {10.1093/mnras/stv1605}, \href
  {https://ui.adsabs.harvard.edu/abs/2015MNRAS.452.4307P} {452, 4307}

\bibitem[\protect\citeauthoryear{{Parker}, {Amorim}, {Parrent}, {Sand},
  {Valenti}, {Graham}  \& {Howell}}{{Parker}
  et~al.}{2013}]{2013CBET.3416....1P}
{Parker} S.,  {Amorim} A.,  {Parrent} J.~T.,  {Sand} D.,  {Valenti} S.,
  {Graham} M.~L.,   {Howell} D.~A.,  2013, Central Bureau Electronic Telegrams,
  \href {https://ui.adsabs.harvard.edu/abs/2013CBET.3416....1P} {3416, 1}

\bibitem[\protect\citeauthoryear{{Pastorello} et~al.,}{{Pastorello}
  et~al.}{2007}]{2007MNRAS.377.1531P}
{Pastorello} A.,  et~al., 2007, \mn@doi [\mnras]
  {10.1111/j.1365-2966.2007.11700.x}, \href
  {https://ui.adsabs.harvard.edu/abs/2007MNRAS.377.1531P} {377, 1531}

\bibitem[\protect\citeauthoryear{{Pellegrino} et~al.,}{{Pellegrino}
  et~al.}{2020}]{2020ApJ...897..159P}
{Pellegrino} C.,  et~al., 2020, \mn@doi [\apj] {10.3847/1538-4357/ab8e3f},
  \href {https://ui.adsabs.harvard.edu/abs/2020ApJ...897..159P} {897, 159}

\bibitem[\protect\citeauthoryear{{Pereira} et~al.,}{{Pereira}
  et~al.}{2013}]{2013A&A...554A..27P}
{Pereira} R.,  et~al., 2013, \mn@doi [\aap] {10.1051/0004-6361/201221008},
  \href {https://ui.adsabs.harvard.edu/abs/2013A&A...554A..27P} {554, A27}

\bibitem[\protect\citeauthoryear{{Perley}}{{Perley}}{2019}]{2019PASP..131h4503P}
{Perley} D.~A.,  2019, \mn@doi [\pasp] {10.1088/1538-3873/ab215d}, \href
  {https://ui.adsabs.harvard.edu/abs/2019PASP..131h4503P} {131, 084503}

\bibitem[\protect\citeauthoryear{{Phillips}}{{Phillips}}{1993}]{1993ApJ...413L.105P}
{Phillips} M.~M.,  1993, \mn@doi [\apjl] {10.1086/186970}, \href
  {https://ui.adsabs.harvard.edu/abs/1993ApJ...413L.105P} {413, L105}

\bibitem[\protect\citeauthoryear{{Phillips} et~al.,}{{Phillips}
  et~al.}{1987}]{1987PASP...99..592P}
{Phillips} M.~M.,  et~al., 1987, \mn@doi [\pasp] {10.1086/132020}, \href
  {https://ui.adsabs.harvard.edu/abs/1987PASP...99..592P} {99, 592}

\bibitem[\protect\citeauthoryear{{Phillips}, {Wells}, {Suntzeff}, {Hamuy},
  {Leibundgut}, {Kirshner}  \& {Foltz}}{{Phillips}
  et~al.}{1992}]{1992AJ....103.1632P}
{Phillips} M.~M.,  {Wells} L.~A.,  {Suntzeff} N.~B.,  {Hamuy} M.,  {Leibundgut}
  B.,  {Kirshner} R.~P.,   {Foltz} C.~B.,  1992, \mn@doi [\aj]
  {10.1086/116177}, \href
  {https://ui.adsabs.harvard.edu/abs/1992AJ....103.1632P} {103, 1632}

\bibitem[\protect\citeauthoryear{{Phillips}, {Lira}, {Suntzeff}, {Schommer},
  {Hamuy}  \& {Maza}}{{Phillips} et~al.}{1999}]{1999AJ....118.1766P}
{Phillips} M.~M.,  {Lira} P.,  {Suntzeff} N.~B.,  {Schommer} R.~A.,  {Hamuy}
  M.,   {Maza} J.,  1999, \mn@doi [\aj] {10.1086/301032}, \href
  {https://ui.adsabs.harvard.edu/abs/1999AJ....118.1766P} {118, 1766}

\bibitem[\protect\citeauthoryear{{Pignata} et~al.,}{{Pignata}
  et~al.}{2008}]{2008MNRAS.388..971P}
{Pignata} G.,  et~al., 2008, \mn@doi [\mnras]
  {10.1111/j.1365-2966.2008.13434.x}, \href
  {https://ui.adsabs.harvard.edu/abs/2008MNRAS.388..971P} {388, 971}

\bibitem[\protect\citeauthoryear{{Raskin}, {Timmes}, {Scannapieco}, {Diehl}  \&
  {Fryer}}{{Raskin} et~al.}{2009}]{2009MNRAS.399L.156R}
{Raskin} C.,  {Timmes} F.~X.,  {Scannapieco} E.,  {Diehl} S.,   {Fryer} C.,
  2009, \mn@doi [\mnras] {10.1111/j.1745-3933.2009.00743.x}, \href
  {https://ui.adsabs.harvard.edu/abs/2009MNRAS.399L.156R} {399, L156}

\bibitem[\protect\citeauthoryear{{Sai} et~al.,}{{Sai}
  et~al.}{2022}]{2022MNRAS.514.3541S}
{Sai} H.,  et~al., 2022, \mn@doi [\mnras] {10.1093/mnras/stac1525}, \href
  {https://ui.adsabs.harvard.edu/abs/2022MNRAS.514.3541S} {514, 3541}

\bibitem[\protect\citeauthoryear{{Schlafly} \& {Finkbeiner}}{{Schlafly} \&
  {Finkbeiner}}{2011}]{2011ApJ...737..103S}
{Schlafly} E.~F.,  {Finkbeiner} D.~P.,  2011, \mn@doi [\apj]
  {10.1088/0004-637X/737/2/103}, \href
  {https://ui.adsabs.harvard.edu/abs/2011ApJ...737..103S} {737, 103}

\bibitem[\protect\citeauthoryear{{Seitenzahl} et~al.,}{{Seitenzahl}
  et~al.}{2013}]{2013MNRAS.429.1156S}
{Seitenzahl} I.~R.,  et~al., 2013, \mn@doi [\mnras] {10.1093/mnras/sts402},
  \href {https://ui.adsabs.harvard.edu/abs/2013MNRAS.429.1156S} {429, 1156}

\bibitem[\protect\citeauthoryear{{Shen} \& {Bildsten}}{{Shen} \&
  {Bildsten}}{2009}]{2009ApJ...699.1365S}
{Shen} K.~J.,  {Bildsten} L.,  2009, \mn@doi [\apj]
  {10.1088/0004-637X/699/2/1365}, \href
  {https://ui.adsabs.harvard.edu/abs/2009ApJ...699.1365S} {699, 1365}

\bibitem[\protect\citeauthoryear{{Shen}, {Kasen}, {Miles}  \&
  {Townsley}}{{Shen} et~al.}{2018}]{2018ApJ...854...52S}
{Shen} K.~J.,  {Kasen} D.,  {Miles} B.~J.,   {Townsley} D.~M.,  2018, \mn@doi
  [\apj] {10.3847/1538-4357/aaa8de}, \href
  {https://ui.adsabs.harvard.edu/abs/2018ApJ...854...52S} {854, 52}

\bibitem[\protect\citeauthoryear{{Shingles} et~al.,}{{Shingles}
  et~al.}{2020}]{2020MNRAS.492.2029S}
{Shingles} L.~J.,  et~al., 2020, \mn@doi [\mnras] {10.1093/mnras/stz3412},
  \href {https://ui.adsabs.harvard.edu/abs/2020MNRAS.492.2029S} {492, 2029}

\bibitem[\protect\citeauthoryear{{Shivvers} et~al.,}{{Shivvers}
  et~al.}{2019}]{2019MNRAS.482.1545S}
{Shivvers} I.,  et~al., 2019, \mn@doi [\mnras] {10.1093/mnras/sty2719}, \href
  {https://ui.adsabs.harvard.edu/abs/2019MNRAS.482.1545S} {482, 1545}

\bibitem[\protect\citeauthoryear{{Silverman}, {Ganeshalingam}, {Li},
  {Filippenko}, {Miller}  \& {Poznanski}}{{Silverman}
  et~al.}{2011}]{2011MNRAS.410..585S}
{Silverman} J.~M.,  {Ganeshalingam} M.,  {Li} W.,  {Filippenko} A.~V.,
  {Miller} A.~A.,   {Poznanski} D.,  2011, \mn@doi [\mnras]
  {10.1111/j.1365-2966.2010.17474.x}, \href
  {https://ui.adsabs.harvard.edu/abs/2011MNRAS.410..585S} {410, 585}

\bibitem[\protect\citeauthoryear{{Silverman} et~al.,}{{Silverman}
  et~al.}{2012a}]{2012MNRAS.425.1789S}
{Silverman} J.~M.,  et~al., 2012a, \mn@doi [\mnras]
  {10.1111/j.1365-2966.2012.21270.x}, \href
  {https://ui.adsabs.harvard.edu/abs/2012MNRAS.425.1789S} {425, 1789}

\bibitem[\protect\citeauthoryear{{Silverman} et~al.,}{{Silverman}
  et~al.}{2012b}]{2012ApJ...756L...7S}
{Silverman} J.~M.,  et~al., 2012b, \mn@doi [\apjl]
  {10.1088/2041-8205/756/1/L7}, \href
  {https://ui.adsabs.harvard.edu/abs/2012ApJ...756L...7S} {756, L7}

\bibitem[\protect\citeauthoryear{{Silverman}, {Ganeshalingam}  \&
  {Filippenko}}{{Silverman} et~al.}{2013}]{2013MNRAS.430.1030S}
{Silverman} J.~M.,  {Ganeshalingam} M.,   {Filippenko} A.~V.,  2013, \mn@doi
  [\mnras] {10.1093/mnras/sts674}, \href
  {https://ui.adsabs.harvard.edu/abs/2013MNRAS.430.1030S} {430, 1030}

\bibitem[\protect\citeauthoryear{{Sim}, {R{\"o}pke}, {Hillebrandt}, {Kromer},
  {Pakmor}, {Fink}, {Ruiter}  \& {Seitenzahl}}{{Sim}
  et~al.}{2010}]{2010ApJ...714L..52S}
{Sim} S.~A.,  {R{\"o}pke} F.~K.,  {Hillebrandt} W.,  {Kromer} M.,  {Pakmor} R.,
   {Fink} M.,  {Ruiter} A.~J.,   {Seitenzahl} I.~R.,  2010, \mn@doi [\apjl]
  {10.1088/2041-8205/714/1/L52}, \href
  {https://ui.adsabs.harvard.edu/abs/2010ApJ...714L..52S} {714, L52}

\bibitem[\protect\citeauthoryear{{Simon} et~al.,}{{Simon}
  et~al.}{2007}]{2007ApJ...671L..25S}
{Simon} J.~D.,  et~al., 2007, \mn@doi [\apjl] {10.1086/524707}, \href
  {https://ui.adsabs.harvard.edu/abs/2007ApJ...671L..25S} {671, L25}

\bibitem[\protect\citeauthoryear{{Simon} et~al.,}{{Simon}
  et~al.}{2009}]{2009ApJ...702.1157S}
{Simon} J.~D.,  et~al., 2009, \mn@doi [\apj] {10.1088/0004-637X/702/2/1157},
  \href {https://ui.adsabs.harvard.edu/abs/2009ApJ...702.1157S} {702, 1157}

\bibitem[\protect\citeauthoryear{{Srivastav}, {Ninan}, {Kumar}, {Anupama},
  {Sahu}, {Ojha}  \& {Prabhu}}{{Srivastav} et~al.}{2016}]{2016MNRAS.457.1000S}
{Srivastav} S.,  {Ninan} J.~P.,  {Kumar} B.,  {Anupama} G.~C.,  {Sahu} D.~K.,
  {Ojha} D.~K.,   {Prabhu} T.~P.,  2016, \mn@doi [\mnras]
  {10.1093/mnras/stw039}, \href
  {https://ui.adsabs.harvard.edu/abs/2016MNRAS.457.1000S} {457, 1000}

\bibitem[\protect\citeauthoryear{{Stahl} et~al.,}{{Stahl}
  et~al.}{2020}]{2020MNRAS.492.4325S}
{Stahl} B.~E.,  et~al., 2020, \mn@doi [\mnras] {10.1093/mnras/staa102}, \href
  {https://ui.adsabs.harvard.edu/abs/2020MNRAS.492.4325S} {492, 4325}

\bibitem[\protect\citeauthoryear{{Stanishev} et~al.,}{{Stanishev}
  et~al.}{2007}]{2007A&A...469..645S}
{Stanishev} V.,  et~al., 2007, \mn@doi [\aap] {10.1051/0004-6361:20066020},
  \href {https://ui.adsabs.harvard.edu/abs/2007A&A...469..645S} {469, 645}

\bibitem[\protect\citeauthoryear{{Timmes}, {Brown}  \& {Truran}}{{Timmes}
  et~al.}{2003}]{2003ApJ...590L..83T}
{Timmes} F.~X.,  {Brown} E.~F.,   {Truran} J.~W.,  2003, \mn@doi [\apjl]
  {10.1086/376721}, \href
  {https://ui.adsabs.harvard.edu/abs/2003ApJ...590L..83T} {590, L83}

\bibitem[\protect\citeauthoryear{{Tiwari}, {Graur}, {Fisher}, {Seitenzahl},
  {Leung}, {Nomoto}, {Perets}  \& {Shen}}{{Tiwari}
  et~al.}{2022}]{2022MNRAS.tmp.1645T}
{Tiwari} V.,  {Graur} O.,  {Fisher} R.,  {Seitenzahl} I.,  {Leung} S.-C.,
  {Nomoto} K.,  {Perets} H.~B.,   {Shen} K.,  2022, \mn@doi [\mnras]
  {10.1093/mnras/stac1618}, \href
  {https://ui.adsabs.harvard.edu/abs/2022MNRAS.tmp.1645T} {}

\bibitem[\protect\citeauthoryear{{Townsley}, {Miles}, {Shen}  \&
  {Kasen}}{{Townsley} et~al.}{2019}]{2019ApJ...878L..38T}
{Townsley} D.~M.,  {Miles} B.~J.,  {Shen} K.~J.,   {Kasen} D.,  2019, \mn@doi
  [\apjl] {10.3847/2041-8213/ab27cd}, \href
  {https://ui.adsabs.harvard.edu/abs/2019ApJ...878L..38T} {878, L38}

\bibitem[\protect\citeauthoryear{{Tripp}}{{Tripp}}{1998}]{1998A&A...331..815T}
{Tripp} R.,  1998, \aap, \href
  {https://ui.adsabs.harvard.edu/abs/1998A&A...331..815T} {331, 815}

\bibitem[\protect\citeauthoryear{{Tucker}, {Shappee}  \& {Wisniewski}}{{Tucker}
  et~al.}{2019}]{2019ApJ...872L..22T}
{Tucker} M.~A.,  {Shappee} B.~J.,   {Wisniewski} J.~P.,  2019, \mn@doi [\apjl]
  {10.3847/2041-8213/ab0286}, \href
  {https://ui.adsabs.harvard.edu/abs/2019ApJ...872L..22T} {872, L22}

\bibitem[\protect\citeauthoryear{{Tucker} et~al.,}{{Tucker}
  et~al.}{2020}]{2020MNRAS.493.1044T}
{Tucker} M.~A.,  et~al., 2020, \mn@doi [\mnras] {10.1093/mnras/stz3390}, \href
  {https://ui.adsabs.harvard.edu/abs/2020MNRAS.493.1044T} {493, 1044}

\bibitem[\protect\citeauthoryear{{Tucker}, {Ashall}, {Shappee}, {Kochanek},
  {Stanek}  \& {Garnavich}}{{Tucker} et~al.}{2022}]{2022ApJ...926L..25T}
{Tucker} M.~A.,  {Ashall} C.,  {Shappee} B.~J.,  {Kochanek} C.~S.,  {Stanek}
  K.~Z.,   {Garnavich} P.,  2022, \mn@doi [\apjl] {10.3847/2041-8213/ac4fbd},
  \href {https://ui.adsabs.harvard.edu/abs/2022ApJ...926L..25T} {926, L25}

\bibitem[\protect\citeauthoryear{{Walker} et~al.,}{{Walker}
  et~al.}{2015}]{2015ApJS..219...13W}
{Walker} E.~S.,  et~al., 2015, \mn@doi [\apjs] {10.1088/0067-0049/219/1/13},
  \href {https://ui.adsabs.harvard.edu/abs/2015ApJS..219...13W} {219, 13}

\bibitem[\protect\citeauthoryear{{Wang}, {Wang}, {Zhou}, {Lou}  \& {Li}}{{Wang}
  et~al.}{2005}]{2005ApJ...620L..87W}
{Wang} X.,  {Wang} L.,  {Zhou} X.,  {Lou} Y.-Q.,   {Li} Z.,  2005, \mn@doi
  [\apjl] {10.1086/428774}, \href
  {https://ui.adsabs.harvard.edu/abs/2005ApJ...620L..87W} {620, L87}

\bibitem[\protect\citeauthoryear{{Wang} et~al.,}{{Wang}
  et~al.}{2008}]{2008ApJ...675..626W}
{Wang} X.,  et~al., 2008, \mn@doi [\apj] {10.1086/526413}, \href
  {https://ui.adsabs.harvard.edu/abs/2008ApJ...675..626W} {675, 626}

\bibitem[\protect\citeauthoryear{{Wang} et~al.,}{{Wang}
  et~al.}{2009a}]{2009ApJ...697..380W}
{Wang} X.,  et~al., 2009a, \mn@doi [\apj] {10.1088/0004-637X/697/1/380}, \href
  {https://ui.adsabs.harvard.edu/abs/2009ApJ...697..380W} {697, 380}

\bibitem[\protect\citeauthoryear{{Wang} et~al.,}{{Wang}
  et~al.}{2009b}]{2009ApJ...699L.139W}
{Wang} X.,  et~al., 2009b, \mn@doi [\apjl] {10.1088/0004-637X/699/2/L139},
  \href {https://ui.adsabs.harvard.edu/abs/2009ApJ...699L.139W} {699, L139}

\bibitem[\protect\citeauthoryear{{Wang}, {Wang}, {Filippenko}, {Zhang}  \&
  {Zhao}}{{Wang} et~al.}{2013}]{2013Sci...340..170W}
{Wang} X.,  {Wang} L.,  {Filippenko} A.~V.,  {Zhang} T.,   {Zhao} X.,  2013,
  \mn@doi [Science] {10.1126/science.1231502}, \href
  {https://ui.adsabs.harvard.edu/abs/2013Sci...340..170W} {340, 170}

\bibitem[\protect\citeauthoryear{{Wang}, {Chen}, {Wang}, {Hu}, {Xi}, {Yang},
  {Zhao}  \& {Li}}{{Wang} et~al.}{2019}]{2019ApJ...882..120W}
{Wang} X.,  {Chen} J.,  {Wang} L.,  {Hu} M.,  {Xi} G.,  {Yang} Y.,  {Zhao} X.,
   {Li} W.,  2019, \mn@doi [\apj] {10.3847/1538-4357/ab26b5}, \href
  {https://ui.adsabs.harvard.edu/abs/2019ApJ...882..120W} {882, 120}

\bibitem[\protect\citeauthoryear{{Wang} et~al.,}{{Wang}
  et~al.}{2020}]{2020ApJ...904...14W}
{Wang} L.,  et~al., 2020, \mn@doi [\apj] {10.3847/1538-4357/abba82}, \href
  {https://ui.adsabs.harvard.edu/abs/2020ApJ...904...14W} {904, 14}

\bibitem[\protect\citeauthoryear{{Webbink}}{{Webbink}}{1984}]{1984ApJ...277..355W}
{Webbink} R.~F.,  1984, \mn@doi [\apj] {10.1086/161701}, \href
  {https://ui.adsabs.harvard.edu/abs/1984ApJ...277..355W} {277, 355}

\bibitem[\protect\citeauthoryear{{Whelan} \& {Iben}}{{Whelan} \&
  {Iben}}{1973}]{1973ApJ...186.1007W}
{Whelan} J.,  {Iben} Icko J.,  1973, \mn@doi [\apj] {10.1086/152565}, \href
  {https://ui.adsabs.harvard.edu/abs/1973ApJ...186.1007W} {186, 1007}

\bibitem[\protect\citeauthoryear{{Xi} et~al.,}{{Xi}
  et~al.}{2022}]{2022MNRAS.517.4098X}
{Xi} G.,  et~al., 2022, \mn@doi [\mnras] {10.1093/mnras/stac2848}, \href
  {https://ui.adsabs.harvard.edu/abs/2022MNRAS.517.4098X} {517, 4098}

\bibitem[\protect\citeauthoryear{{Yaron} \& {Gal-Yam}}{{Yaron} \&
  {Gal-Yam}}{2012}]{2012PASP..124..668Y}
{Yaron} O.,  {Gal-Yam} A.,  2012, \mn@doi [\pasp] {10.1086/666656}, \href
  {https://ui.adsabs.harvard.edu/abs/2012PASP..124..668Y} {124, 668}

\bibitem[\protect\citeauthoryear{{Zeng} et~al.,}{{Zeng}
  et~al.}{2021}]{2021ApJ...919...49Z}
{Zeng} X.,  et~al., 2021, \mn@doi [\apj] {10.3847/1538-4357/ac0e9c}, \href
  {https://ui.adsabs.harvard.edu/abs/2021ApJ...919...49Z} {919, 49}

\bibitem[\protect\citeauthoryear{{Zhang}, {Wang}, {Bai}, {Zhang}, {Wang},
  {Liu}, {Zhao}  \& {Chen}}{{Zhang} et~al.}{2014}]{2014AJ....148....1Z}
{Zhang} J.-J.,  {Wang} X.-F.,  {Bai} J.-M.,  {Zhang} T.-M.,  {Wang} B.,  {Liu}
  Z.-W.,  {Zhao} X.-L.,   {Chen} J.-C.,  2014, \mn@doi [\aj]
  {10.1088/0004-6256/148/1/1}, \href
  {https://ui.adsabs.harvard.edu/abs/2014AJ....148....1Z} {148, 1}

\bibitem[\protect\citeauthoryear{{Zhang} et~al.,}{{Zhang}
  et~al.}{2016}]{2016ApJ...820...67Z}
{Zhang} K.,  et~al., 2016, \mn@doi [\apj] {10.3847/0004-637X/820/1/67}, \href
  {https://ui.adsabs.harvard.edu/abs/2016ApJ...820...67Z} {820, 67}

\makeatother
\end{thebibliography}




\appendix

\section{Overview of SNe~Ia and Fitting Results} \label{sec:spectra}

\begin{table*}
\caption{SNe~Ia light  curve, spectral, and host-galaxy information.\label{tab:base_para}}
\begin{tabular}{ccccccccccc}
\hline
\hline
   {Name} &    {Type$^{a}$} &    {Host galaxy} &    {Redshift} &    {$A_V$$^{b}$} &
   {$\Delta m_{15}(B)$} &    {\textcolor{black}{Si vel., Phase}} & Date of max.$^{c}$ &    {Ref.$^{d}$} &    {Ref.$^{e}$} &    {Ref.} \\
   {} &    {} &    {} &    {} &    {[mag]} &
   {[mag]} &    {[1000 $\rm km\ s^{-1}$, d]} & {UTC} &    {Spec.} &    {LC} &    {$A_V$} \\
\hline
SN 2019ein & pec & NGC 5353 & 0.007755 & 0.033 & 1.35$\pm$0.01 & 13.10$\pm$0.26, +0 & 20190516 & 1 & 2 & 3 \\
SN 2019np & norm & NGC 3254 & 0.00452 & 0.054 & 1.04$\pm$0.07 & 10.20$\pm$0.20, +0 & 20190127 & 4 & 4 & 3 \\
SN 2021hpr & norm & NGC 3147 & 0.009346 & 0.065 & 1.0$\pm$0.06 & 11.98$\pm$0.36, +1 & 20210418 & 5 & 5 & 3 \\
SN 2021wuf & norm & NGC 6500/NGC 6501 & 0.01 & 0.241 & 1.11$\pm$0.06 & 12.52$\pm$0.37, +1 & 20210905 & 6 & 6 & 3 \\
SN 2022hrs & norm & NGC 4647 & 0.0047 & 0.744 & 1.11$\pm$0.08 & 13.75$\pm$0.20, +0 & 20220429 & 7 & 7 & 7,3 \\
\hline
SN 1986G & 91bg & NGC 5128 & 0.001825 & 2.79 & 1.73$\pm$0.07 & 10.11$\pm$0.23, +0 & 19860511 & 8 & 9,10 & 9 \\
SN 1990N & norm & NGC 4639 & 0.003369 & 0.069 & 1.03$\pm$0.06 & 10.20$\pm$0.20, +2 & 19900710 & 11 & 12 & 3 \\
SN 1998bu & norm & NGC 3368 & 0.002992 & 1.054 & 1.02$\pm$0.04 & 10.82$\pm$0.20, -1 & 19980519 & 13 & 14 & 14 \\
SN 1999aa & 91T & NGC 2595 & 0.014907 & 0.106 & 0.75$\pm$0.03 & 10.54$\pm$0.31, -1 & 19990222 & 15 & 16 & 3 \\
SN 2002bo & norm & NGC 3190 & 0.0043 & 1.333 & 1.13$\pm$0.05 & 13.41$\pm$0.31, +0 & 20020323 & 17 & 17 & 17 \\
SN 2002dj & norm & NGC 5018 & 0.009393 & 0.254 & 1.08$\pm$0.05 & 14.11$\pm$0.56, -3 & 20020624 & 18 & 18 & 3 \\
SN 2003du & norm & NGC 9391 & 0.006408 & 0.027 & 1.04$\pm$0.03 & 10.80$\pm$0.20, +0 & 20030506 & 19 & 20 & 3 \\
SN 2003gs & 91bg & NGC 936 & 0.00477 & 0.094 & 1.83$\pm$0.02 & 11.00$\pm$0.20, +2 & 20030728 & 21 & 22 & 3 \\
SN 2003hv & norm & NGC 1201 & 0.005624 & 0.041 & 1.61$\pm$0.02 & 11.14$\pm$0.20, +1 & 20030909 & 23 & 23 & 3 \\
SN 2003kf & norm & MCG -02-16-002 & 0.0074 & 0.833 & 0.98$\pm$0.07 & 11.39$\pm$0.21, -3 & 20031207 & 24 & 25 & 3 \\
SN 2004eo & norm & NGC 6928 & 0.015718 & 0.288 & 1.45$\pm$0.04 & 10.79$\pm$0.20, -3 & 20040930 & 26 & 26 & 3 \\
SN 2005cf & norm & MCG -01-39-003 & 0.006461 & 0.261 & 1.05$\pm$0.03 & 10.51$\pm$0.20, +0 & 20050612 & 27 & 28 & 3 \\
SN 2006X & norm & NGC 4321 & 0.005294 & 2.172 & 1.17$\pm$0.05 & 17.49$\pm$0.20, +0 & 20060219 & 29 & 29 & 29,3 \\
SN 2007af & norm & NGC 5584 & 0.005464 & 0.104 & 1.2$\pm$0.05 & 11.07$\pm$0.20, +0 & 20070314 & 24 & 30,25 & 3 \\
SN 2007le & norm & NGC 7721 & 0.006721 & 0.71 & 1.02$\pm$0.04 & 12.36$\pm$0.20, +3 & 20071025 & 31 & 32,33 & 32,3 \\
SN 2008Q & norm & NGC 524 & 0.0081 & 0.221 & 1.39$\pm$0.1 & 11.82$\pm$0.26, +0 & 20080209 & 24 & 34 & 3 \\
SN 2011by & norm & NGC 3972 & 0.002843 & 0.037 & 1.14$\pm$0.03 & 10.27$\pm$0.21, +1 & 20110509 & 33 & 33 & 3 \\
SN 2011fe & norm & M 101 & 0.000804 & 0.024 & 1.18$\pm$0.03 & 10.54$\pm$0.20, +0 & 20110910 & 35 & 36 & 3 \\
SN 2012cg & 91T & NGC 4424 & 0.001458 & 0.62 & 0.86$\pm$0.02 & 10.39$\pm$0.20, +3 & 20120603 & 37 & 37 & 38,3 \\
SN 2012fr & norm & NGC 1365 & 0.004 & 0.054 & 0.85$\pm$0.05 & 11.65$\pm$0.20, +0 & 20121112 & 39 & 40 & 3 \\
SN 2012hr & norm & ESO 121-G026 & 0.008 & 0.121 & 1.1$\pm$0.04 & 11.38$\pm$0.22, +5 & 20121228 & 41 & 42 & 3 \\
SN 2013aa & norm & NGC 5643 & 0.003999 & 0.453 & 0.95$\pm$0.01 & 10.20$\pm$0.20, +0 & 20130220 & 41 & 43 & 3 \\
SN 2013cs & norm & ESO 576-G017 & 0.00924 & 0.248 & 1.07$\pm$0.05 & 12.86$\pm$0.20, +1 & 20130525 & 44 & 45 & 3 \\
SN 2013dy & norm & NGC 7250 & 0.00389 & 0.409 & 0.89$\pm$0.01 & 10.51$\pm$0.20, +1 & 20130728 & 46 & 46 & 3 \\
SN 2013gy & norm & NGC 1418 & 0.014023 & 0.154 & 1.23$\pm$0.06 & 10.79$\pm$0.20, -1 & 20131222 & 47 & 47 & 3 \\
SN 2014J & norm & NGC 3034 & 0.000677 & 2.421 & 1.1$\pm$0.02 & 12.08$\pm$0.20, +0 & 20140201 & 48 & 49 & 50,3 \\
SN 2015F & norm & NGC 2442 & 0.0049 & 0.542 & 1.35$\pm$0.03 & 10.44$\pm$0.23, -3 & 20150325 & 51 & 52 & 3 \\
SN 2017cbv & norm & NGC 5643 & 0.003999 & 0.45 & 0.99$\pm$0.01 & 9.75$\pm$0.20, -1 & 20170328 & 53 & 54 & 3 \\
SN 2017fgc & norm & NGC 0474 & 0.008 & 0.091 & 1.05$\pm$0.07 & 15.56$\pm$0.26, +1 & 20170725 & 55 & 56 & 3 \\
SN 2018oh & norm & UGC 04780 & 0.012 & 0.12 & 0.96$\pm$0.03 & 10.55$\pm$0.24, +0 & 20180213 & 57 & 57 & 3 \\
ASASSN-14jg & norm & PGC 128348 & 0.0148 & 0.04 & 0.92$\pm$0.01 & 10.76$\pm$0.20, +6 & 20141031 & 44 & 41 & 3 \\
\hline
\end{tabular}
\begin{flushleft}
\textcolor{black}{Reference: (1) \citet{2020ApJ...897..159P}; (2) \citet{2022MNRAS.517.4098X}; (3) \citet{2011ApJ...737..103S}; (4) \citet{2022MNRAS.514.3541S}; (5) Iskandar et al. (in prep.); (6) Zeng et al. (in prep.); (7) Liu et al. (in prep.); (8) \citet{1992A&A...259...63C}; (9) \citet{1987PASP...99..592P}; (10) \citet{1999AJ....118.1766P}; (11) \citet{1993A&A...269..423M}; (12) \citet{1998AJ....115..234L}; (13) \citet{2008AJ....135.1598M}; (14) \citet{1999ApJS..125...73J}; (15) \citet{2012MNRAS.425.1789S}; (16) \citet{2000ApJ...539..658K}; (17) \citet{2004MNRAS.348..261B}; (18) \citet{2008MNRAS.388..971P}; (19) \citet{2005ASPC..342..250G}; (20) \citet{2005A&A...429..667A}; (21) \citet{2003IAUC.8171....1E}; (22) \citet{2009AJ....138.1584K}; (23) \citet{2009A&A...505..265L}; (24) \citet{2012AJ....143..126B}; (25) \citet{2009ApJ...700..331H}; (26) \citet{2007MNRAS.377.1531P}; (27) \citet{2007A&A...471..527G}; (28) \citet{2009ApJ...697..380W}; (29) \citet{2008ApJ...675..626W}; (30) \citet{2007ApJ...671L..25S}; (31) \citet{2013ApJ...773...53F}; (32) \citet{2009ApJ...702.1157S}; (33) \citet{2013MNRAS.430.1030S}; (34) \citet{2010ApJ...721.1627M}; (35) \citet{2013A&A...554A..27P}; (36) \citet{2016ApJ...820...67Z}; (37) \citet{2016ApJ...820...92M}; (38) \citet{2012ApJ...756L...7S}; (39) \citet{2013ApJ...770...29C}; (40) \citet{2014AJ....148....1Z}; (41) \citet{2017MNRAS.472.3437G}; (42) \citet{2014Ap&SS.354...89B}; (43) \citet{2020ApJ...895..118B}; (44) \citet{2016PASA...33...55C}; (45) \citet{2015ApJS..219...13W}; (46) \citet{2015MNRAS.452.4307P}; (47) \citet{2019A&A...627A.174H}; (48) \citet{2016MNRAS.457.1000S}; (49) \citet{2019ApJ...882...30L}; (50) \citet{2014MNRAS.443.2887F}; (51) \citet{2016MNRAS.461.1308F}; (52) \citet{2017MNRAS.464.4476C}; (53) \citet{2017ApJ...845L..11H}; (54) \citet{2020ApJ...904...14W}; (55) \citet{2020MNRAS.492.4325S}; (56) \citet{2021ApJ...919...49Z}; (57) \citet{2019ApJ...870...12L}. The sample from our own project are put at the top of the table and separated from the public sample by a line.\\
$^{a}$The type is estimated with SNID \citep{2007ApJ...666.1024B} using the spectrum around the maximum light except for SN~2003gs, SN~2013aa, and SN~2019ein, whose classifications are taken from \citet{2009AJ....138.1584K}, \citet{2013CBET.3416....1P}, and \citet{2019TNSCR.701....1B}, respectively. \\
$^{b}$We assume host $R_V=3.1$ except for SN~2006X, SN~2007le, and SN~2014J, whose host $R_V$ values are 1.48, 2.56, and 1.6, respectively.\\
$^{c}$The date of $B$-band maximum light used in this work. \\
$^{d}$Sources for the spectra used to calculate Si~II $\lambda$6355 velocities around the time of $B$-band maximum and identify the type. The Si~II $\lambda$6355 velocities of SN~2003gs and SN~2013aa are taken from \citet{2003IAUC.8171....1E} and \citet{2017MNRAS.472.3437G}, respectively.\\
$^{e}$References for the date of $B$-band maximum and $\Delta m_{15}(B)$. For SN~2003kf, SN~2012hr, and SN~2013cs, these two parameters are measured in this work with SALT2 using the light curves from the references. \\}
\end{flushleft}
\end{table*}

\begin{table*}
\caption{Multicomponent Gaussian-fit parameters of nebular-phase emission lines, and implications.
\label{tab:Multi}}
\begin{tabular}{ccccccccccccc}
\hline
   {Name} &    {Phase} &    {[Fe~II] Vel.} &    {[Ni~II] Vel.} & 
   {[Fe~II] FWHM} &    {[Ni~II] FWHM} &    {Flux ratio} &    {Nebular Vel.} &    {$M_{\rm Ni}$/$M_{\rm Fe}$} &     {Ref.} \\
   {} &    {[days]} &    {[km s$^{-1}$]} &    {[km s$^{-1}$]} & 
   {[km s$^{-1}$]} &    {[km s$^{-1}$]} &    {$\lambda$7155/$\lambda$7378}  &    {[km s$^{-1}$]} &    {$t \rightarrow \infty$} &    {Spec.} \\
\hline
SN 2019ein & +313 & -11$\pm$623 & 2436$\pm$326 & 12531$\pm$434 & 4368$\pm$379 & 0.370$\pm$0.023 & 2436$\pm$326 & 0.065$\pm$0.026 & 1 \\
SN 2019np & +303 & -1700$\pm$227 & -3258$\pm$412 & 6980$\pm$80 & 5520$\pm$240 & 0.233$\pm$0.008 & -1700$\pm$227 & 0.041$\pm$0.016 & 2 \\
SN 2019np & +368 & -1526$\pm$318 & -3057$\pm$404 & 7128$\pm$331 & 5639$\pm$79 & 0.305$\pm$0.018 & -1526$\pm$318 & 0.055$\pm$0.022 & 2 \\
SN 2021hpr & +263 & 790$\pm$249 & 125$\pm$325 & 7453$\pm$626 & 5366$\pm$236 & 0.338$\pm$0.036 & 790$\pm$249 & 0.057$\pm$0.024 & 3 \\
SN 2021hpr & +288 & 623$\pm$231 & 676$\pm$295 & 8363$\pm$222 & 5998$\pm$82 & 0.438$\pm$0.028 & 623$\pm$231 & 0.076$\pm$0.031 & 3 \\
SN 2021wuf & +208 & -3226$\pm$351 & -2776$\pm$242 & 7691$\pm$162 & 4597$\pm$207 & 0.494$\pm$0.038 & -2776$\pm$242 & 0.079$\pm$0.032 & 4 \\
SN 2021wuf & +299 & -3040$\pm$467 & -2503$\pm$362 & 9977$\pm$144 & 4880$\pm$266 & 0.488$\pm$0.022 & -2503$\pm$362 & 0.085$\pm$0.034 & 4 \\
SN 2022hrs & +297 & 628$\pm$554 & 777$\pm$403 & 9064$\pm$213 & 6097$\pm$359 & 0.367$\pm$0.057 & 628$\pm$554 & 0.064$\pm$0.027 & 5 \\
SN 2022hrs & +327 & 890$\pm$498 & 739$\pm$657 & 8212$\pm$515 & 6510$\pm$357 & 0.491$\pm$0.044 & 890$\pm$498 & 0.087$\pm$0.035 & 5 \\
\hline
SN 1986G$^a$ & +256 & -965$\pm$237 & -965$\pm$237 & 8109$\pm$163 & 3569$\pm$393 & 0.293$\pm$0.005 & -965$\pm$237 & 0.049$\pm$0.020 & 6 \\
SN 1990N & +227 & -1696$\pm$236 & -3606$\pm$323 & 8875$\pm$192 & 6472$\pm$103 & 0.108$\pm$0.007 & -1696$\pm$236 & 0.018$\pm$0.007 & 7 \\
SN 1990N & +255 & -1354$\pm$274 & -1827$\pm$569 & 9268$\pm$272 & 6655$\pm$806 & 0.158$\pm$0.042 & -1354$\pm$274 & 0.027$\pm$0.013 & 7 \\
SN 1990N & +280 & -1315$\pm$229 & -1968$\pm$418 & 9017$\pm$114 & 8165$\pm$358 & 0.191$\pm$0.018 & -1315$\pm$229 & 0.033$\pm$0.014 & 7 \\
SN 1998bu & +236 & -1322$\pm$227 & -1604$\pm$272 & 8198$\pm$68 & 5533$\pm$319 & 0.471$\pm$0.020 & -1322$\pm$227 & 0.078$\pm$0.031 & 8 \\
SN 1998bu & +280 & -1429$\pm$265 & -1686$\pm$289 & 8393$\pm$437 & 4824$\pm$303 & 0.353$\pm$0.009 & -1429$\pm$265 & 0.061$\pm$0.024 & 8 \\
SN 1999aa & +260 & 114$\pm$263 & -209$\pm$277 & 8030$\pm$100 & 3392$\pm$211 & 0.138$\pm$0.009 & 114$\pm$263 & 0.023$\pm$0.009 & 8 \\
SN 2002bo & +312 & 1251$\pm$278 & 1859$\pm$324 & 7505$\pm$579 & 8452$\pm$668 & 1.105$\pm$0.143 & 1251$\pm$278 & 0.194$\pm$0.081 & 9 \\
SN 2002dj & +275 & 1224$\pm$276 & 2202$\pm$334 & 9535$\pm$210 & 5445$\pm$208 & 0.222$\pm$0.015 & 1224$\pm$276 & 0.038$\pm$0.015 & 10 \\
SN 2003du & +221 & -2001$\pm$292 & -3024$\pm$741 & 8112$\pm$355 & 8075$\pm$803 & 0.323$\pm$0.085 & -2001$\pm$292 & 0.052$\pm$0.025 & 11 \\
SN 2003gs$^a$ & +201 & 807$\pm$258 & 807$\pm$258 & 9687$\pm$212 & 4932$\pm$87 & 0.303$\pm$0.021 & 807$\pm$258 & 0.048$\pm$0.019 & 8 \\
SN 2003hv & +320 & -2575$\pm$218 & -3688$\pm$257 & 8525$\pm$225 & 4984$\pm$107 & 0.541$\pm$0.030 & -3688$\pm$257 & 0.095$\pm$0.038 & 12 \\
SN 2003kf & +401 & 302$\pm$396 & 115$\pm$728 & 10651$\pm$212 & 4944$\pm$1585 & 0.144$\pm$0.021 & 302$\pm$396 & 0.026$\pm$0.011 & 9 \\
SN 2004eo & +228 & -994$\pm$249 & -2266$\pm$352 & 8329$\pm$149 & 8556$\pm$325 & 0.459$\pm$0.040 & -994$\pm$249 & 0.075$\pm$0.031 & 13 \\
SN 2005cf & +319 & 67$\pm$237 & -84$\pm$1647 & 8574$\pm$1106 & 6790$\pm$461 & 0.208$\pm$0.021 & 67$\pm$237 & 0.036$\pm$0.015 & 14 \\
SN 2006X & +277 & 2298$\pm$277 & 2248$\pm$237 & 8052$\pm$114 & 5434$\pm$117 & 0.472$\pm$0.031 & 2298$\pm$277 & 0.081$\pm$0.033 & 15 \\
SN 2006X & +360 & 2731$\pm$235 & 2488$\pm$244 & 9182$\pm$212 & 5272$\pm$72 & 0.531$\pm$0.024 & 2731$\pm$235 & 0.095$\pm$0.038 & 8 \\
SN 2007af & +303 & 421$\pm$226 & 367$\pm$349 & 7744$\pm$367 & 5775$\pm$223 & 0.265$\pm$0.029 & 421$\pm$226 & 0.046$\pm$0.019 & 9 \\
SN 2007le & +307 & 1497$\pm$246 & 1614$\pm$293 & 8809$\pm$287 & 5238$\pm$77 & 0.255$\pm$0.019 & 1497$\pm$246 & 0.045$\pm$0.018 & 8 \\
SN 2008Q & +201 & -2239$\pm$318 & -1485$\pm$228 & 11131$\pm$225 & 4765$\pm$29 & 0.419$\pm$0.024 & -1485$\pm$228 & 0.066$\pm$0.027 & 8 \\
SN 2011by & +207 & -1536$\pm$277 & -2845$\pm$488 & 7675$\pm$278 & 5547$\pm$435 & 0.246$\pm$0.029 & -1536$\pm$277 & 0.039$\pm$0.016 & 16 \\
SN 2011by & +311 & -1066$\pm$212 & -1716$\pm$243 & 7970$\pm$155 & 5080$\pm$40 & 0.318$\pm$0.013 & -1066$\pm$212 & 0.056$\pm$0.022 & 16 \\
SN 2011fe & +205 & -1143$\pm$210 & -2217$\pm$241 & 8423$\pm$78 & 5606$\pm$56 & 0.222$\pm$0.007 & -1143$\pm$210 & 0.035$\pm$0.014 & 17 \\
SN 2011fe & +226 & -1027$\pm$229 & -2077$\pm$308 & 8148$\pm$128 & 5709$\pm$145 & 0.254$\pm$0.012 & -1027$\pm$229 & 0.041$\pm$0.017 & 18 \\
SN 2011fe & +311 & -1068$\pm$251 & -1077$\pm$332 & 8329$\pm$215 & 7072$\pm$224 & 0.395$\pm$0.022 & -1068$\pm$251 & 0.069$\pm$0.028 & 18 \\
SN 2011fe & +348 & -638$\pm$251 & -1297$\pm$284 & 8507$\pm$107 & 6922$\pm$126 & 0.422$\pm$0.014 & -638$\pm$251 & 0.075$\pm$0.030 & 19 \\
SN 2011fe & +380 & -340$\pm$297 & -1047$\pm$356 & 9051$\pm$291 & 6980$\pm$226 & 0.397$\pm$0.038 & -340$\pm$297 & 0.071$\pm$0.029 & 18 \\
SN 2012cg & +279 & -1070$\pm$223 & -1830$\pm$354 & 7813$\pm$164 & 5603$\pm$94 & 0.187$\pm$0.005 & -1070$\pm$223 & 0.032$\pm$0.013 & 20 \\
SN 2012cg & +339 & -1111$\pm$234 & -841$\pm$314 & 8716$\pm$101 & 7279$\pm$141 & 0.267$\pm$0.010 & -1111$\pm$234 & 0.047$\pm$0.019 & 21 \\
SN 2012fr & +222 & 2041$\pm$264 & 2926$\pm$397 & 8033$\pm$93 & 5650$\pm$155 & 0.145$\pm$0.007 & 2041$\pm$264 & 0.023$\pm$0.009 & 22 \\
SN 2012fr & +261 & 2030$\pm$219 & 2305$\pm$793 & 7338$\pm$785 & 5583$\pm$308 & 0.185$\pm$0.007 & 2030$\pm$219 & 0.031$\pm$0.013 & 22 \\
SN 2012fr & +357 & 2205$\pm$226 & 2633$\pm$226 & 8059$\pm$52 & 5393$\pm$55 & 0.208$\pm$0.006 & 2205$\pm$226 & 0.037$\pm$0.015 & 21 \\
SN 2012fr & +367 & 2085$\pm$248 & 3303$\pm$235 & 8799$\pm$267 & 4567$\pm$40 & 0.234$\pm$0.014 & 2085$\pm$248 & 0.042$\pm$0.017 & 22 \\
SN 2012hr & +282 & 133$\pm$241 & 72$\pm$641 & 8342$\pm$349 & 7553$\pm$863 & 0.168$\pm$0.041 & 133$\pm$241 & 0.029$\pm$0.014 & 23 \\
SN 2013aa & +400 & -691$\pm$229 & -1072$\pm$260 & 8237$\pm$113 & 6666$\pm$171 & 0.268$\pm$0.012 & -691$\pm$229 & 0.048$\pm$0.020 & 23 \\
SN 2013cs & +262 & 1474$\pm$234 & 2622$\pm$246 & 9232$\pm$68 & 4838$\pm$74 & 0.147$\pm$0.003 & 1474$\pm$234 & 0.025$\pm$0.010 & 23 \\
SN 2013cs & +304 & 1343$\pm$245 & 1452$\pm$327 & 8179$\pm$170 & 6069$\pm$140 & 0.220$\pm$0.012 & 1343$\pm$245 & 0.038$\pm$0.015 & 21 \\
SN 2013dy & +333 & -1161$\pm$222 & -1478$\pm$273 & 7664$\pm$97 & 6722$\pm$220 & 0.224$\pm$0.010 & -1161$\pm$222 & 0.040$\pm$0.016 & 24 \\
SN 2013gy & +276 & -367$\pm$213 & -438$\pm$402 & 8200$\pm$186 & 7044$\pm$259 & 0.322$\pm$0.015 & -367$\pm$213 & 0.055$\pm$0.022 & 23 \\
SN 2014J & +231 & 336$\pm$213 & 673$\pm$252 & 8291$\pm$88 & 9370$\pm$229 & 0.349$\pm$0.019 & 336$\pm$213 & 0.057$\pm$0.023 & 22 \\
SN 2014J & +265 & 465$\pm$213 & 354$\pm$280 & 8071$\pm$116 & 7682$\pm$175 & 0.273$\pm$0.016 & 465$\pm$213 & 0.046$\pm$0.019 & 18 \\
SN 2014J & +292 & 581$\pm$232 & 239$\pm$356 & 7826$\pm$108 & 7671$\pm$136 & 0.323$\pm$0.012 & 581$\pm$232 & 0.056$\pm$0.022 & 18 \\
SN 2015F & +280 & -545$\pm$303 & -1233$\pm$254 & 7269$\pm$154 & 9694$\pm$799 & 0.857$\pm$0.136 & -545$\pm$303 & 0.147$\pm$0.063 & 23 \\
SN 2017cbv$^b$ & +318 & -1279$\pm$294 & -1279$\pm$294 & 7826$\pm$293 & 10293$\pm$958 & 0.421$\pm$0.072 & -1279$\pm$294 & 0.074$\pm$0.032 & 25 \\
SN 2017fgc & +384 & 2026$\pm$319 & 2112$\pm$444 & 8603$\pm$798 & 6175$\pm$642 & 0.358$\pm$0.081 & 2026$\pm$319 & 0.064$\pm$0.030 & 26 \\
SN 2018oh & +268 & -1852$\pm$376 & -3934$\pm$813 & 7017$\pm$344 & 5299$\pm$662 & 0.219$\pm$0.048 & -1852$\pm$376 & 0.037$\pm$0.017 & 25 \\
ASASSN-14jg & +221 & 1326$\pm$274 & 1113$\pm$283 & 8065$\pm$78 & 3666$\pm$411 & 0.068$\pm$0.010 & 1326$\pm$274 & 0.011$\pm$0.005 & 27 \\
ASASSN-14jg & +267 & 1612$\pm$244 & 984$\pm$560 & 7662$\pm$62 & 5537$\pm$437 & 0.111$\pm$0.005 & 1612$\pm$244 & 0.019$\pm$0.008 & 23 \\
\hline
\end{tabular}
\end{table*}

\begin{table*}
\contcaption{~}
\label{tab:continued}
\begin{tabular}{ccccccccccccc}
\hline
\hline
   {Name} &    {Phase} &    {[Fe~II] Vel.} &    {[Ni~II] Vel.} & 
   {[Fe~II] FWHM} &    {[Ni~II] FWHM} &    {Flux ratio} &    {Nebular Vel.} &    {$M_{\rm Ni}$/$M_{\rm Fe}$} &     {Ref.} \\
   {} &    {[days]} &    {[km s$^{-1}$]} &    {[km s$^{-1}$]} & 
   {[km s$^{-1}$]} &    {[km s$^{-1}$]} &    {$\lambda$7155/$\lambda$7378}  &    {[km s$^{-1}$]} &    {$t \rightarrow \infty$} &    {Spec.} \\
\hline
ASASSN-14jg & +323 & 1674$\pm$261 & 2236$\pm$596 & 8406$\pm$778 & 4992$\pm$552 & 0.159$\pm$0.021 & 1674$\pm$261 & 0.028$\pm$0.012 & 21 \\
\hline
\end{tabular}
\begin{flushleft}
\textcolor{black}{Reference: (1) \citet{2022MNRAS.517.4098X}; (2) \citet{2022MNRAS.514.3541S}; (3) Iskandar et al. (in prep.); (4) Zeng et al. (in prep.); (5) Liu et al. (in prep.); (6) \citet{1992A&A...259...63C}; (7) \citet{1998AJ....115.1096G}; (8) \citet{2012MNRAS.425.1789S}; (9) \citet{2012AJ....143..126B}; (10) \citet{2008MNRAS.388..971P}; (11) \citet{2007A&A...469..645S}; (12) \citet{2009A&A...505..265L}; (13) \citet{2007MNRAS.377.1531P}; (14) \citet{2009ApJ...697..380W}; (15) \citet{2008ApJ...675..626W}; (16) \citet{2013MNRAS.430.1030S}; (17) \citet{2015MNRAS.450.2631M}; (18) \citet{2020MNRAS.492.4325S}; (19) \citet{2015MNRAS.454.1948G}; (20) \citet{2015MNRAS.453.3300A}; (21) \citet{2018MNRAS.477.3567M}; (22) \citet{2015MNRAS.454.3816C}; (23) \citet{2017MNRAS.472.3437G}; (24) \citet{2015MNRAS.452.4307P}; (25) \citet{2019ApJ...872L..22T}; (26) \citet{2021ApJ...919...49Z}; (27) \citet{2020MNRAS.493.1044T}. The sample from our own project is put at the top of the table and separated from the public sample by a line.\\}
$^{a}$The velocity shifts of iron and nickel features are set to the same values, and [Ca~II] emission lines are added to the fits.\\
$^{b}$The velocity shifts of iron and nickel features are set to the same values.\\
\end{flushleft}
\end{table*}

\begin{figure*}
\centering
\includegraphics[width=0.5\columnwidth]{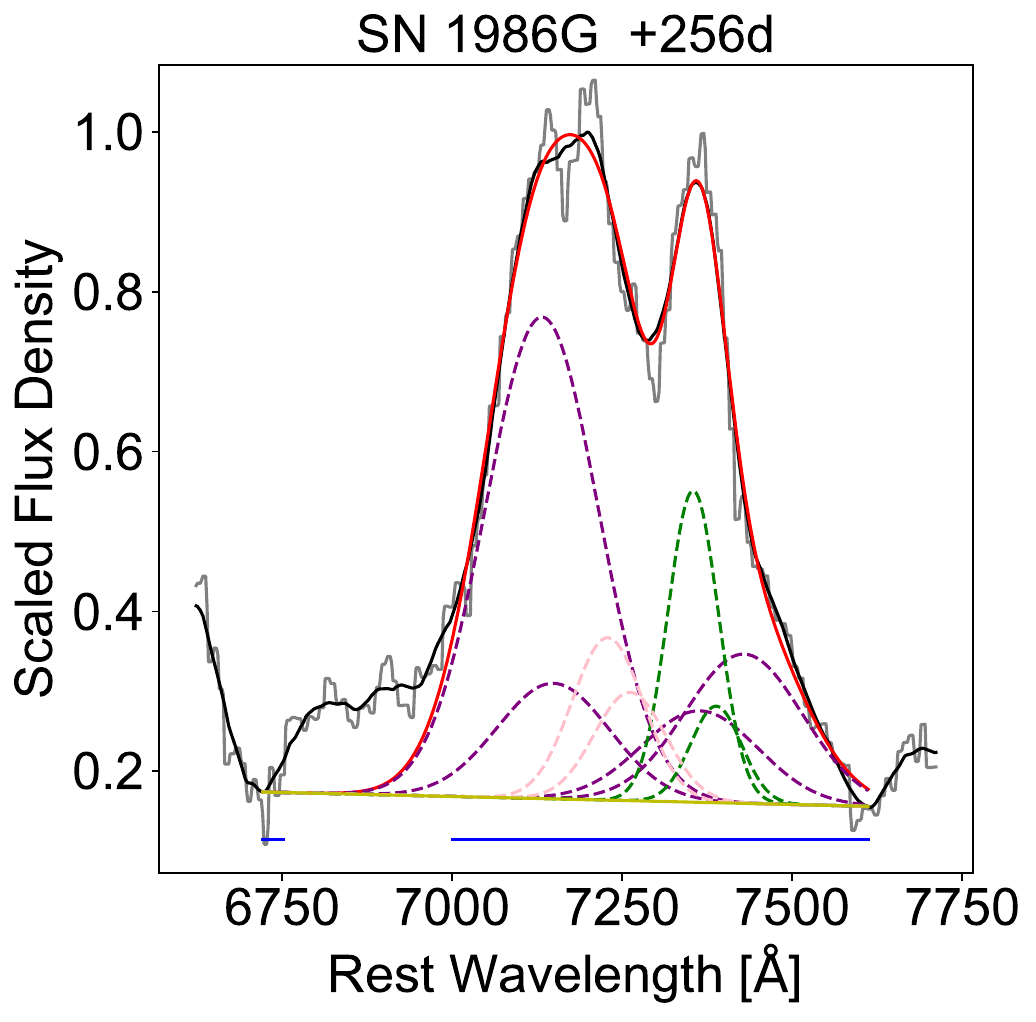}
\includegraphics[width=0.5\columnwidth]{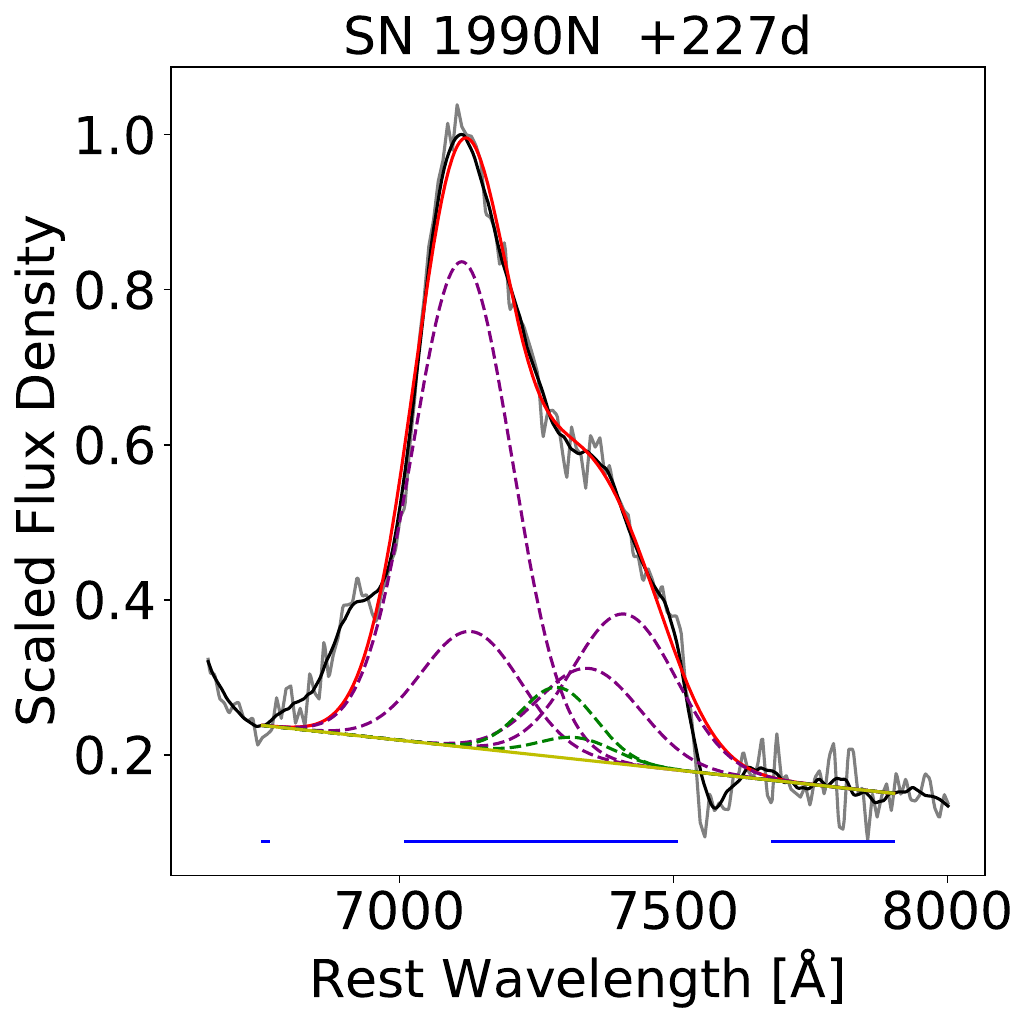}
\includegraphics[width=0.5\columnwidth]{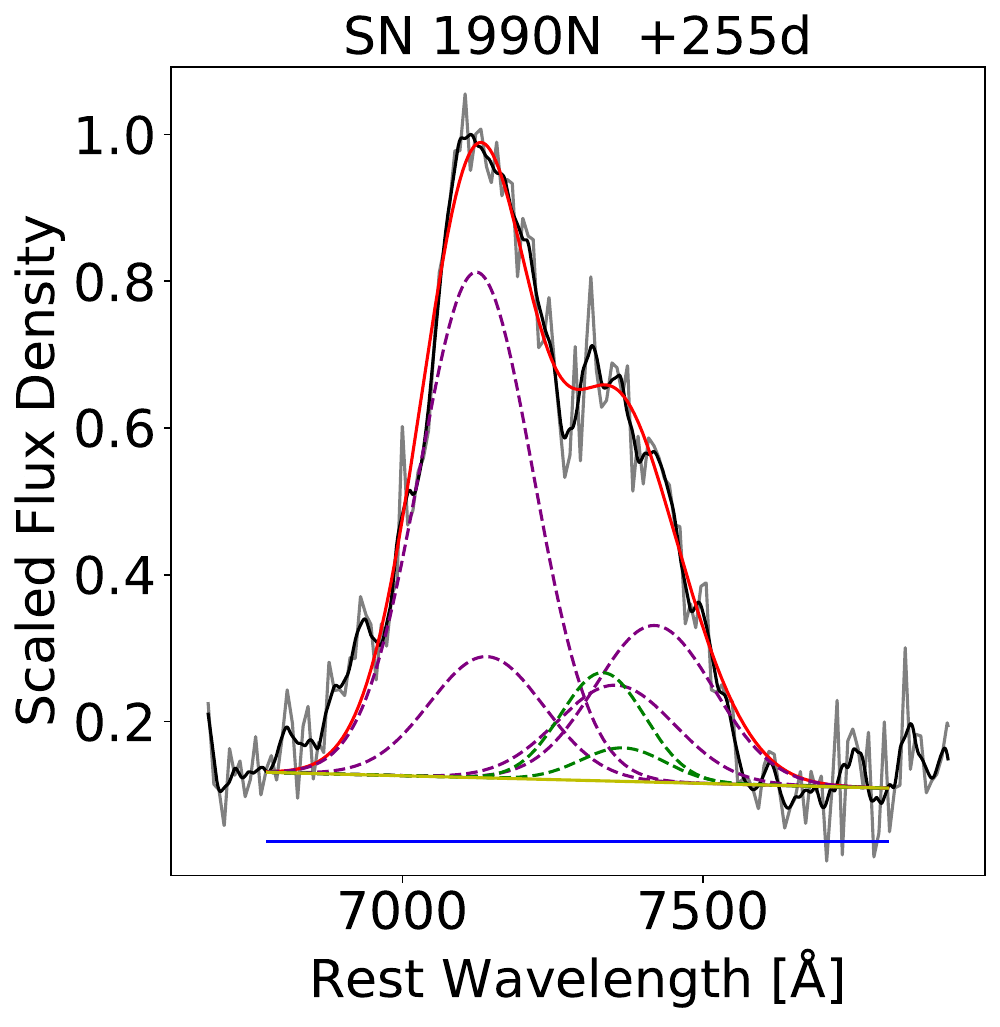}
\includegraphics[width=0.5\columnwidth]{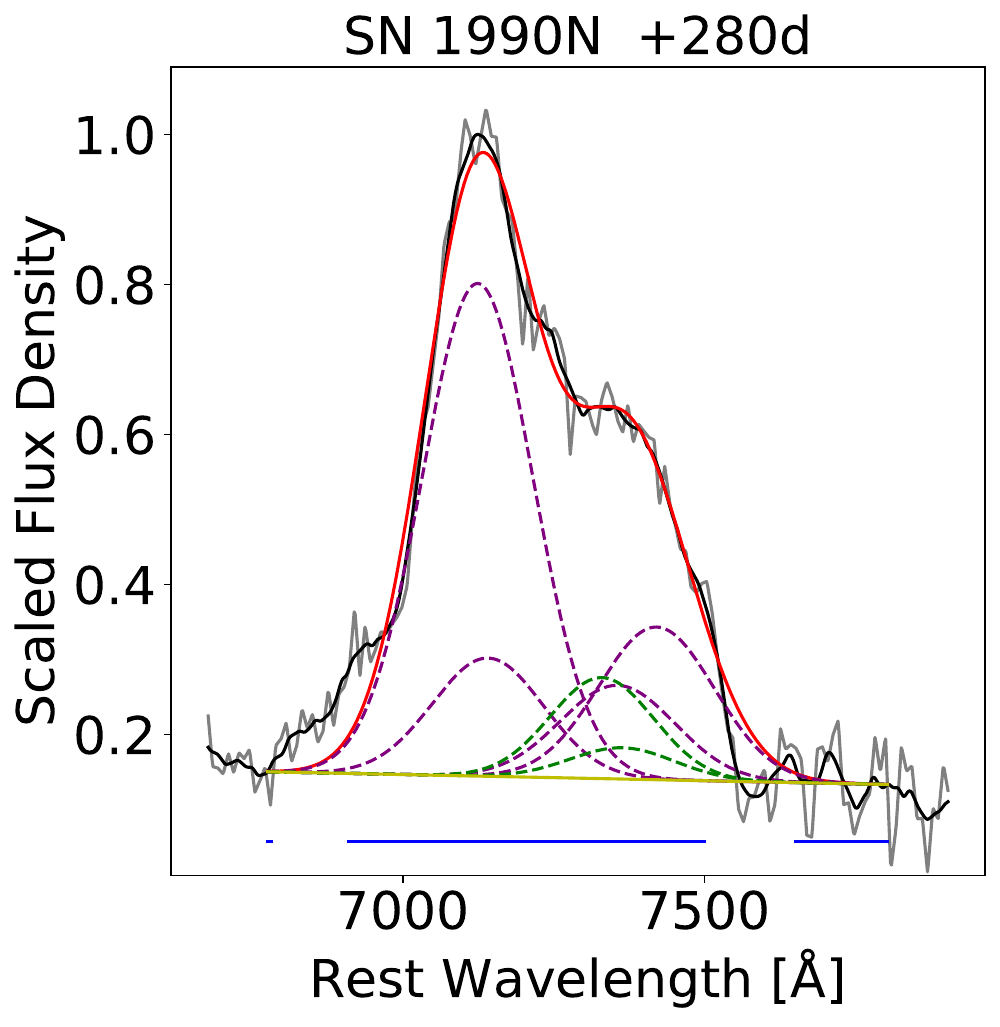}
\includegraphics[width=0.5\columnwidth]{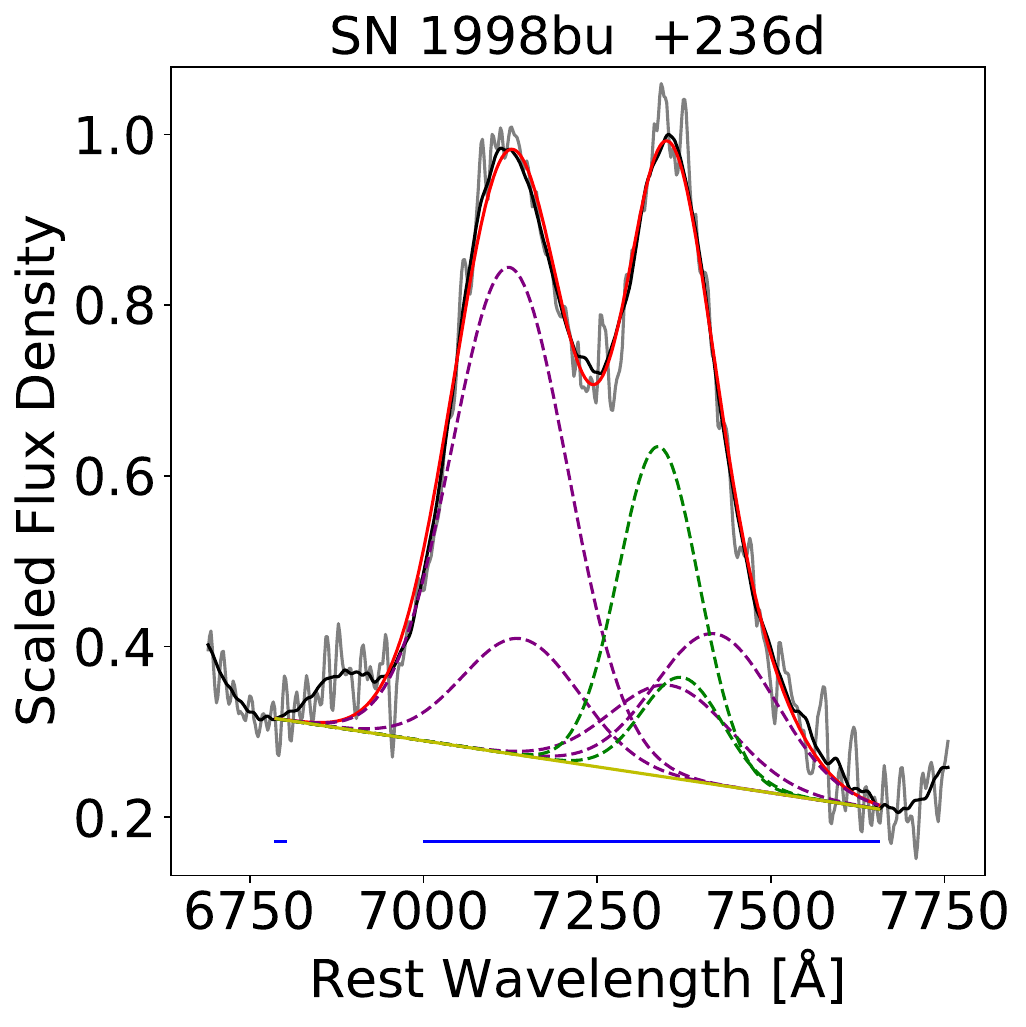}
\includegraphics[width=0.5\columnwidth]{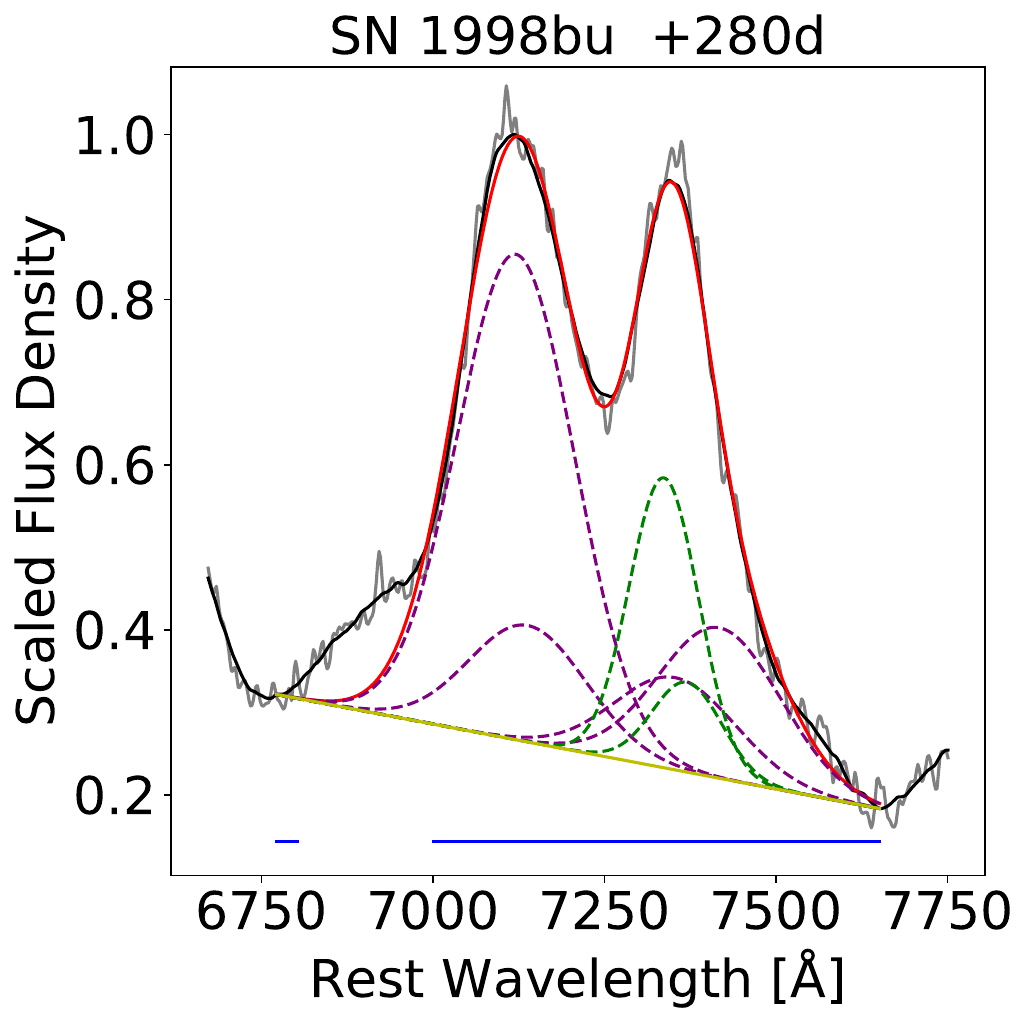}
\includegraphics[width=0.5\columnwidth]{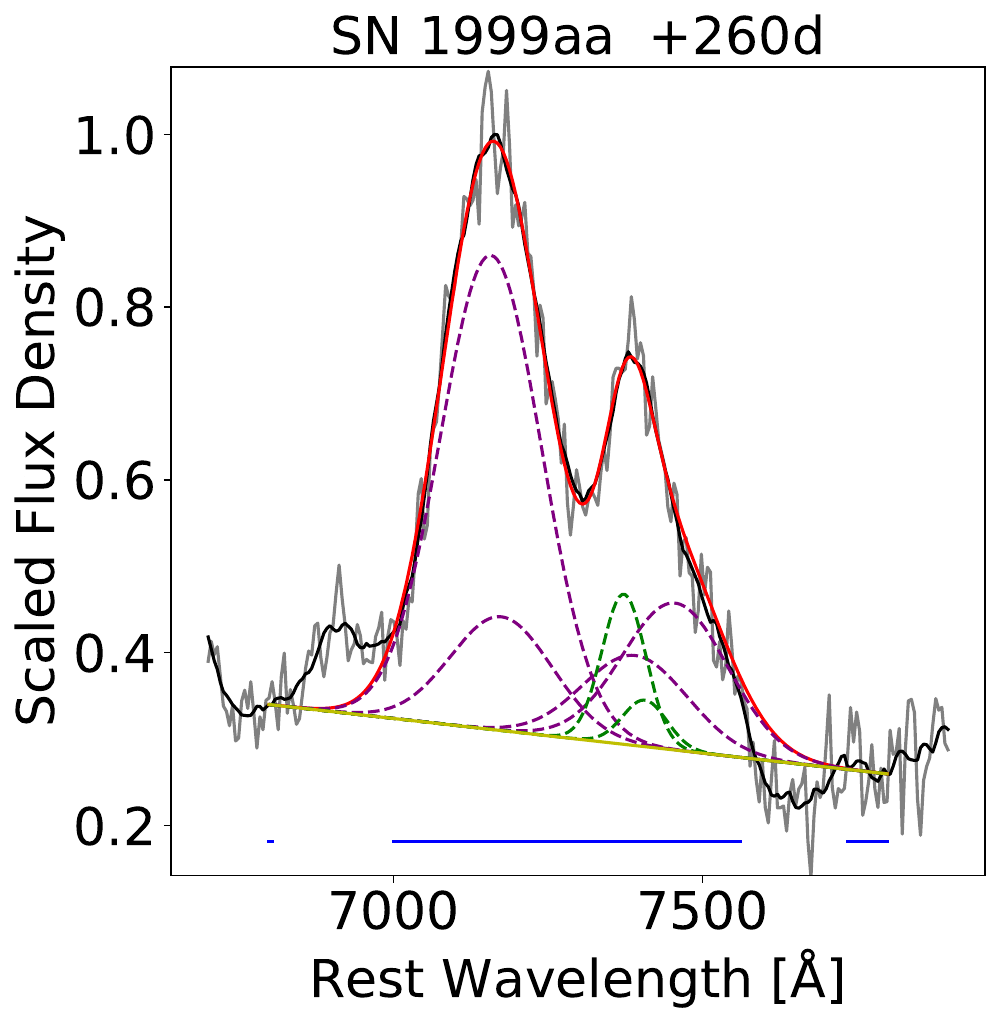}
\includegraphics[width=0.5\columnwidth]{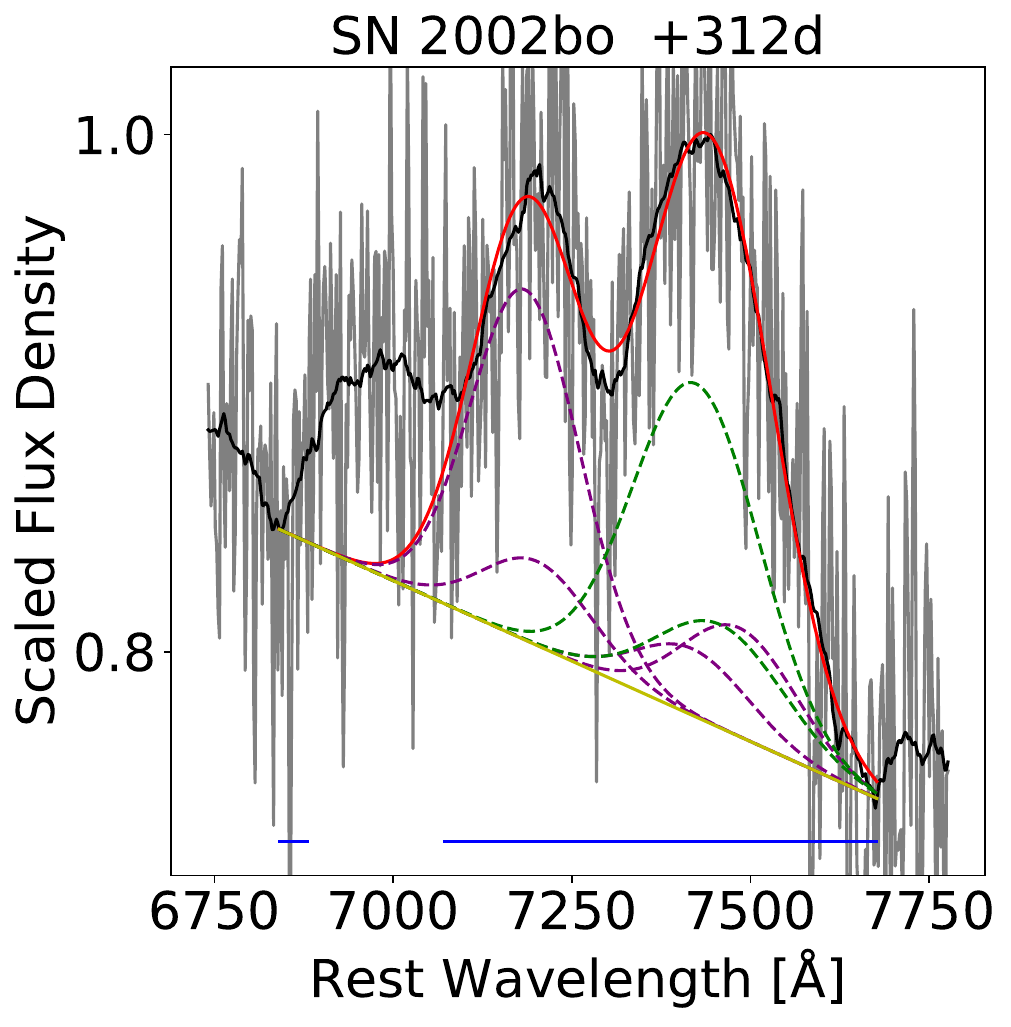}
\includegraphics[width=0.5\columnwidth]{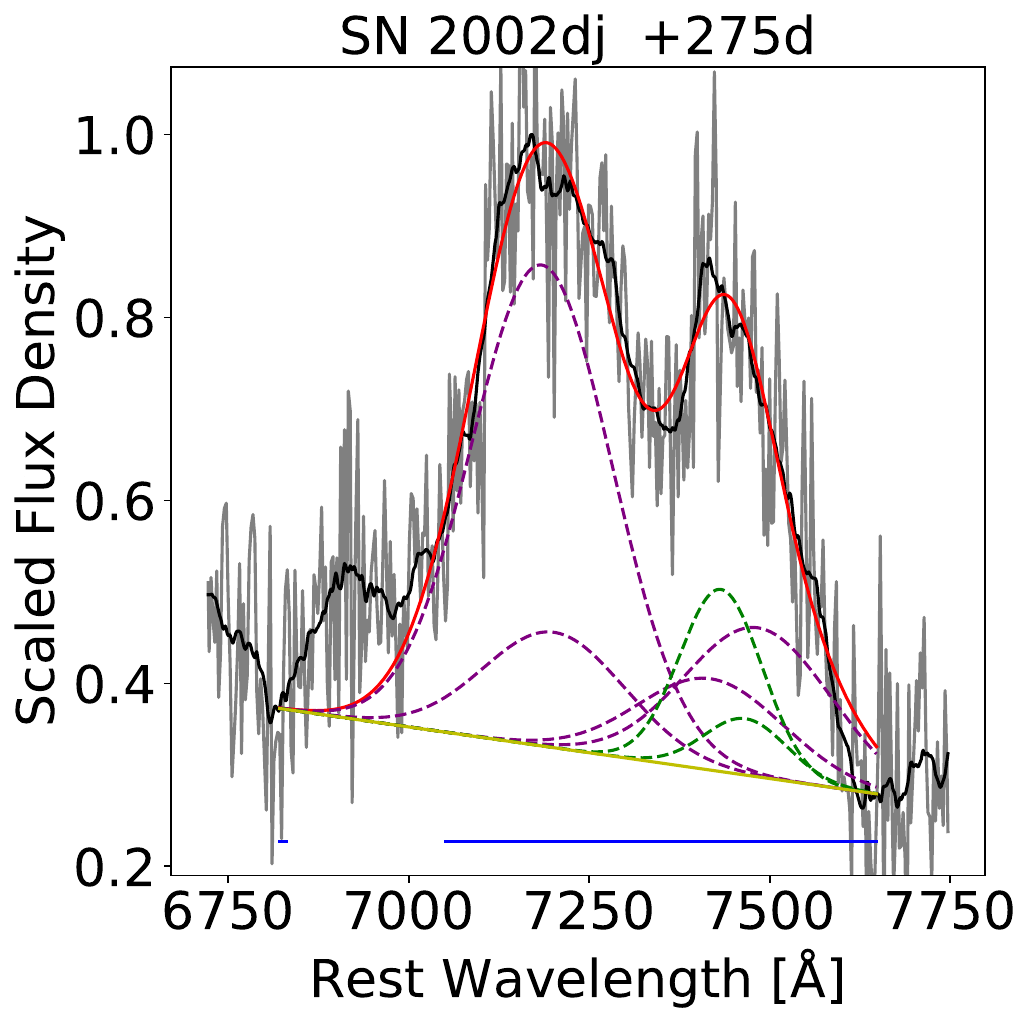}
\includegraphics[width=0.5\columnwidth]{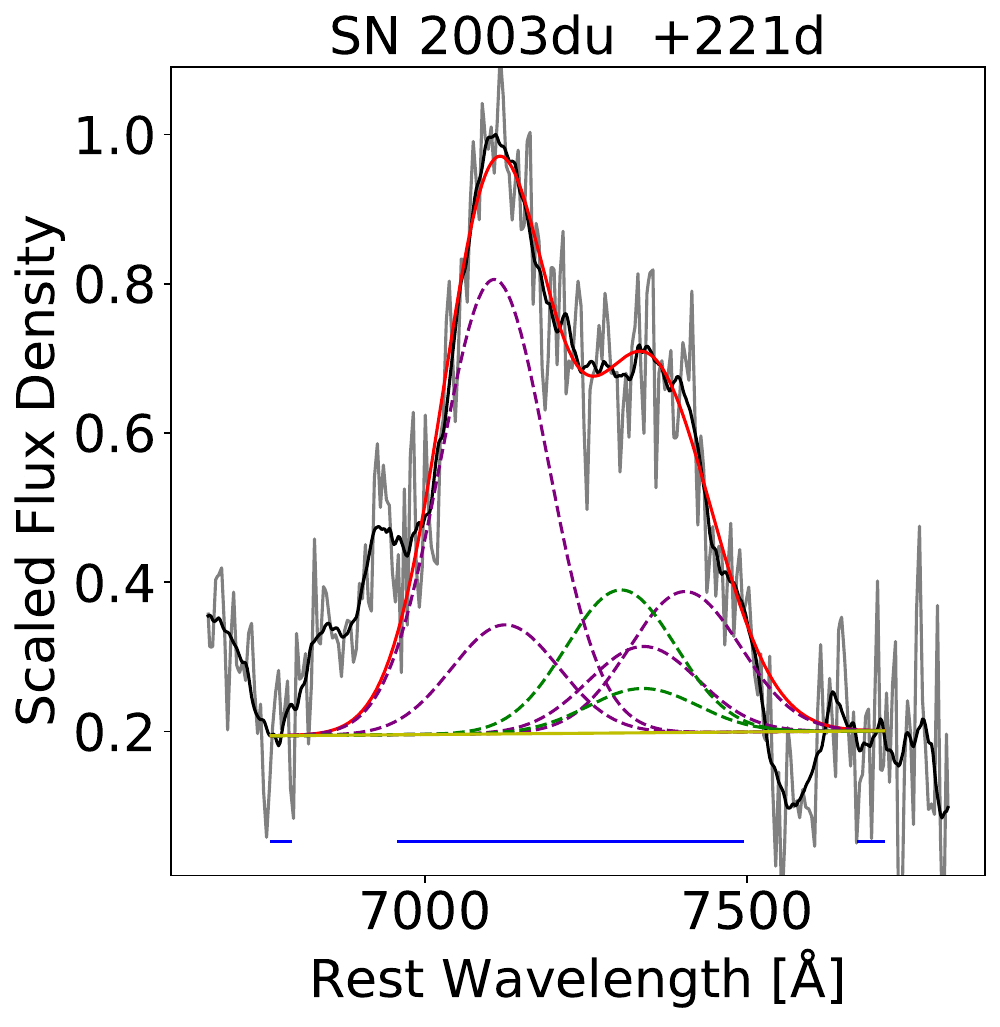}
\includegraphics[width=0.5\columnwidth]{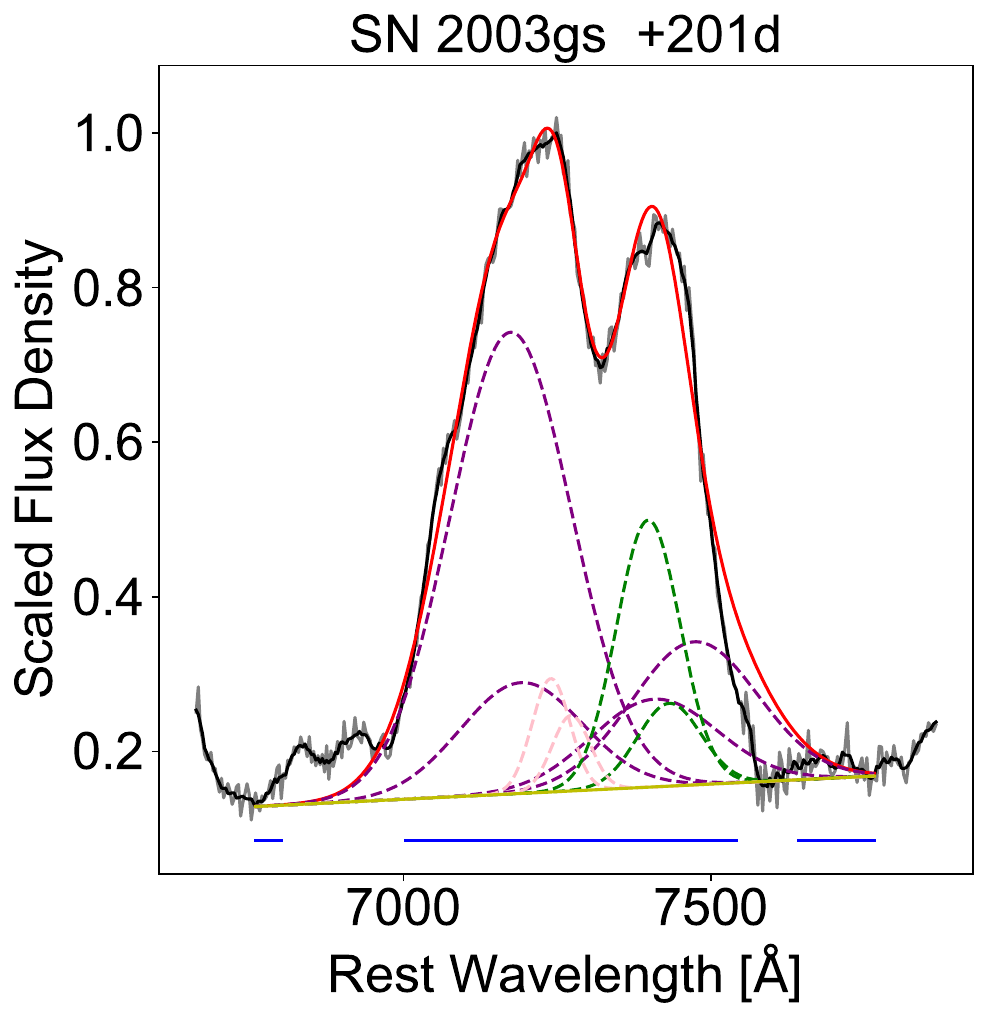}
\includegraphics[width=0.5\columnwidth]{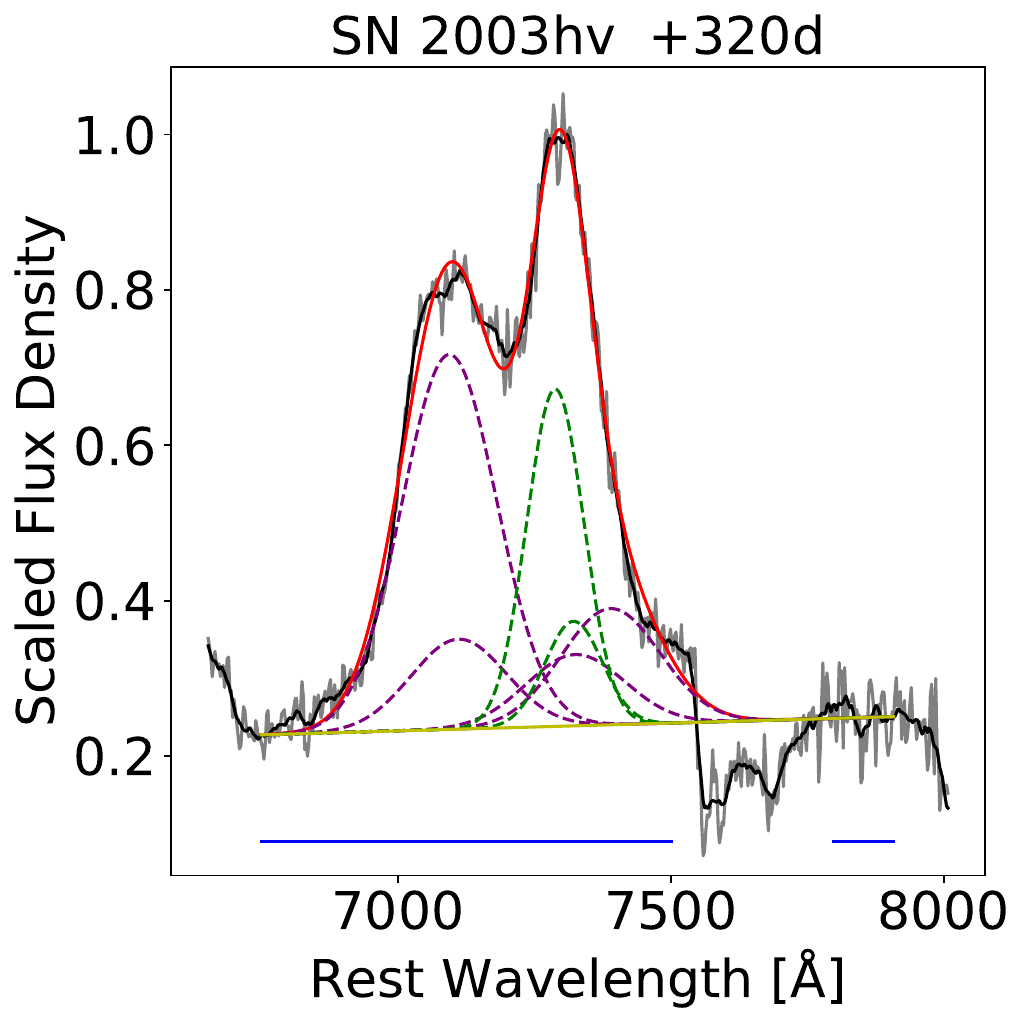}
\includegraphics[width=0.5\columnwidth]{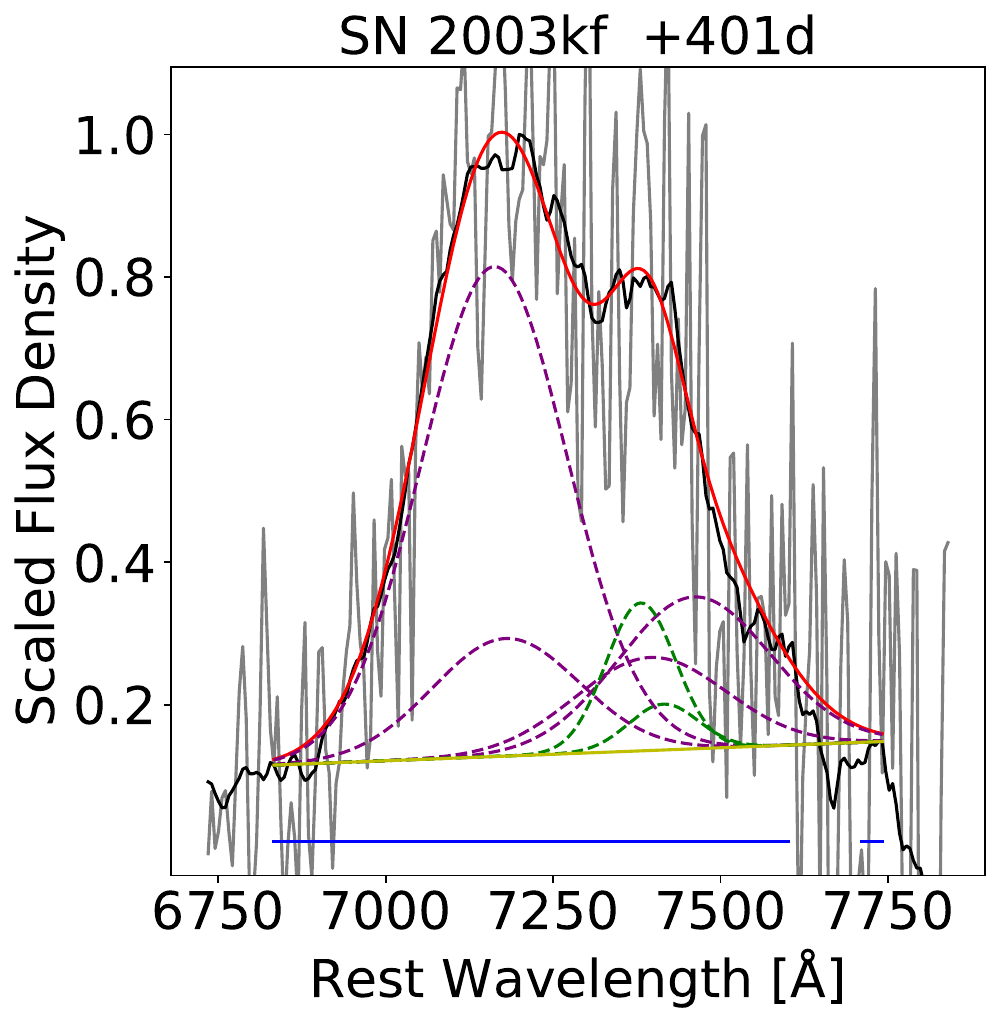}
\includegraphics[width=0.5\columnwidth]{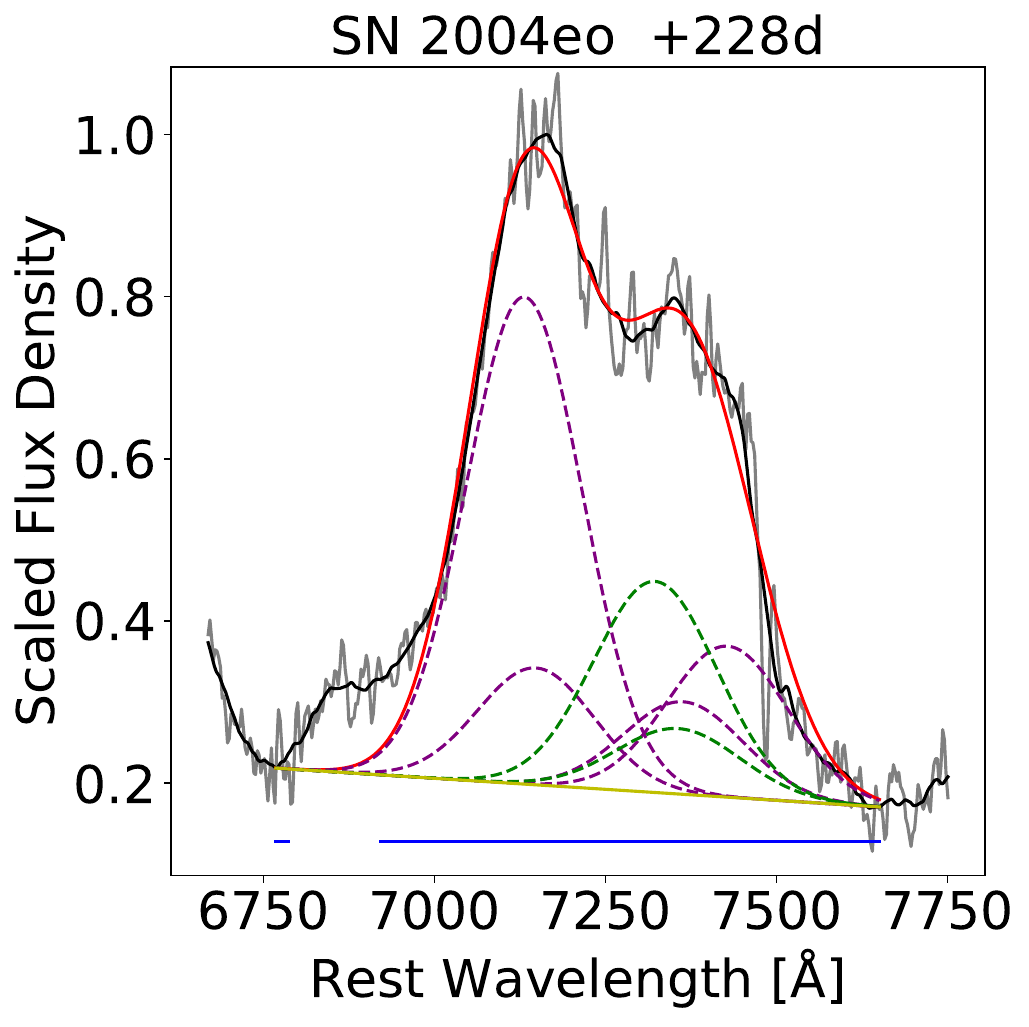}
\includegraphics[width=0.5\columnwidth]{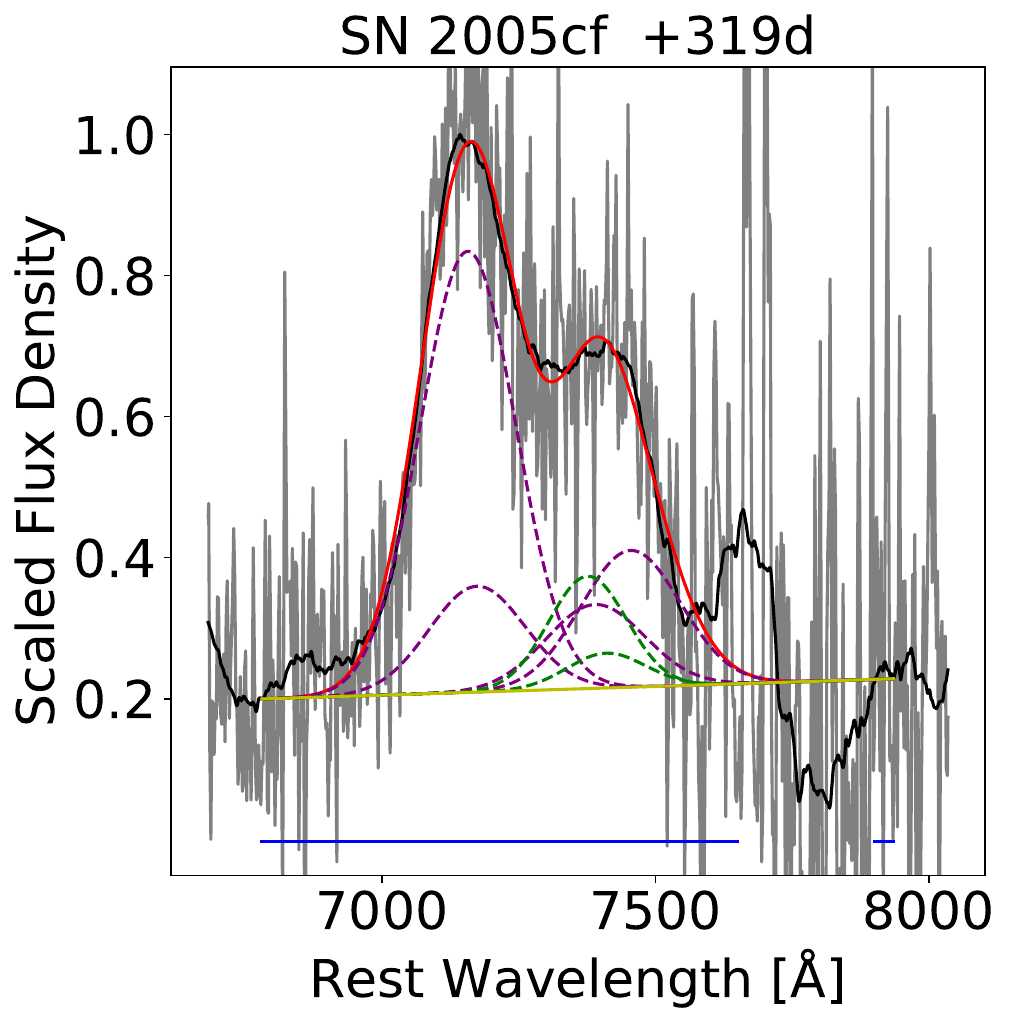}
\includegraphics[width=0.5\columnwidth]{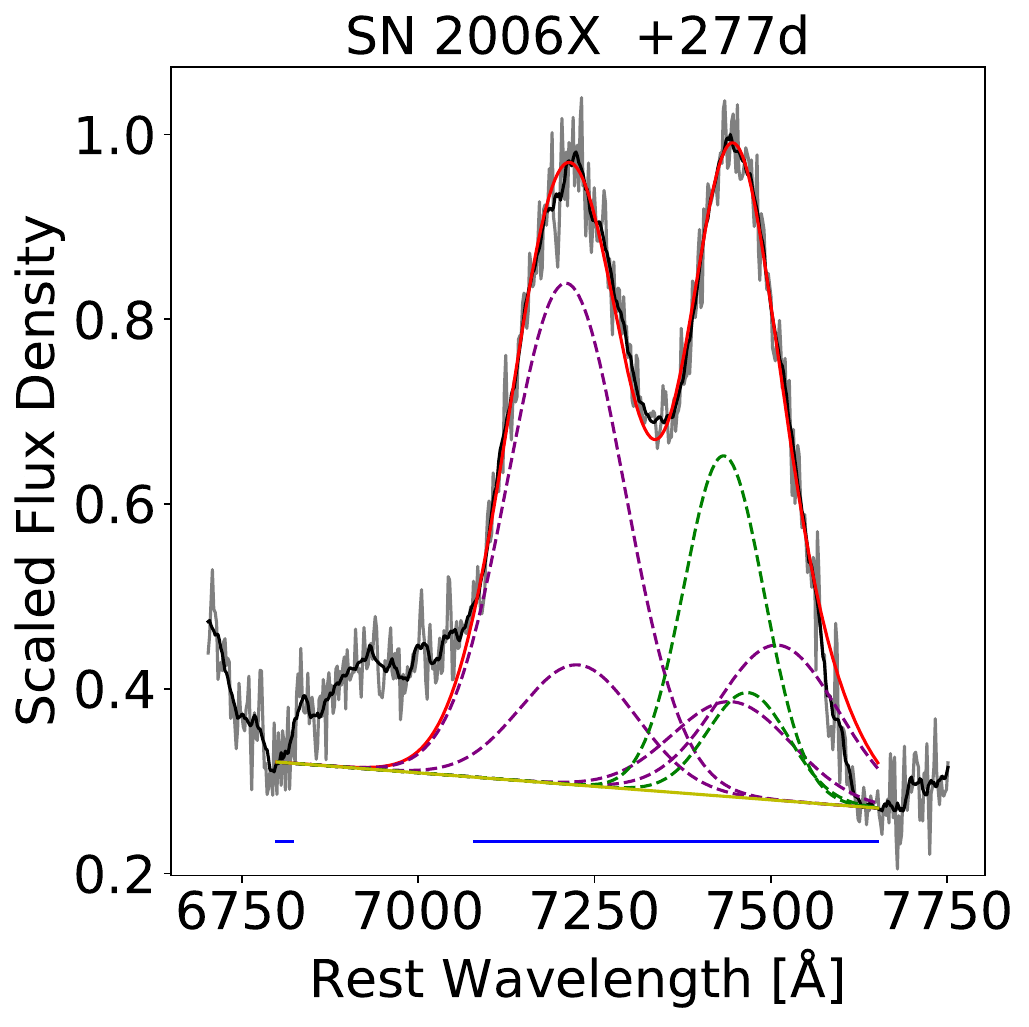}
\end{figure*}

\begin{figure*}
\centering
\includegraphics[width=0.5\columnwidth]{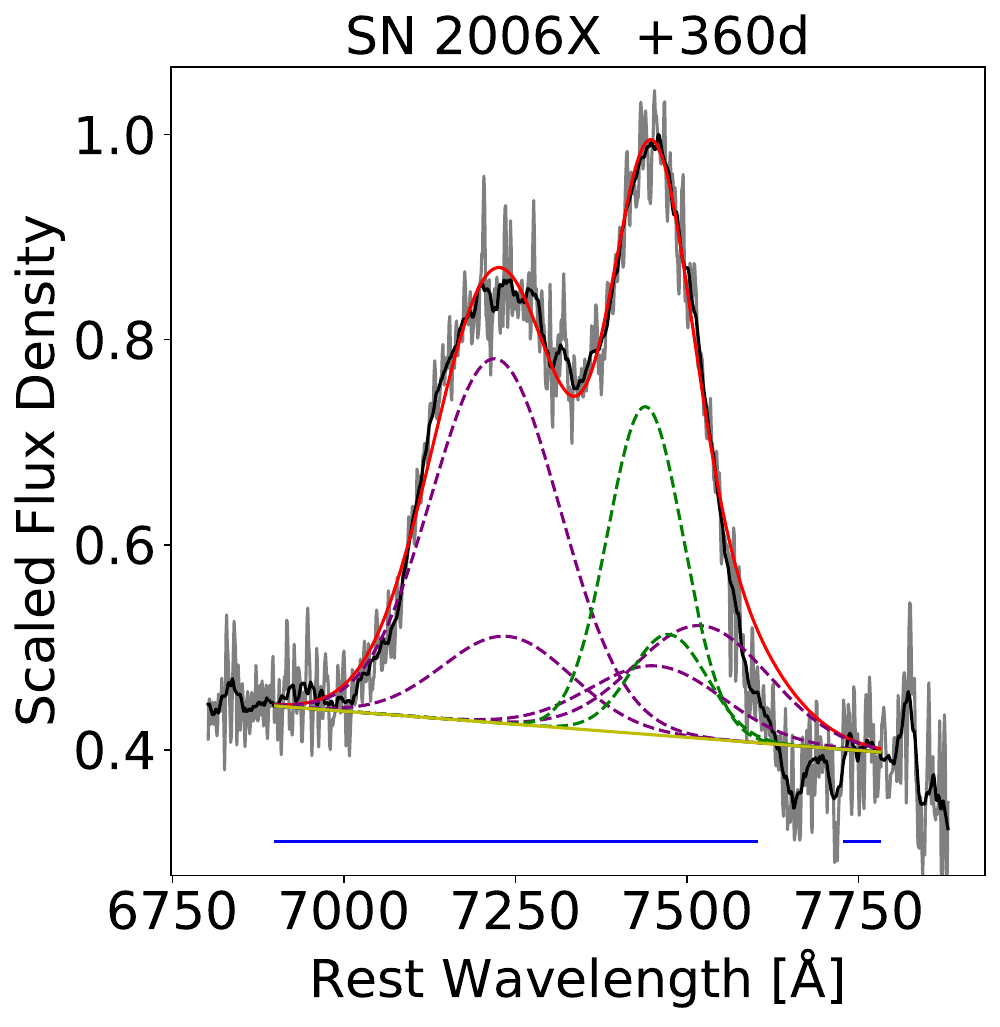}
\includegraphics[width=0.5\columnwidth]{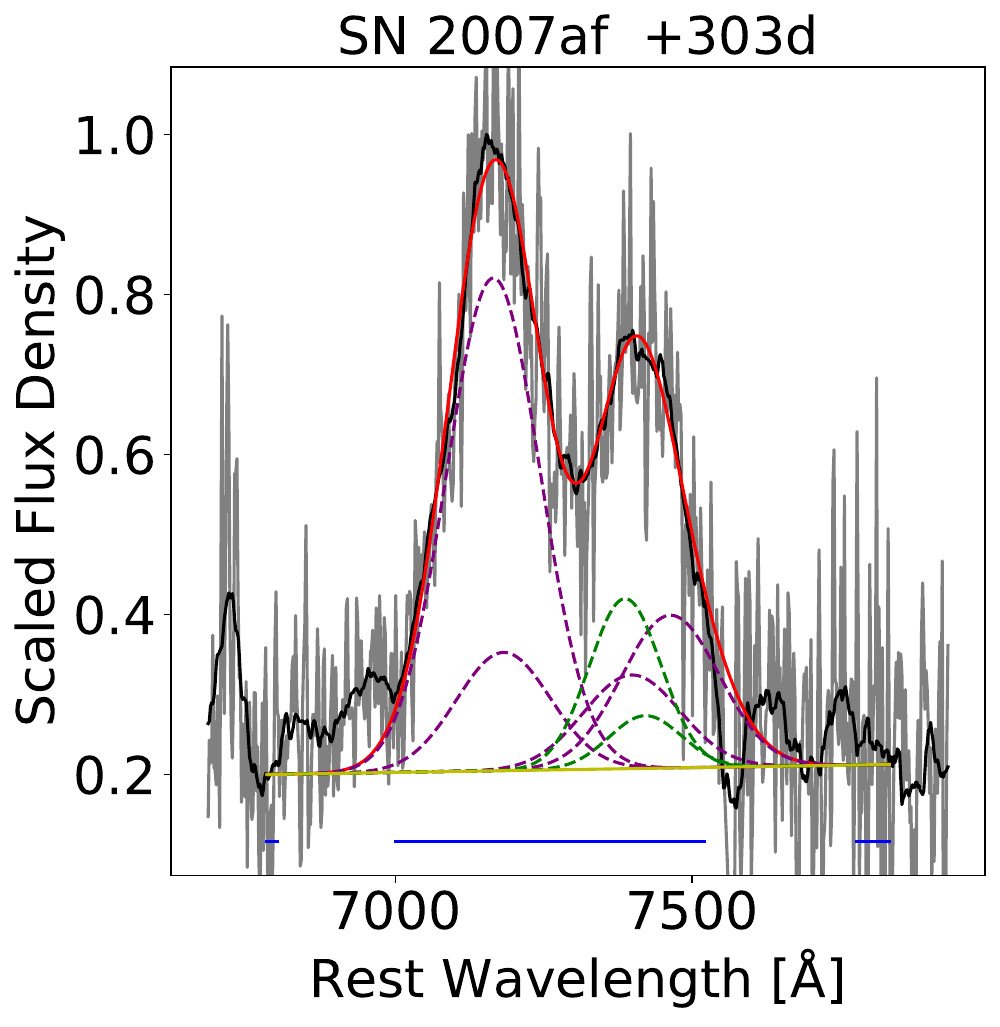}
\includegraphics[width=0.5\columnwidth]{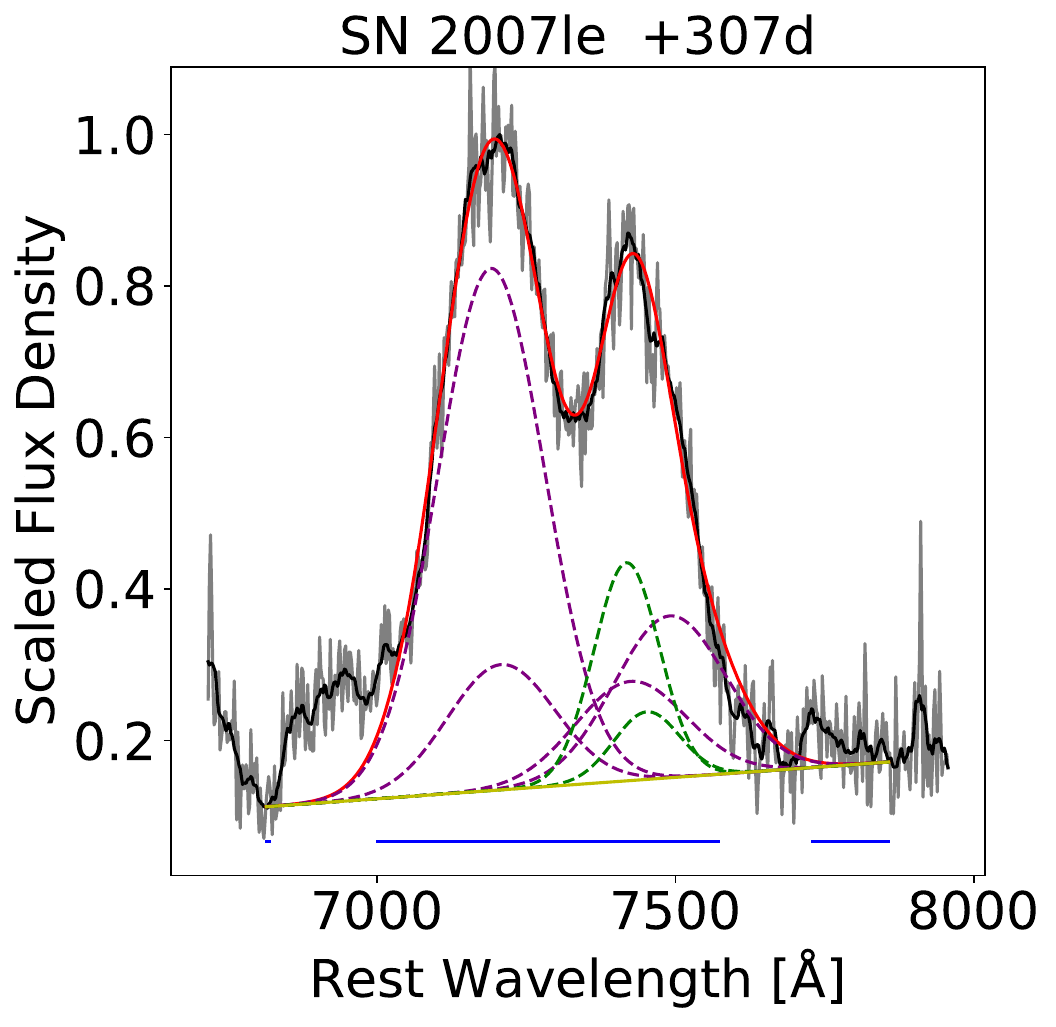}
\includegraphics[width=0.5\columnwidth]{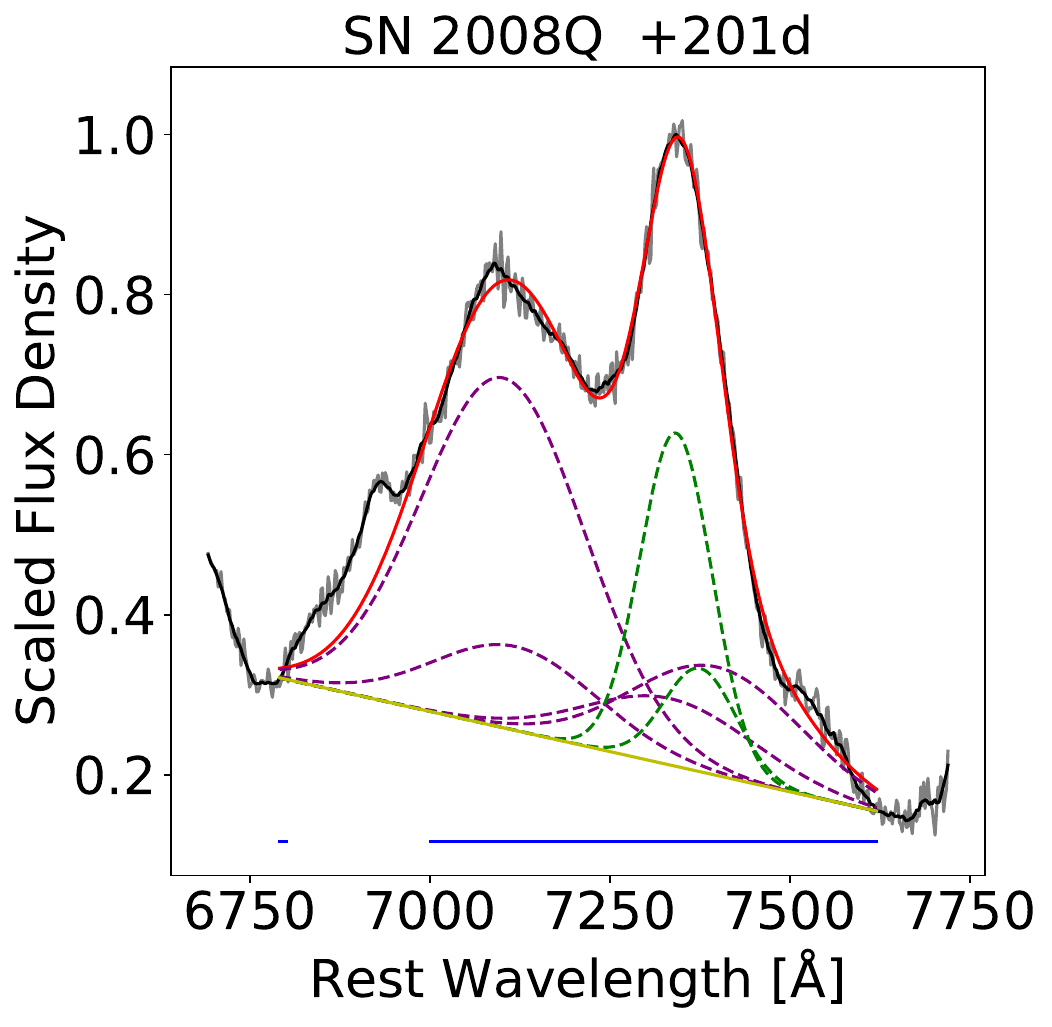}
\includegraphics[width=0.5\columnwidth]{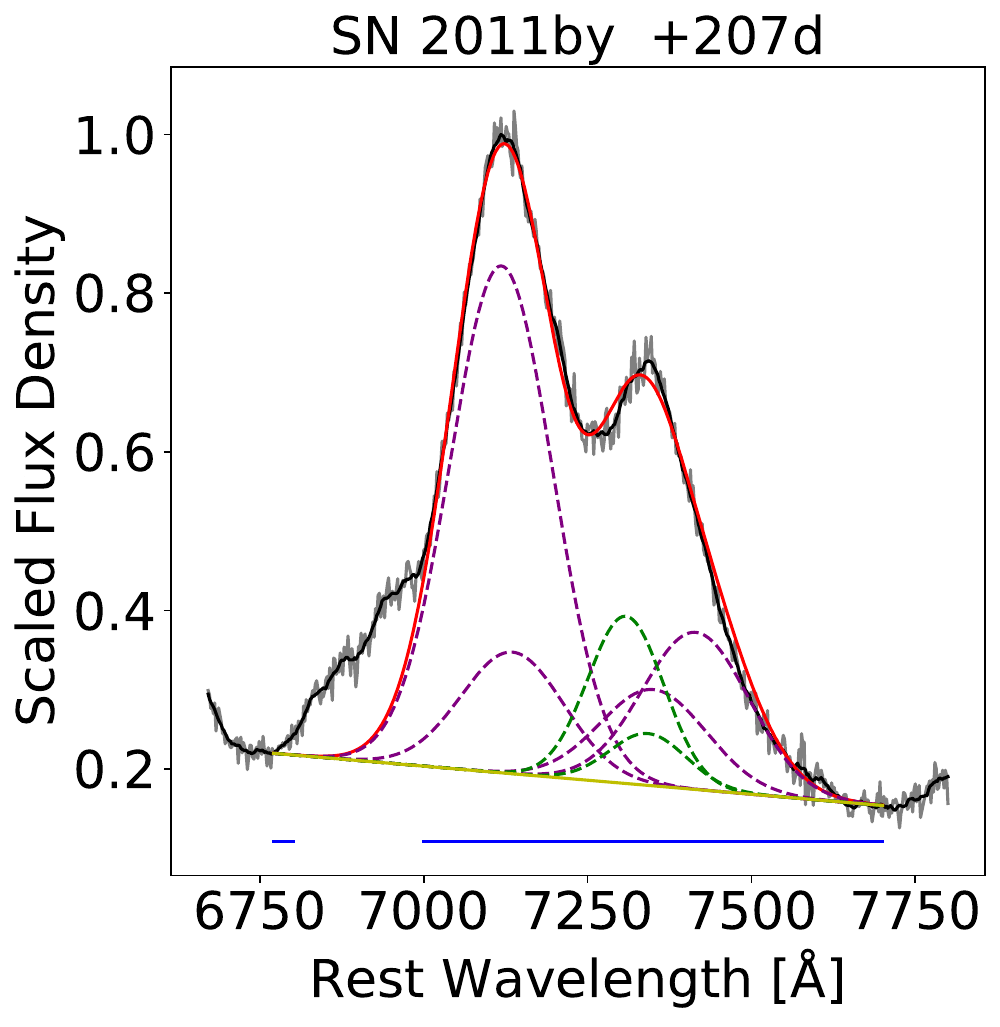}
\includegraphics[width=0.5\columnwidth]{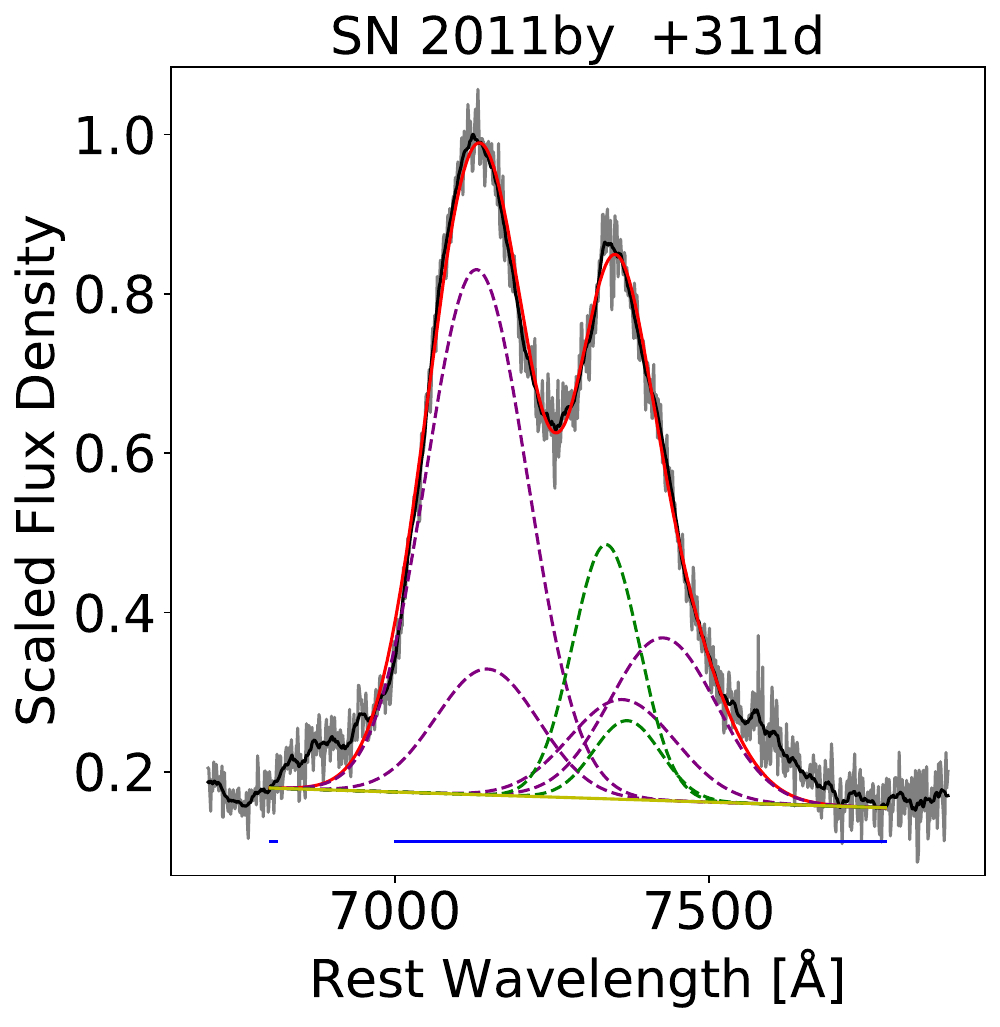}
\includegraphics[width=0.5\columnwidth]{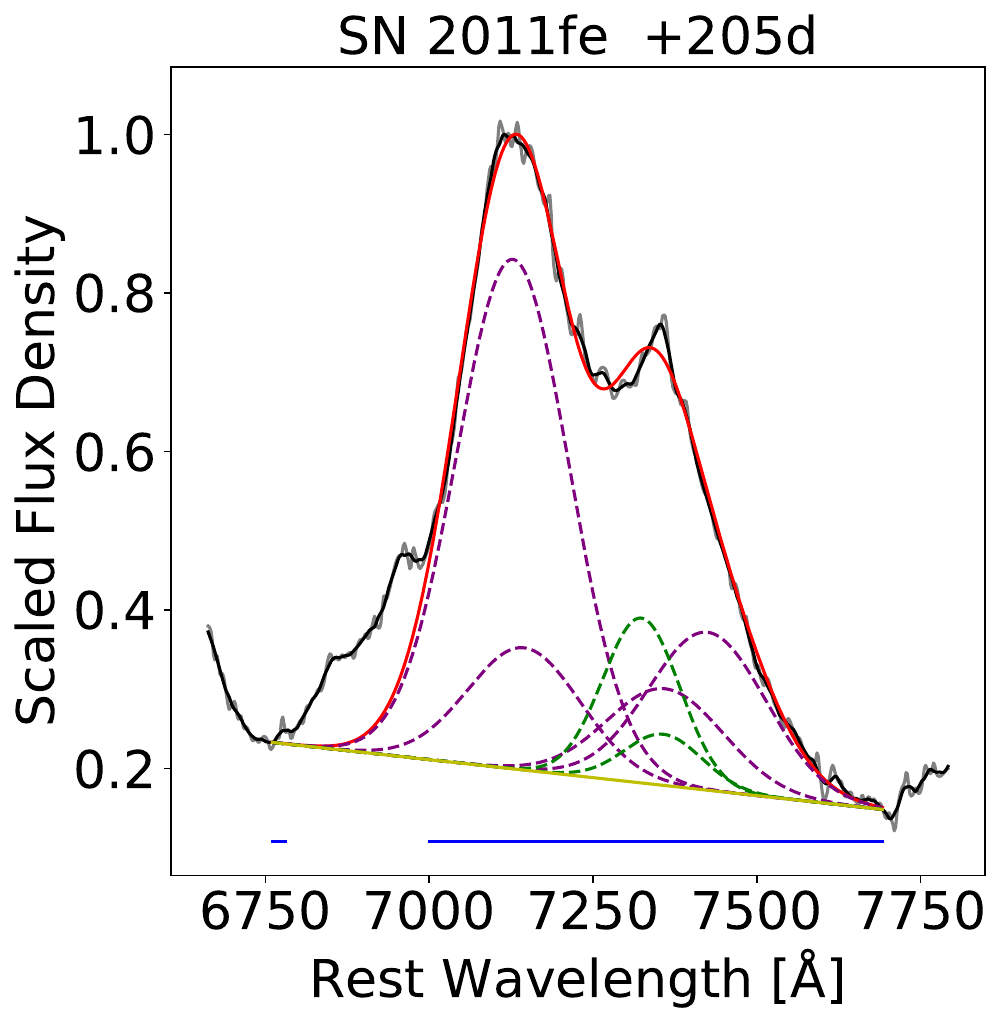}
\includegraphics[width=0.5\columnwidth]{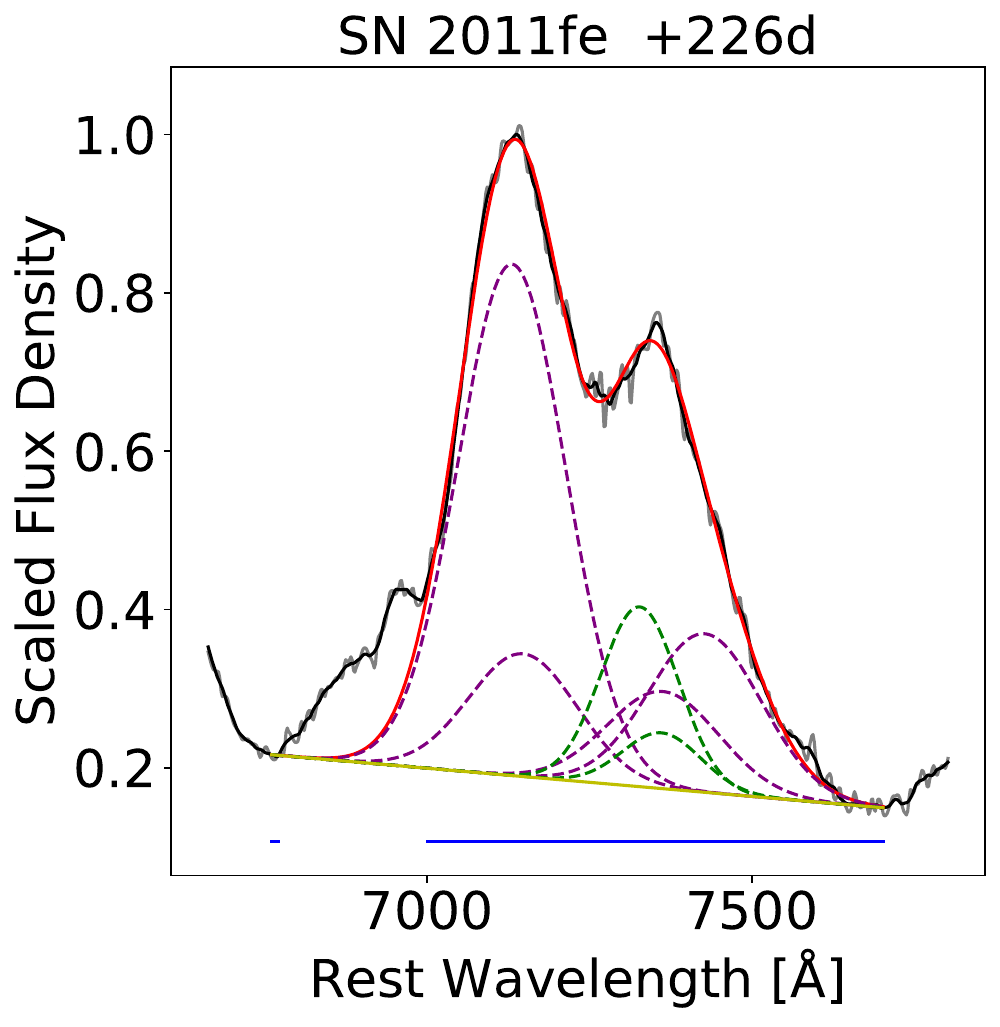}
\includegraphics[width=0.5\columnwidth]{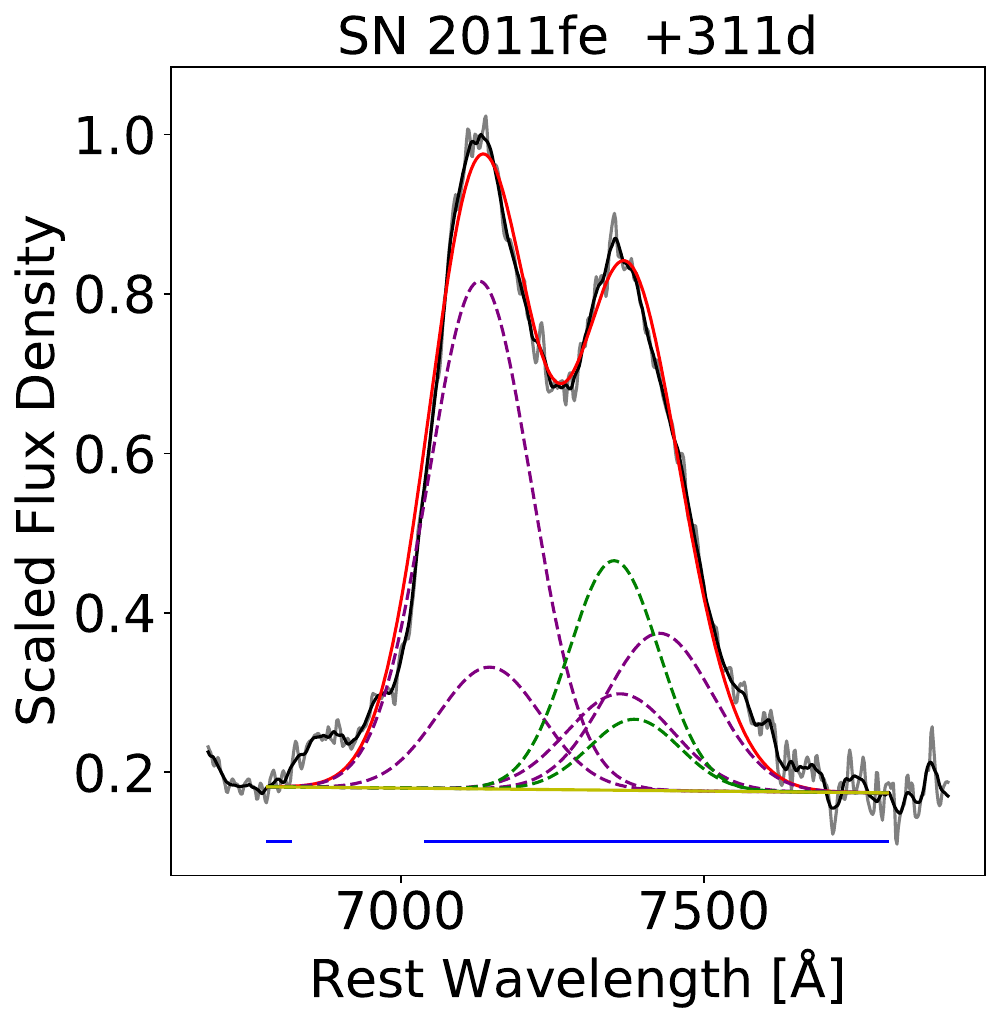}
\includegraphics[width=0.5\columnwidth]{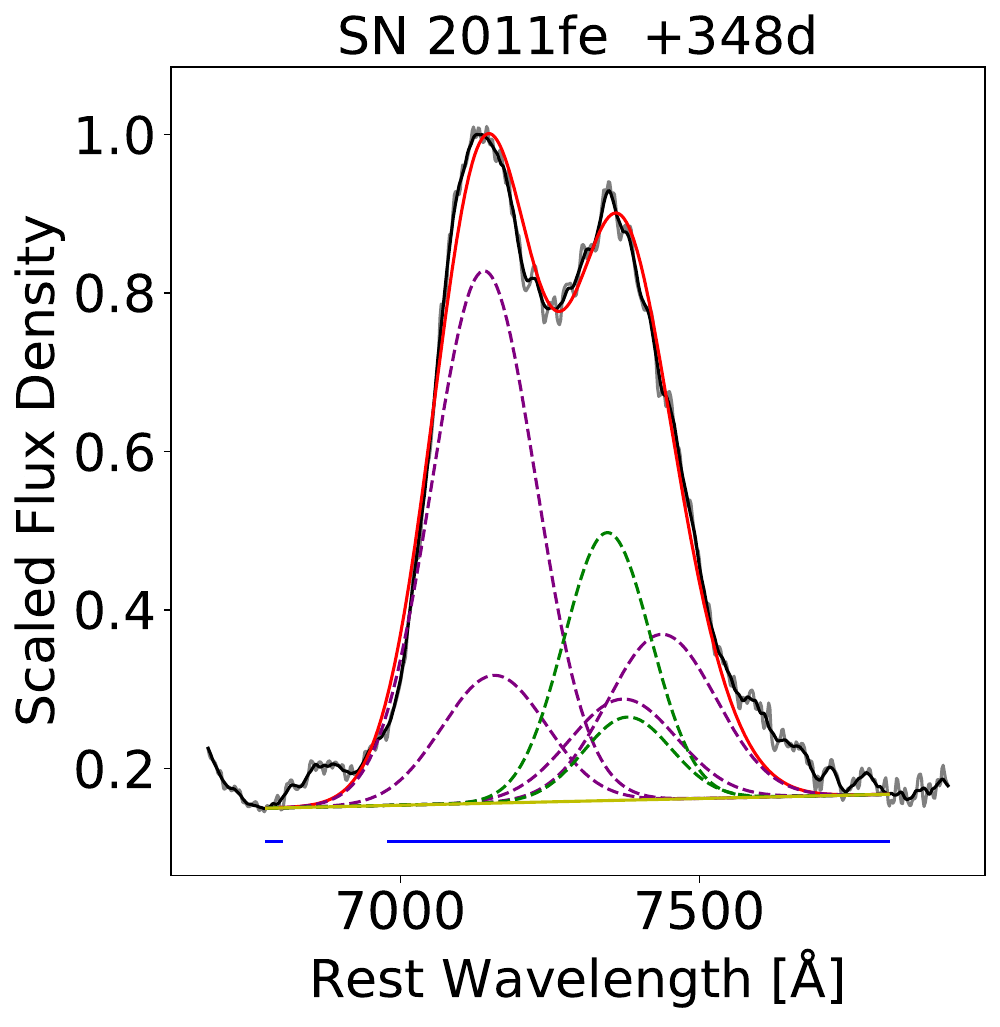}
\includegraphics[width=0.5\columnwidth]{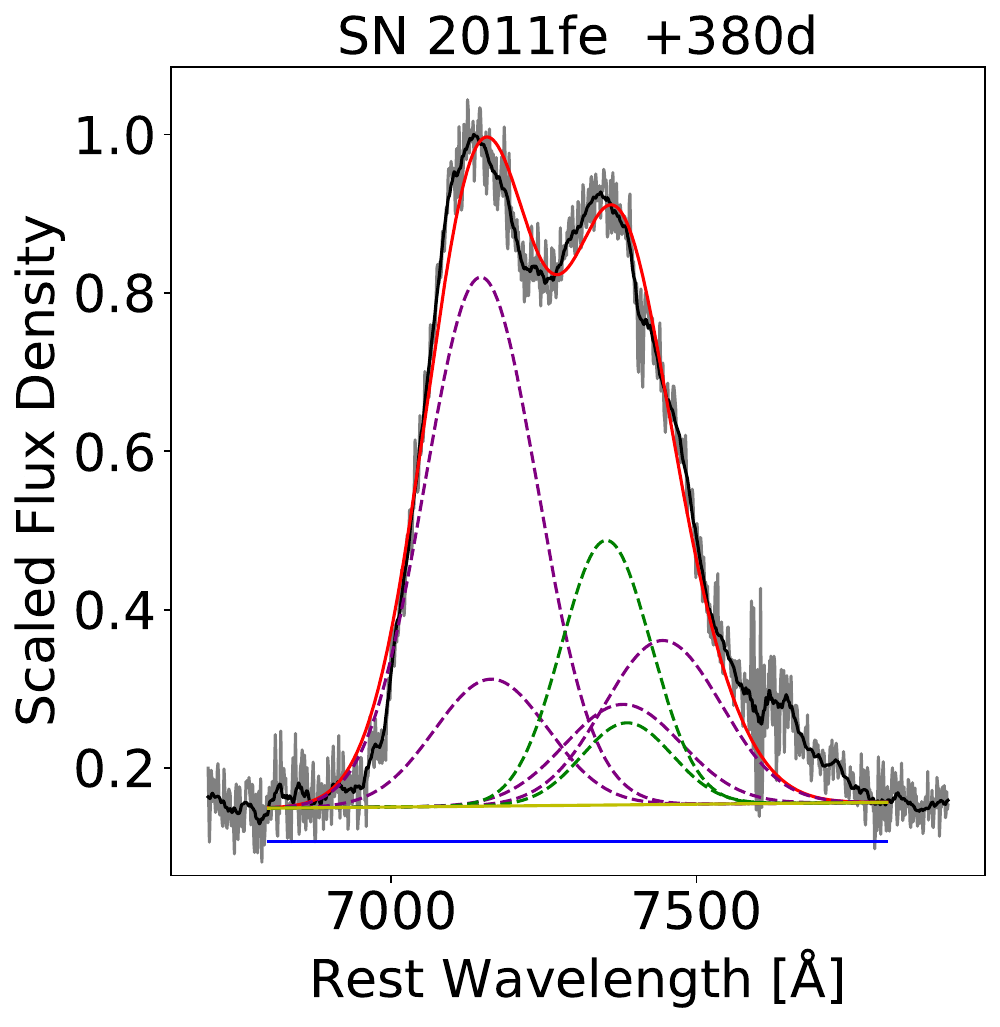}
\includegraphics[width=0.5\columnwidth]{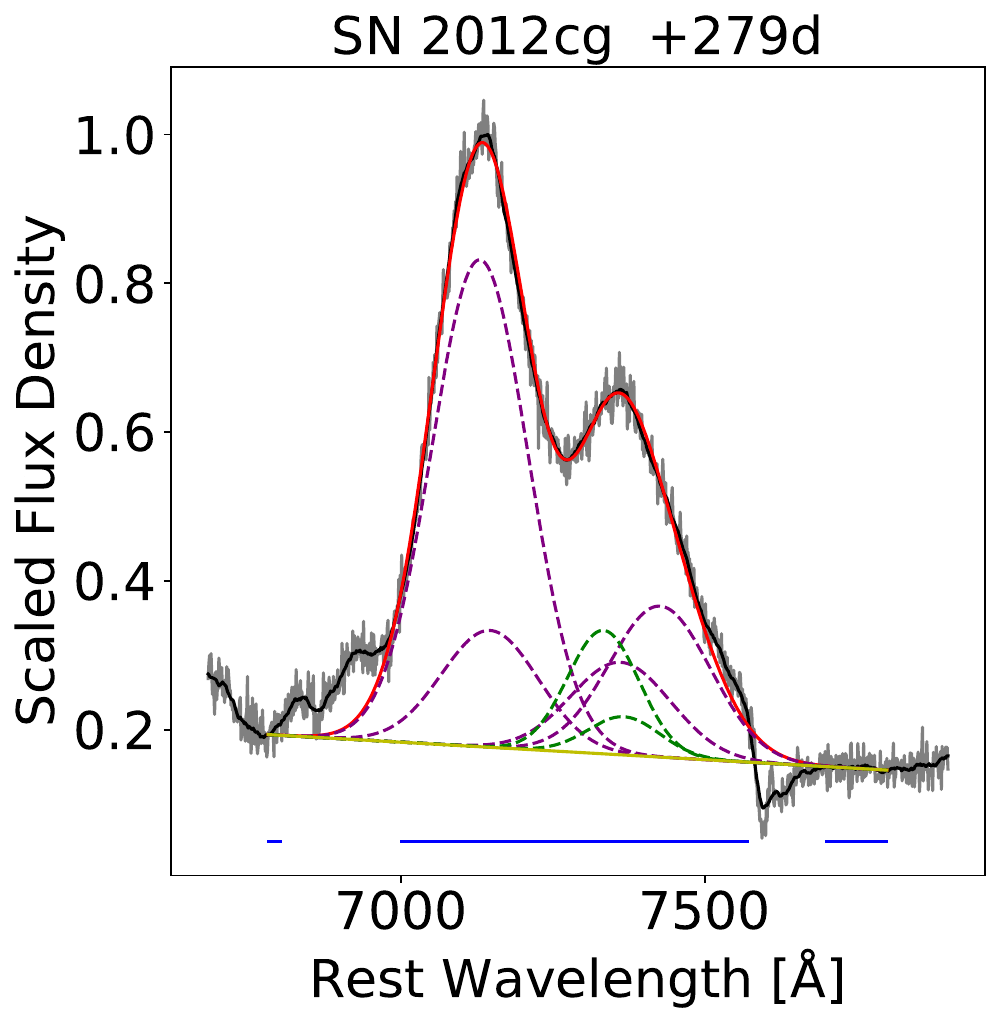}
\includegraphics[width=0.5\columnwidth]{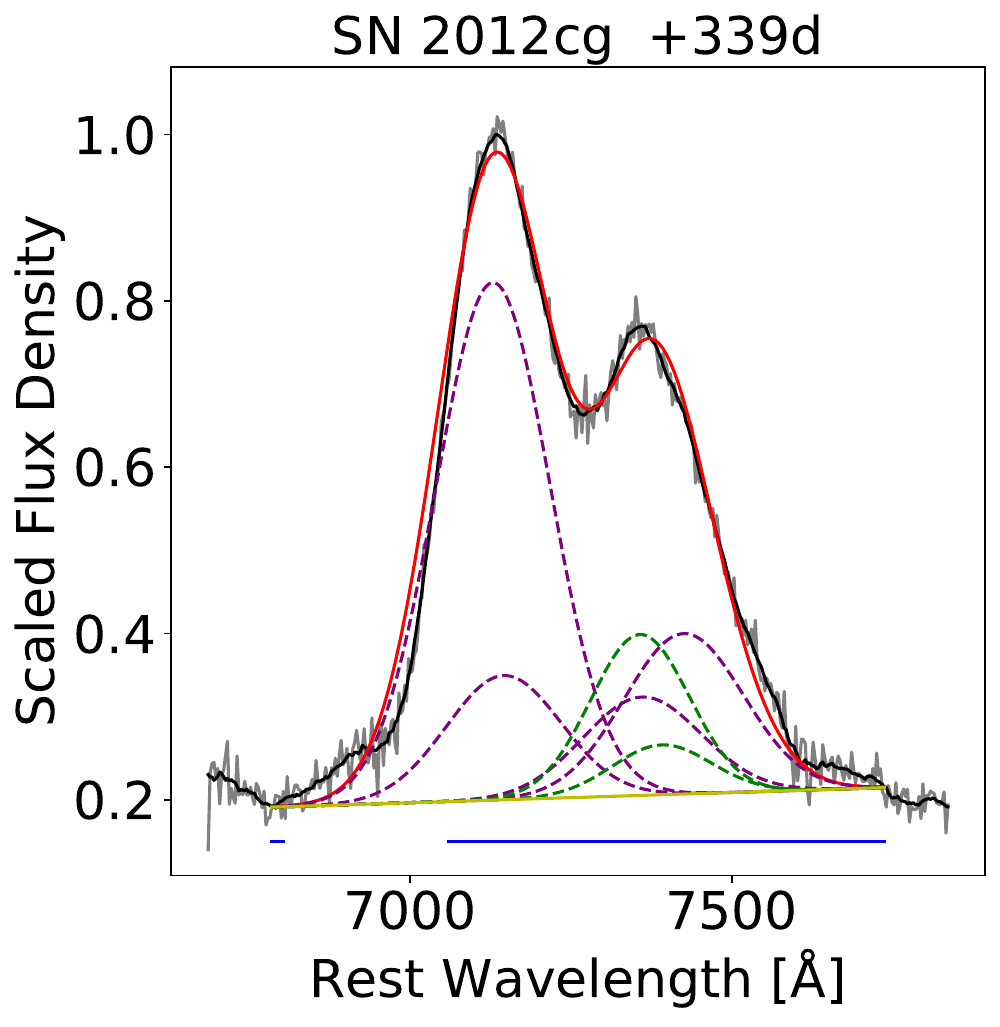}
\includegraphics[width=0.5\columnwidth]{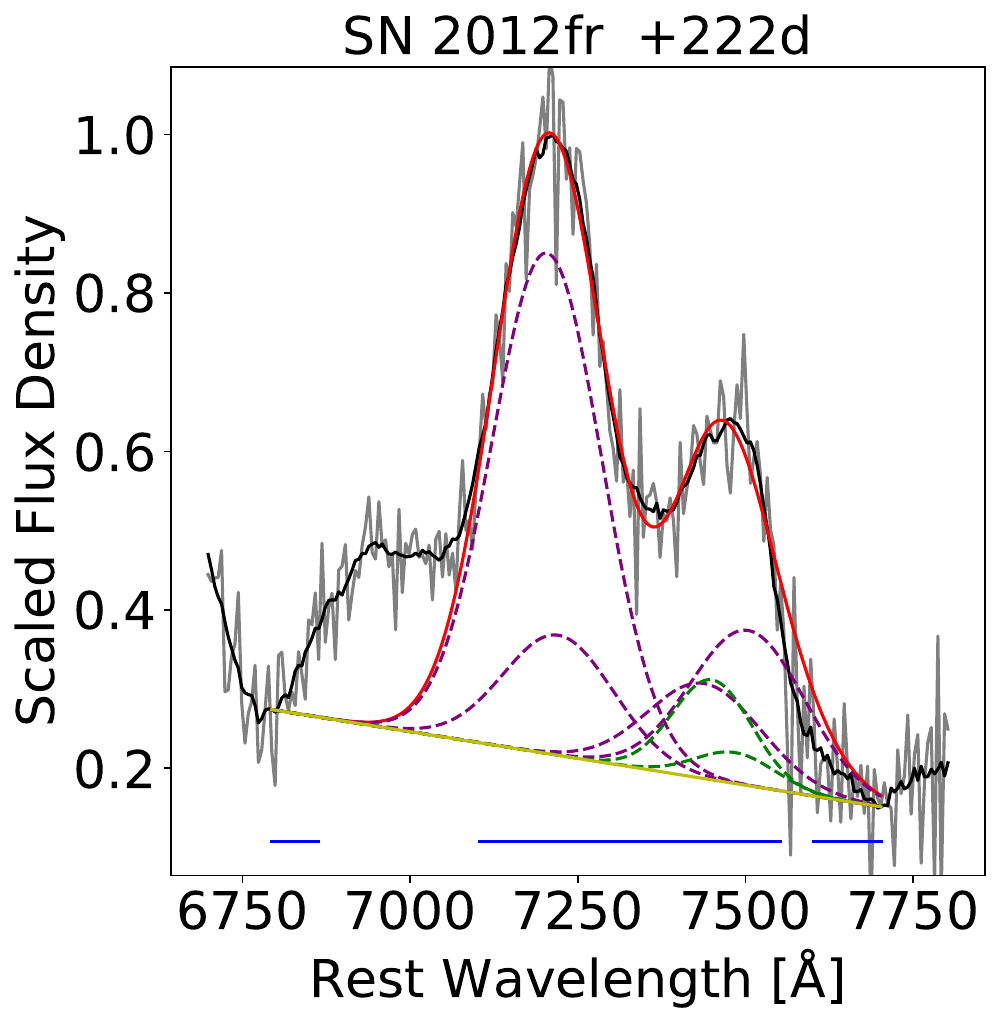}
\includegraphics[width=0.5\columnwidth]{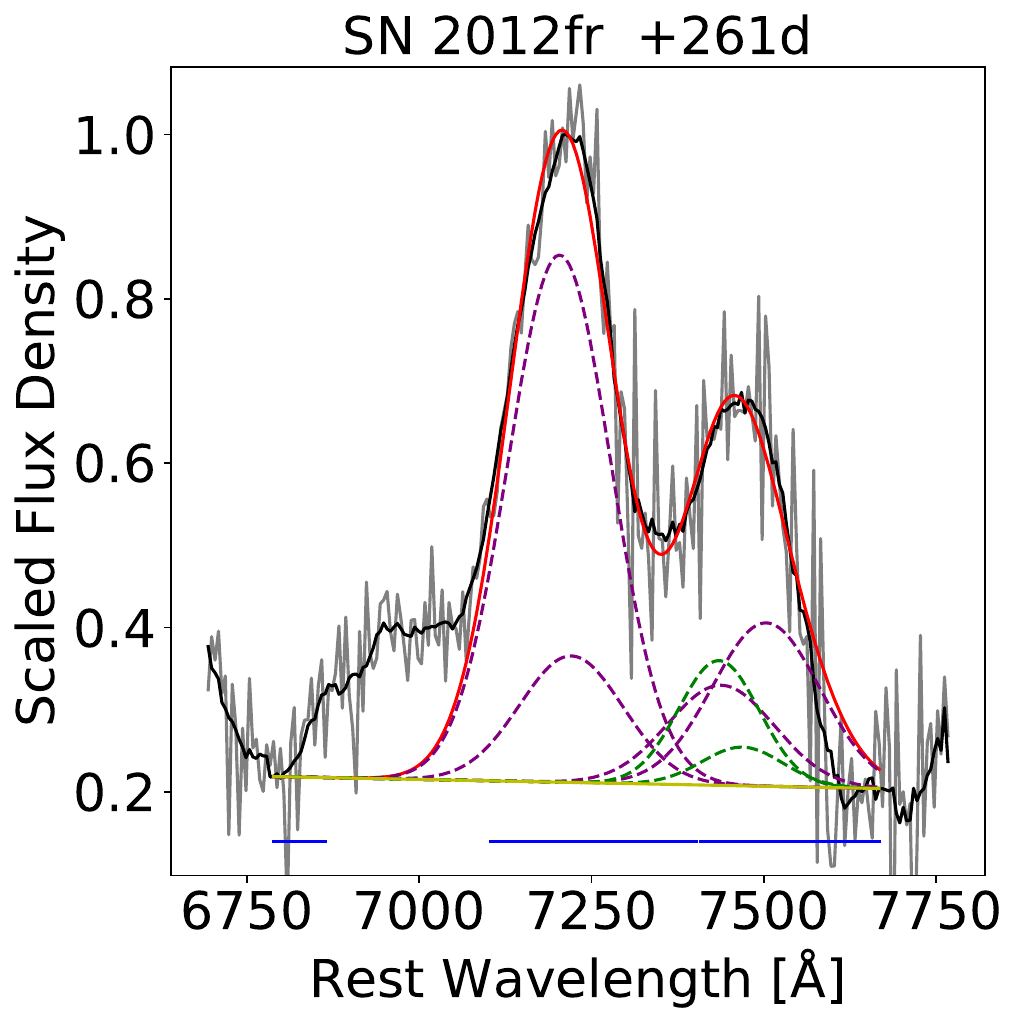}
\includegraphics[width=0.5\columnwidth]{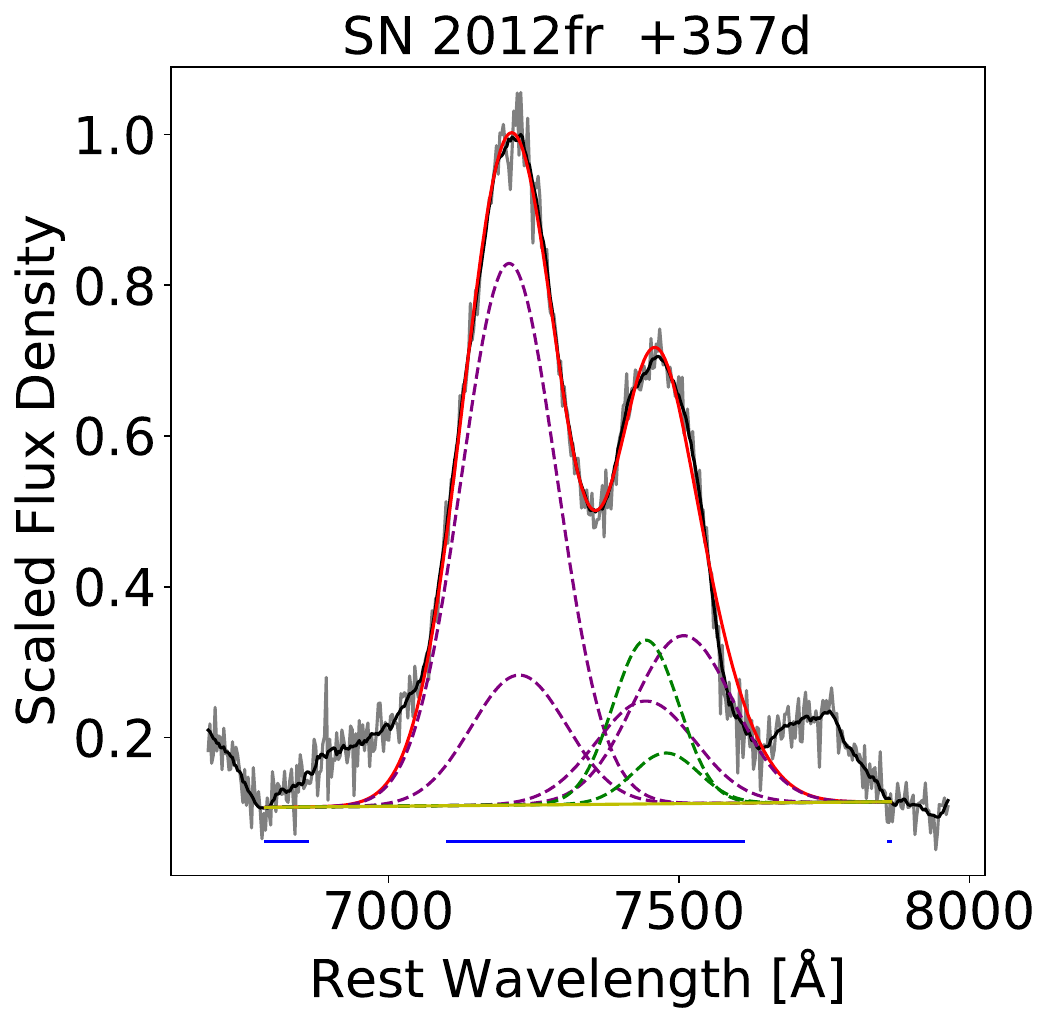}
\includegraphics[width=0.5\columnwidth]{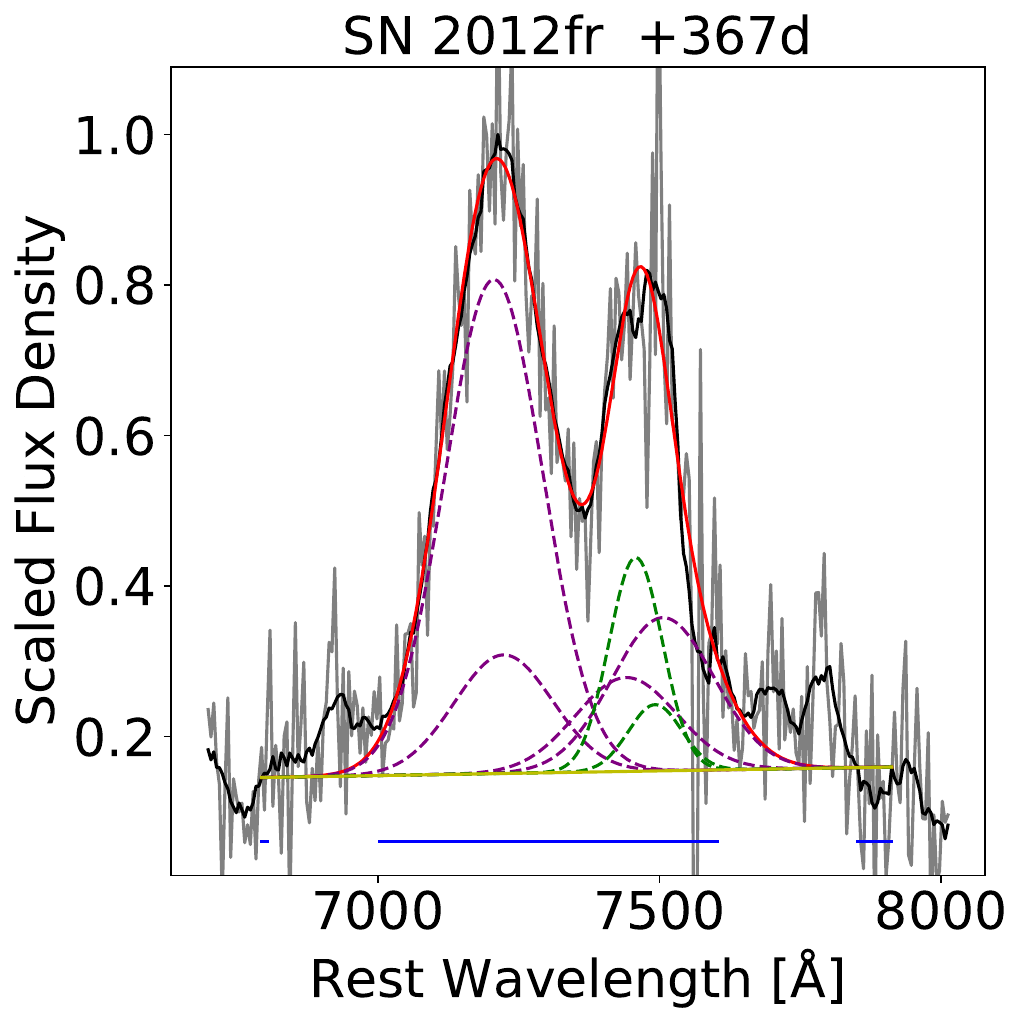}
\includegraphics[width=0.5\columnwidth]{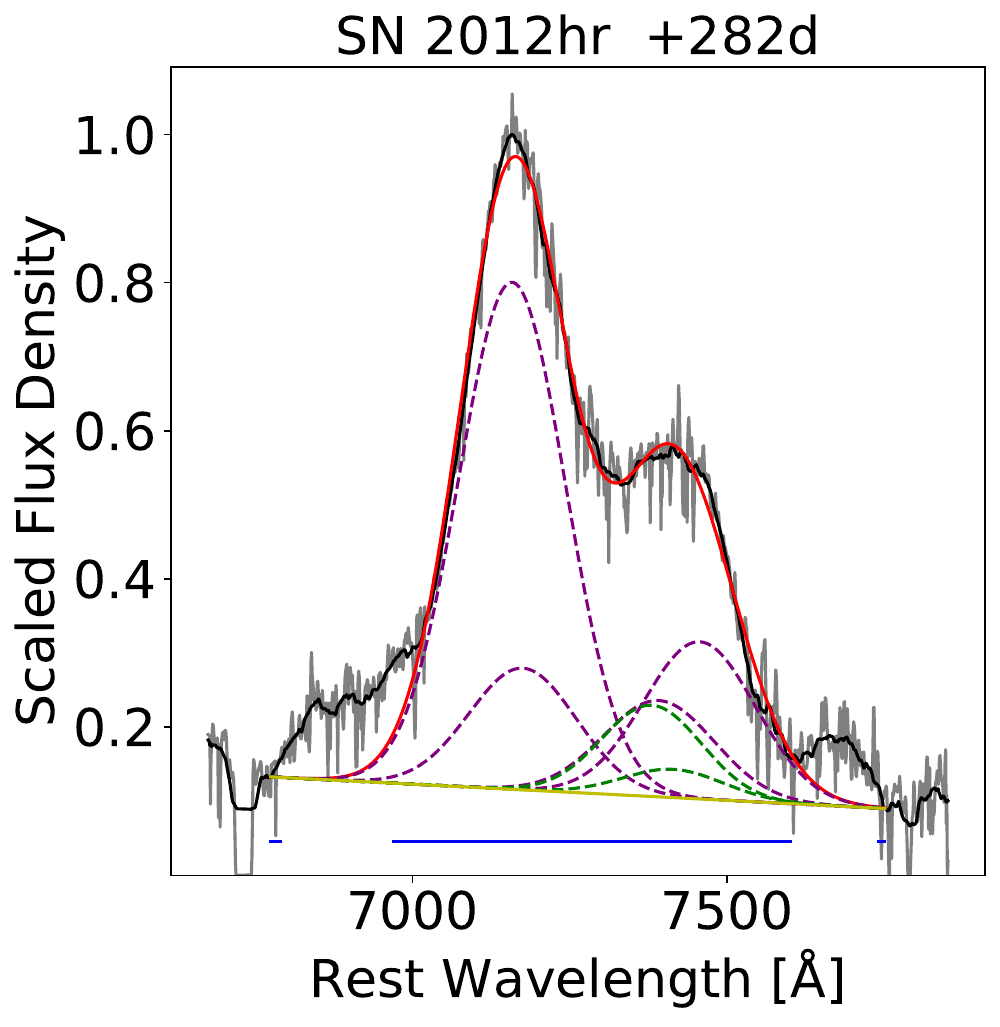}
\includegraphics[width=0.5\columnwidth]{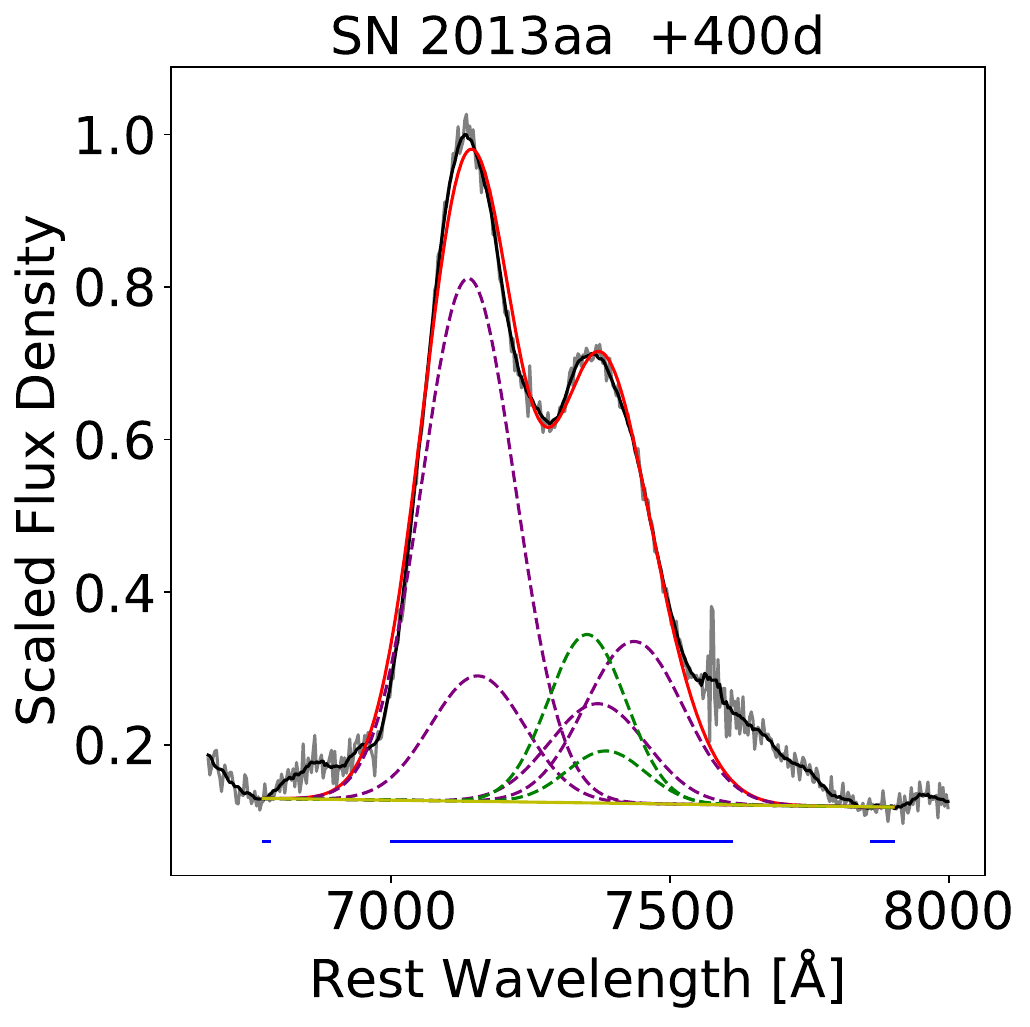}
\includegraphics[width=0.5\columnwidth]{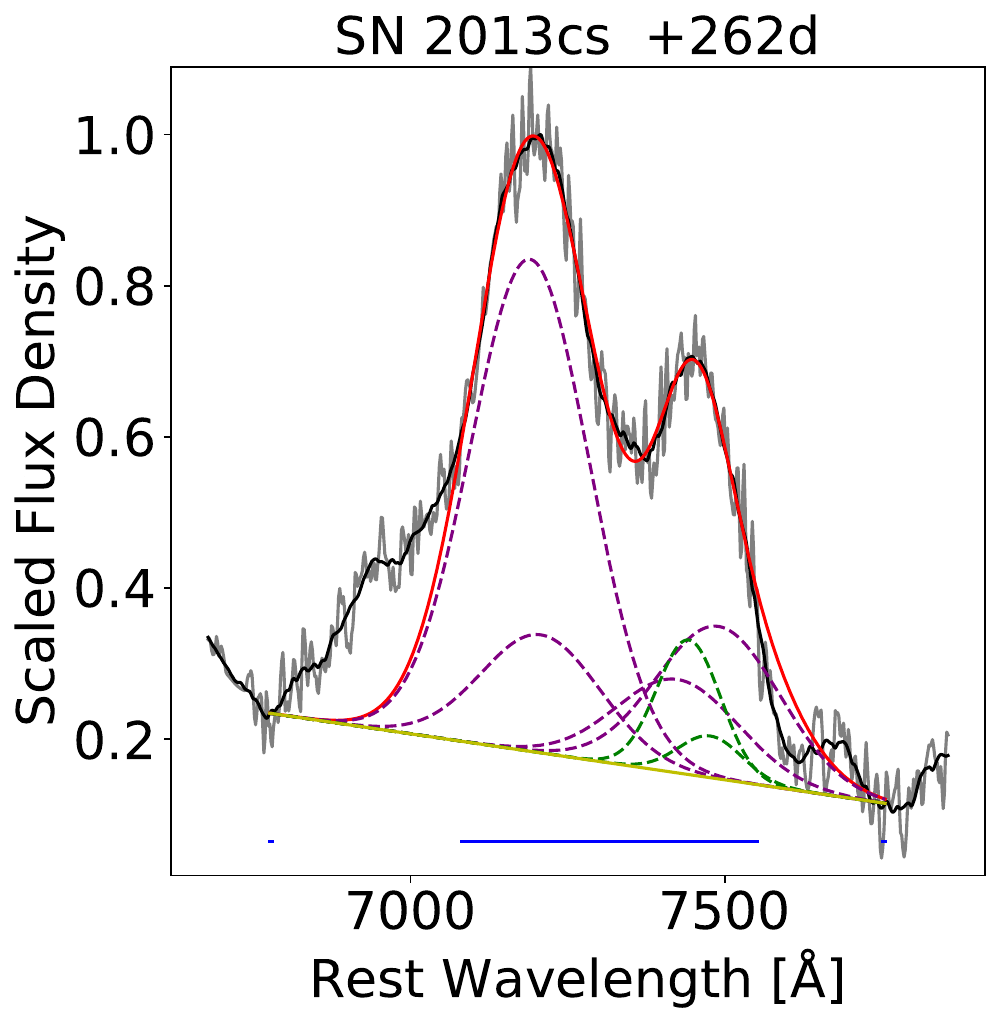}
\end{figure*}

\begin{figure*}
\centering
\includegraphics[width=0.5\columnwidth]{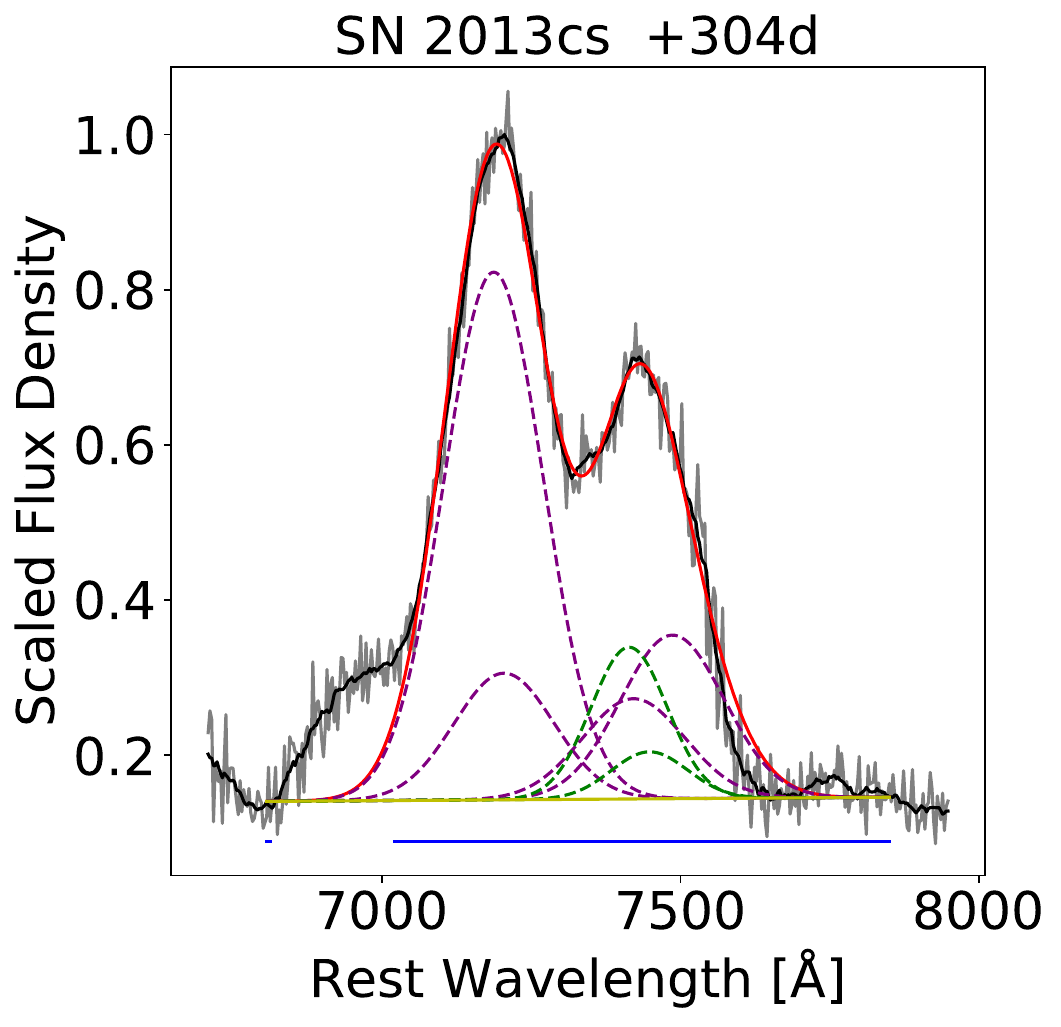}
\includegraphics[width=0.5\columnwidth]{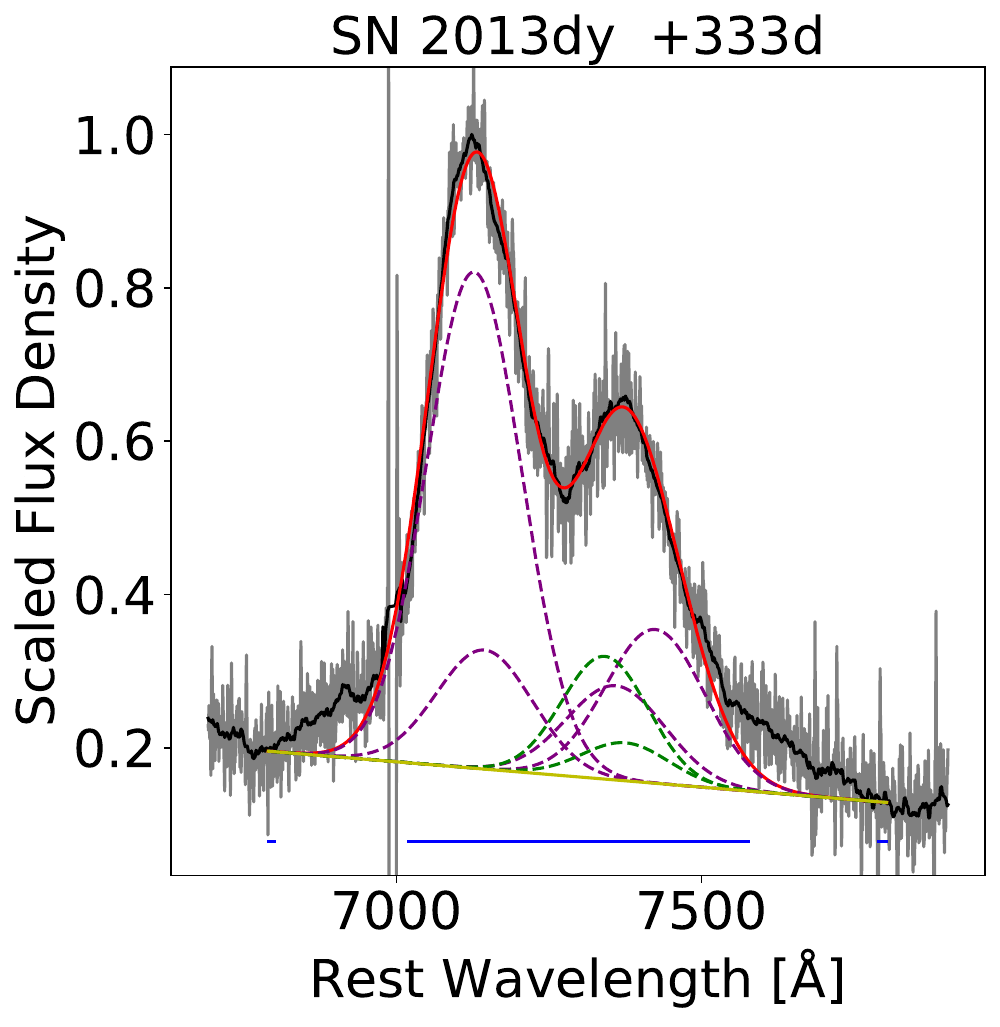}
\includegraphics[width=0.5\columnwidth]{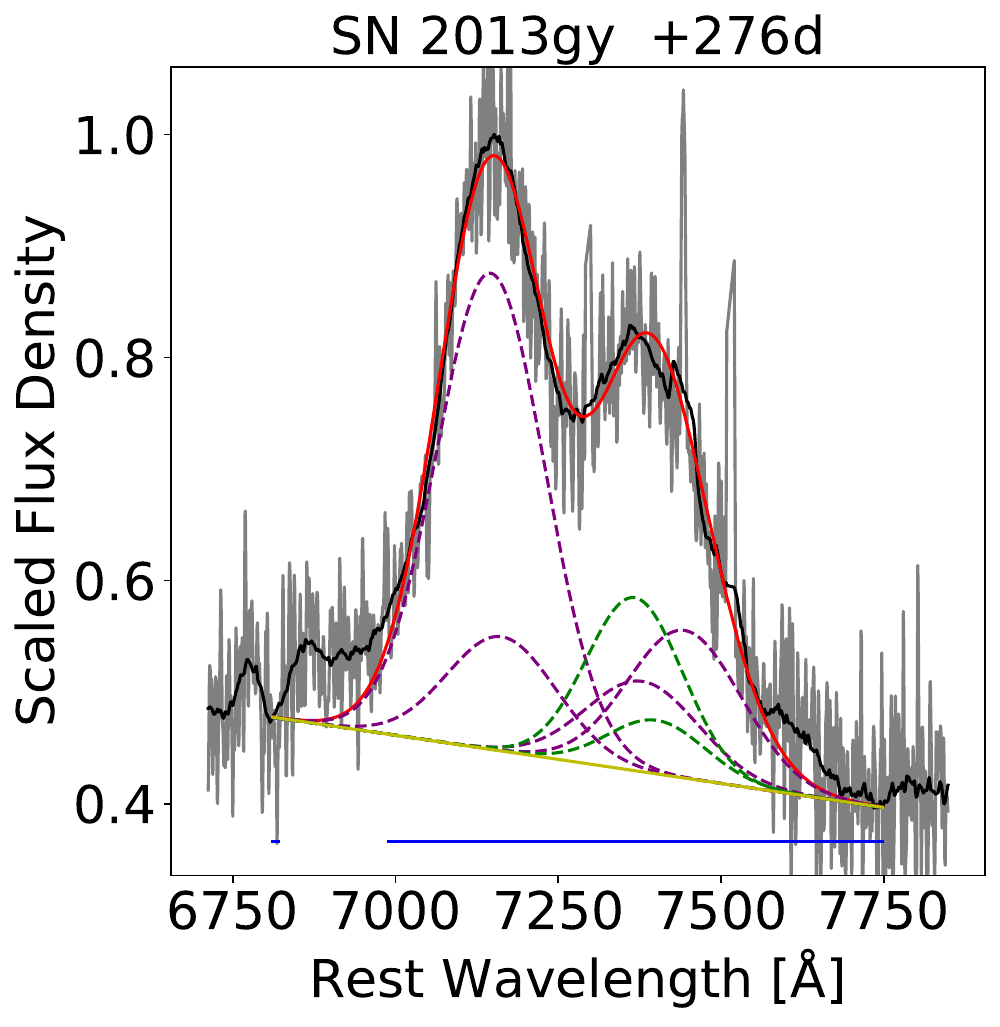}
\includegraphics[width=0.5\columnwidth]{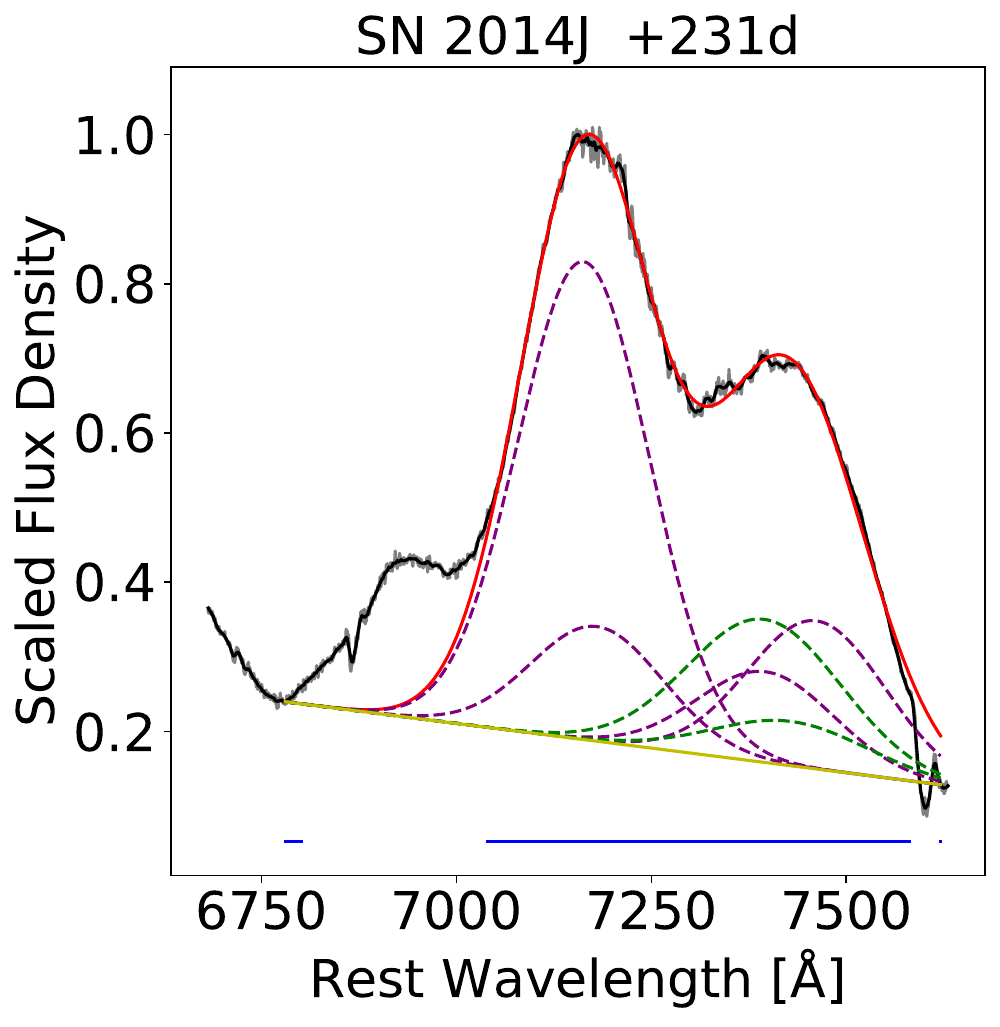}
\includegraphics[width=0.5\columnwidth]{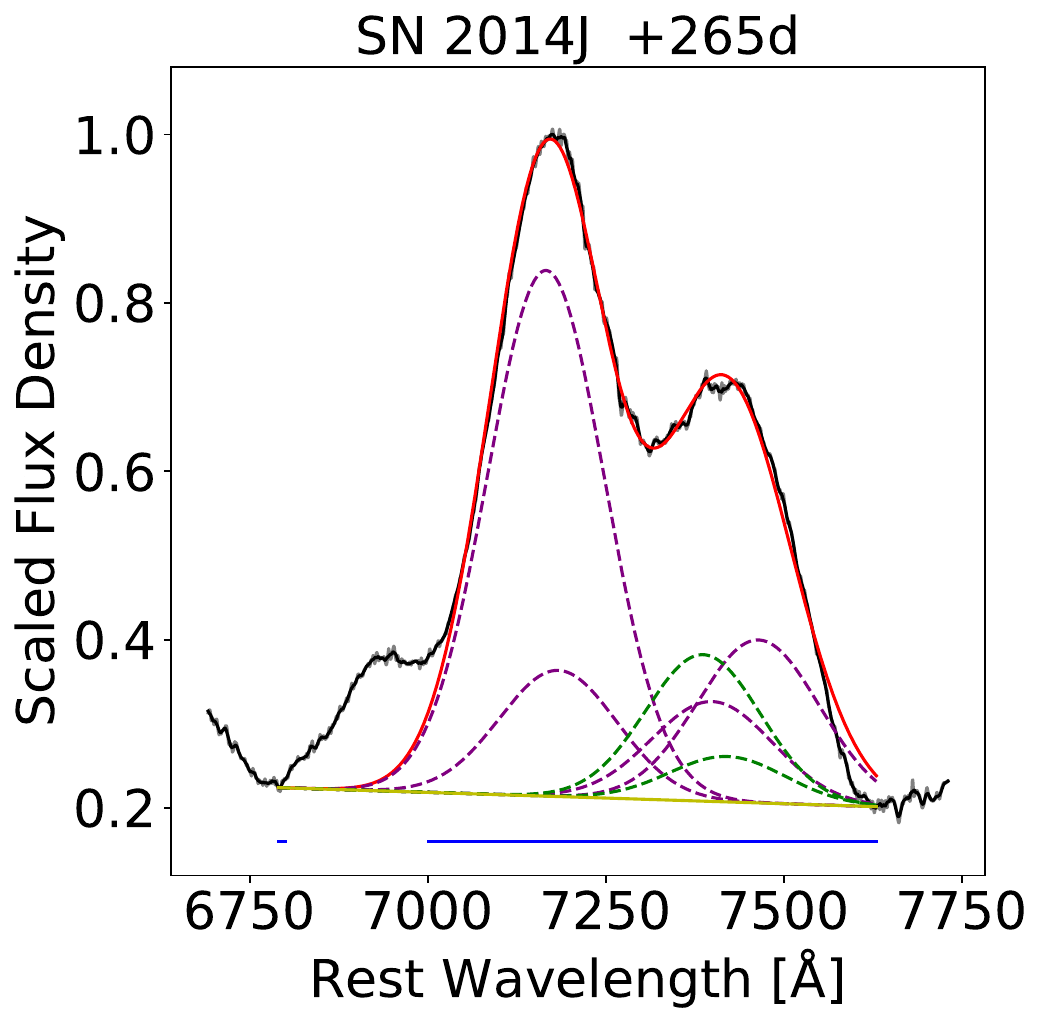}
\includegraphics[width=0.5\columnwidth]{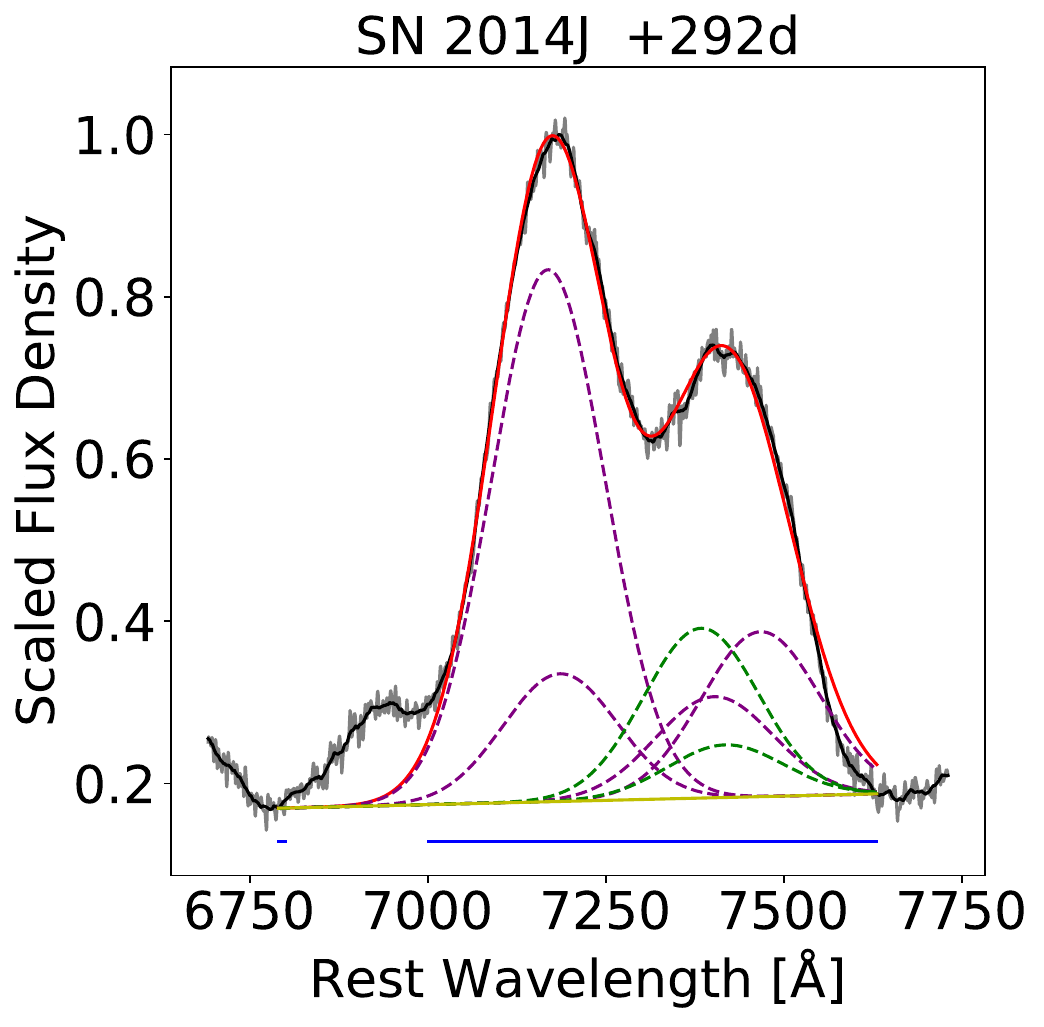}
\includegraphics[width=0.5\columnwidth]{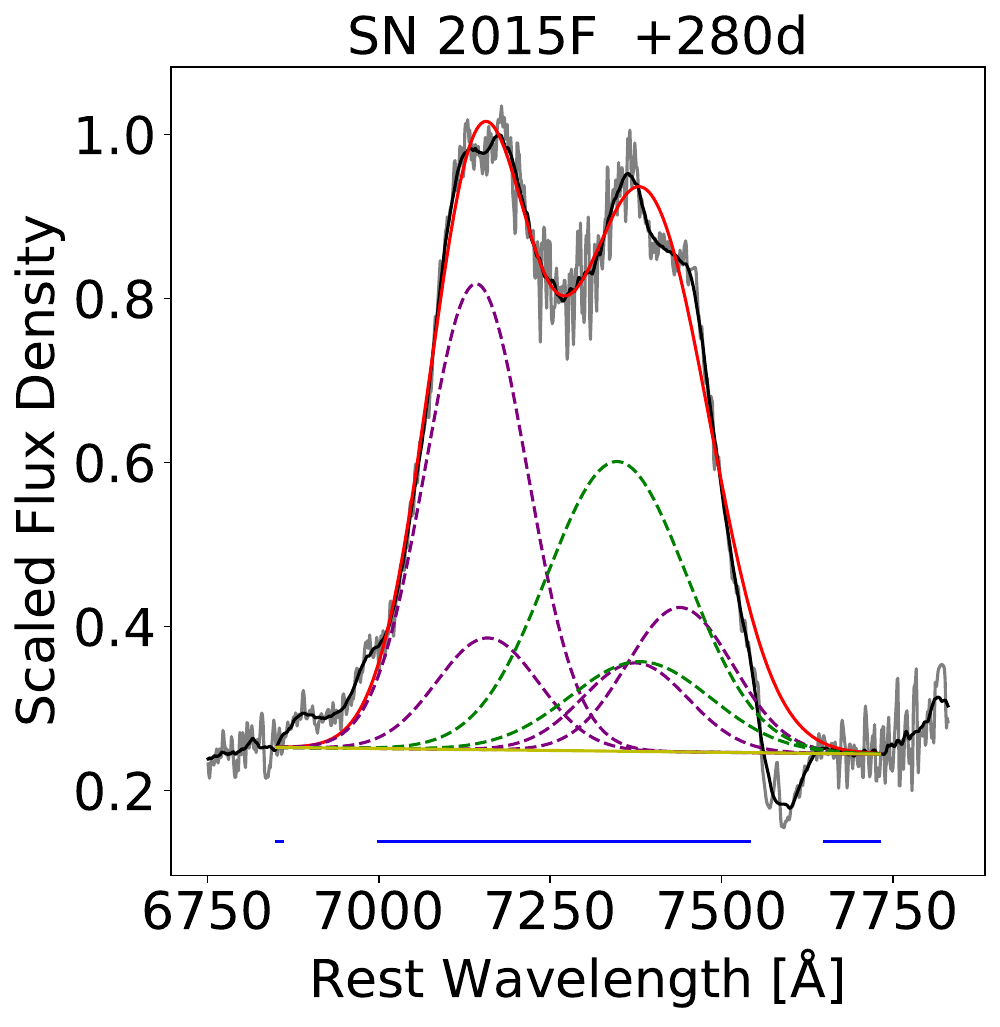}
\includegraphics[width=0.5\columnwidth]{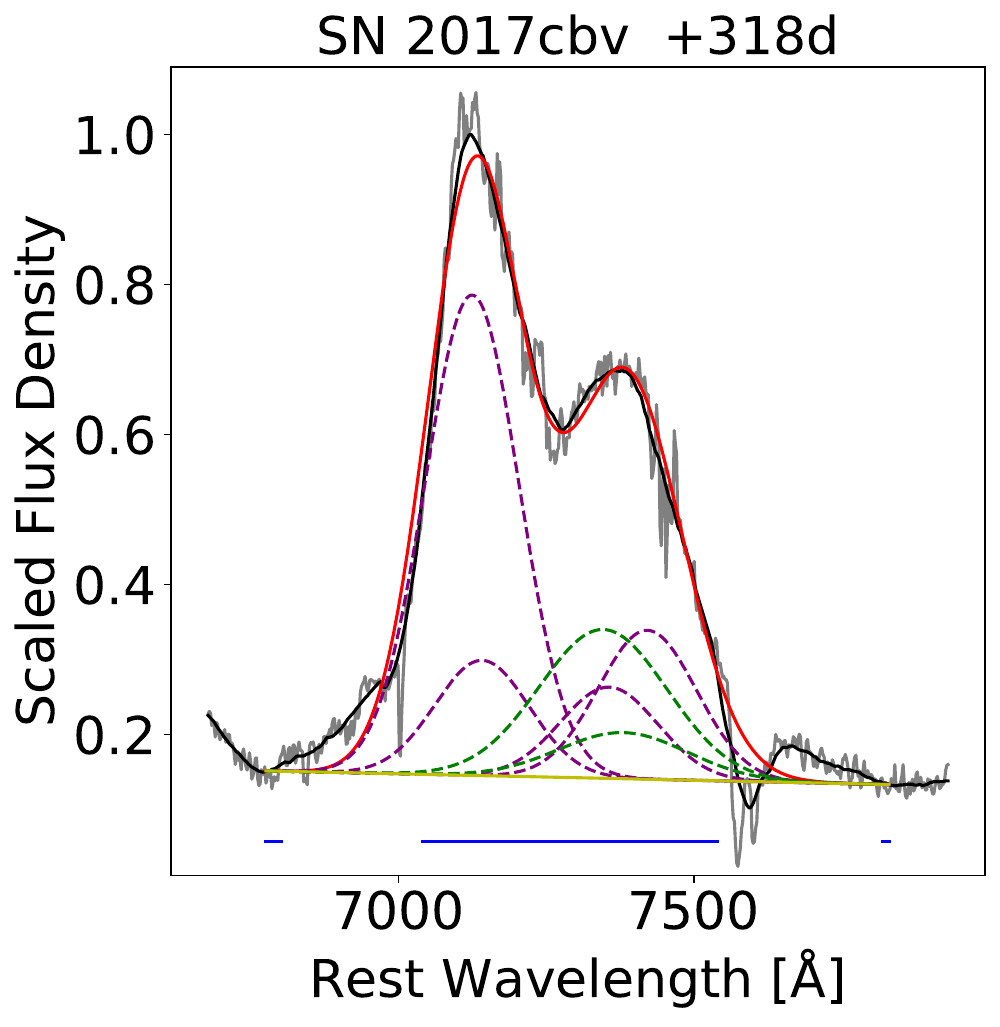}
\includegraphics[width=0.5\columnwidth]{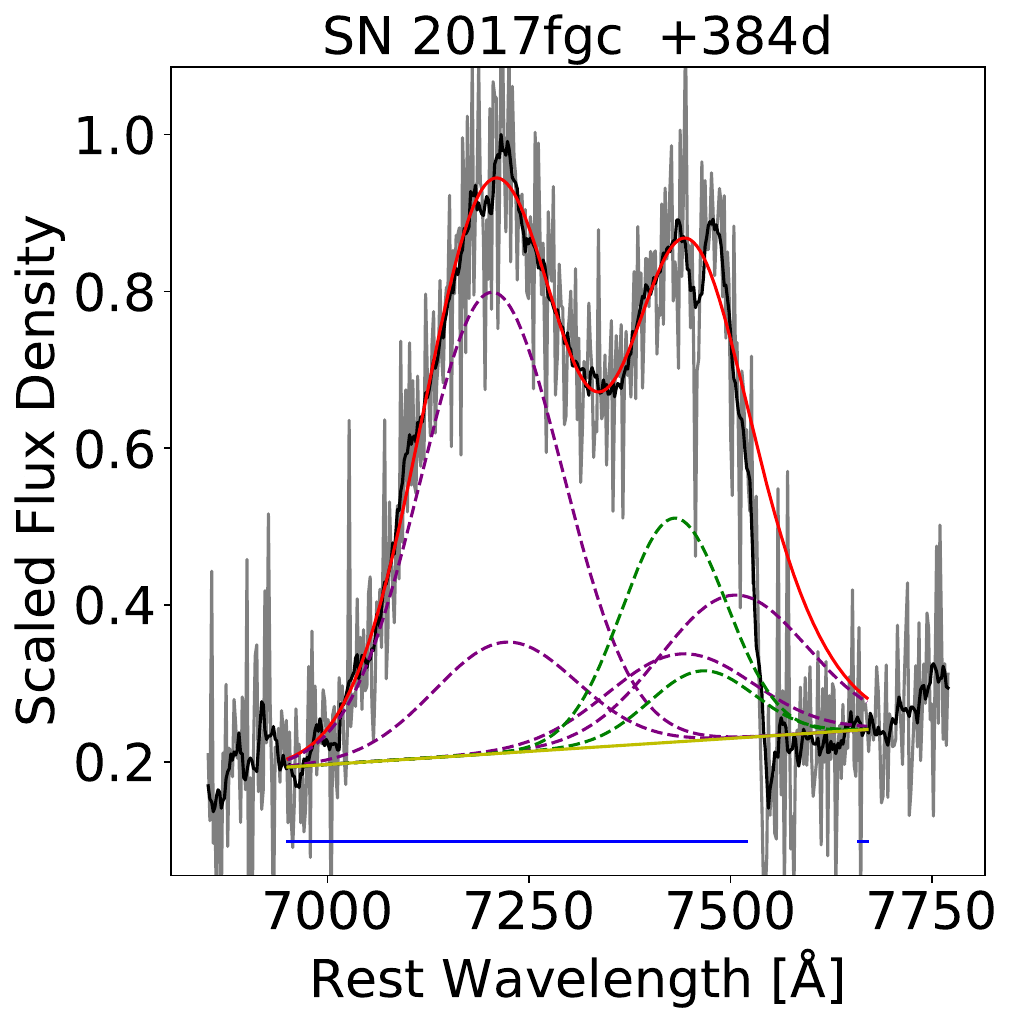}
\includegraphics[width=0.5\columnwidth]{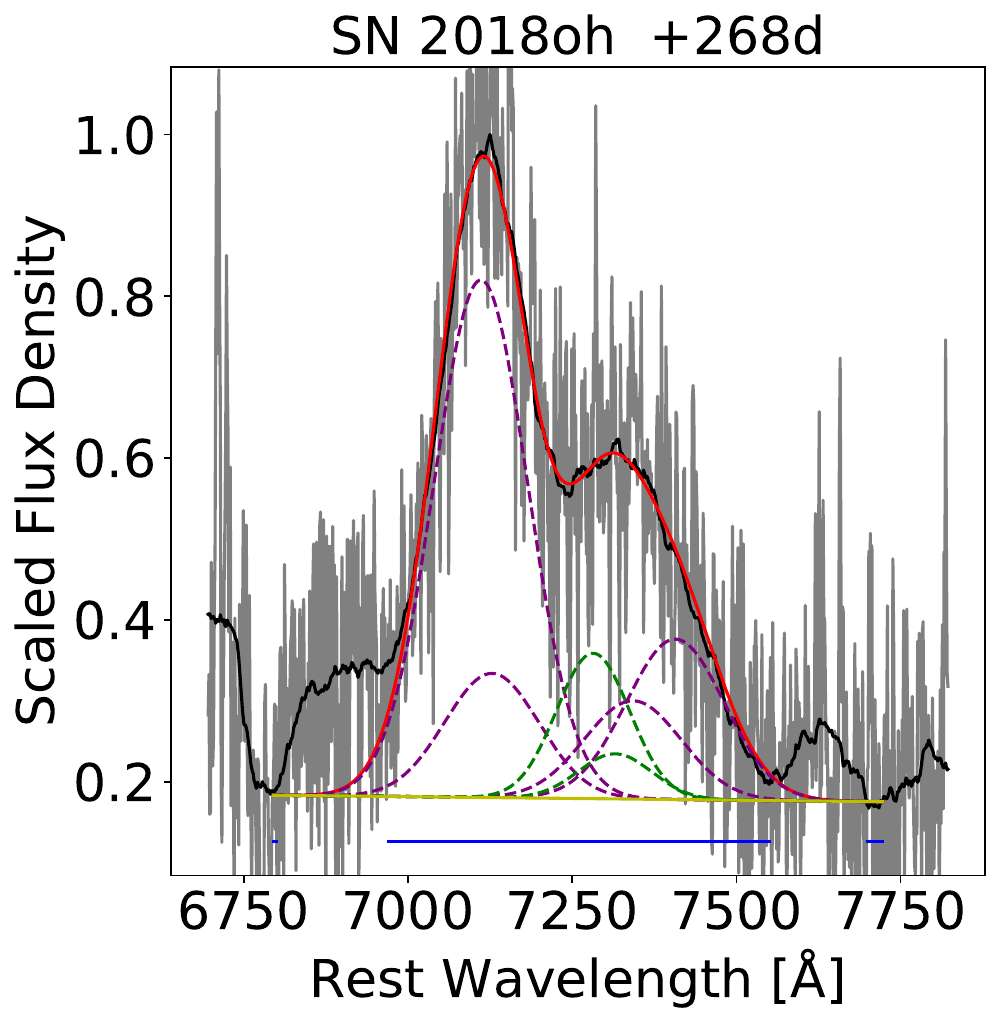}
\includegraphics[width=0.5\columnwidth]{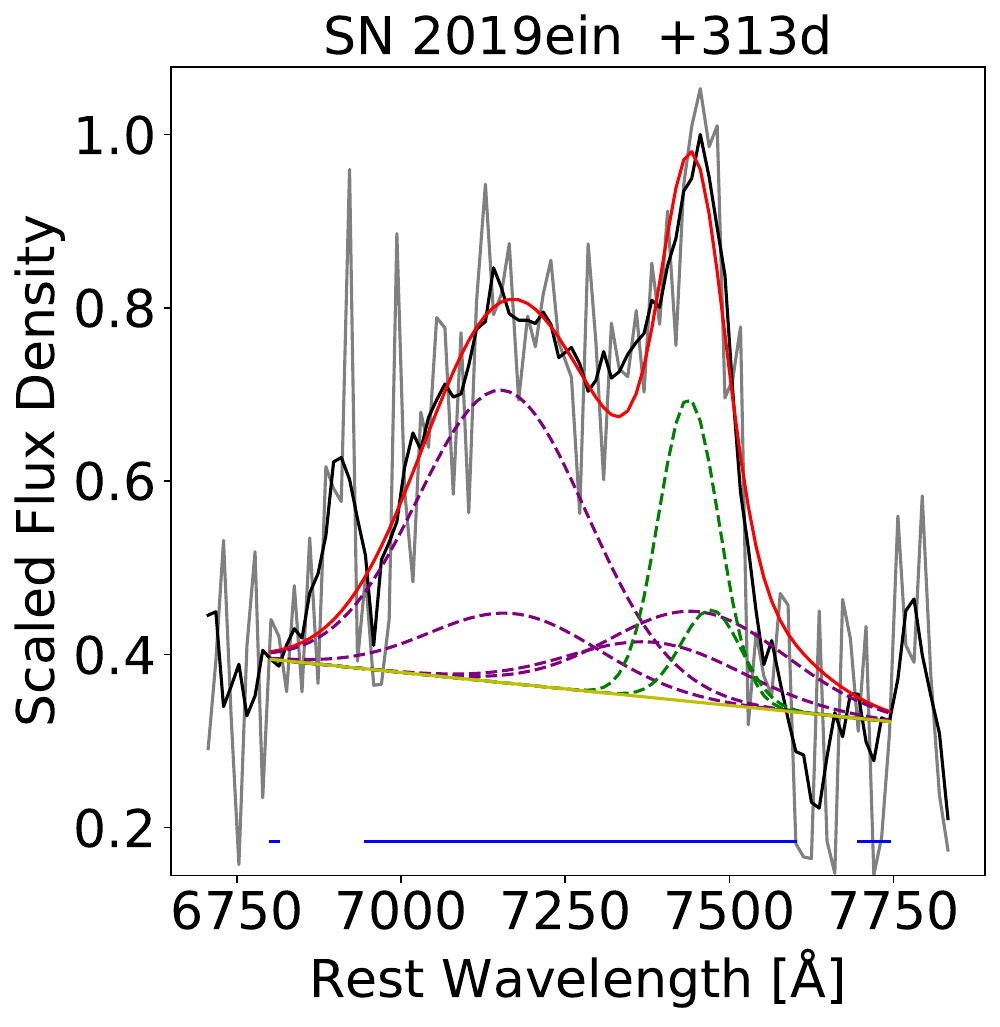}
\includegraphics[width=0.5\columnwidth]{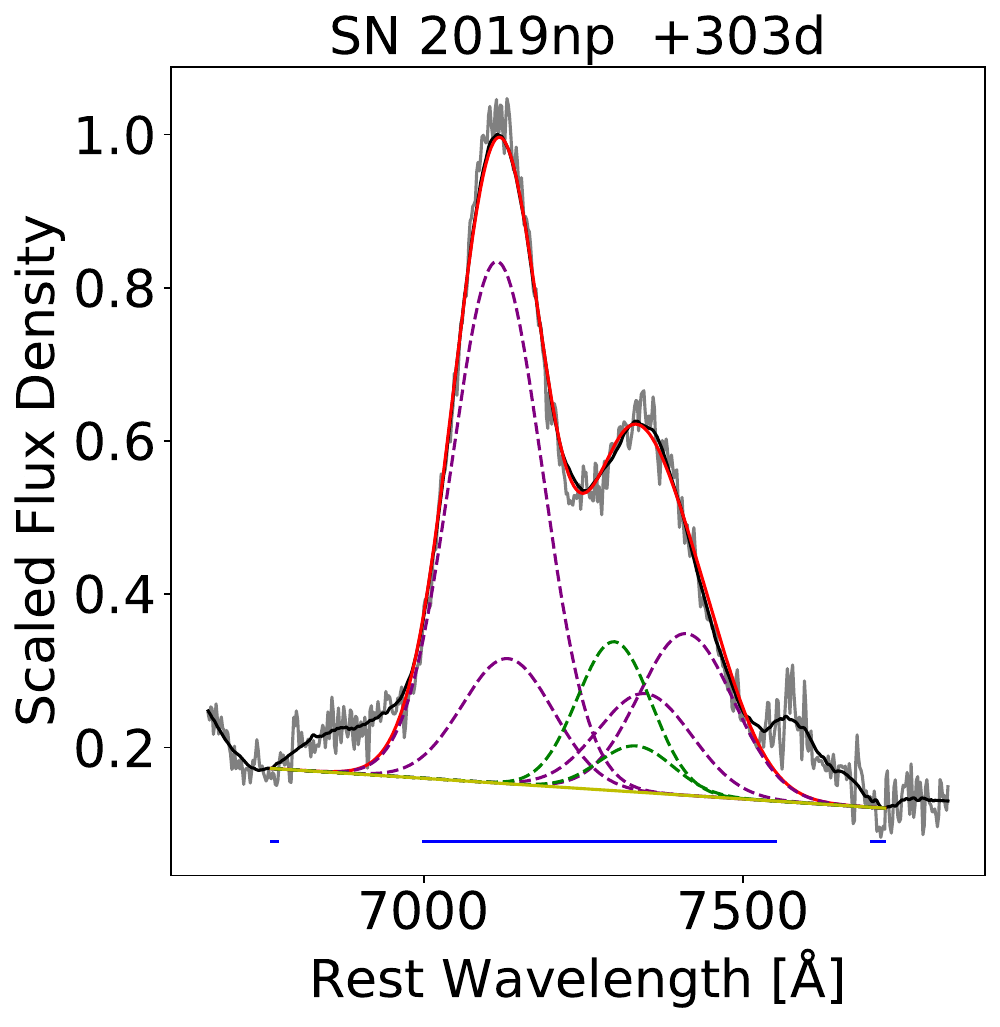}
\includegraphics[width=0.5\columnwidth]{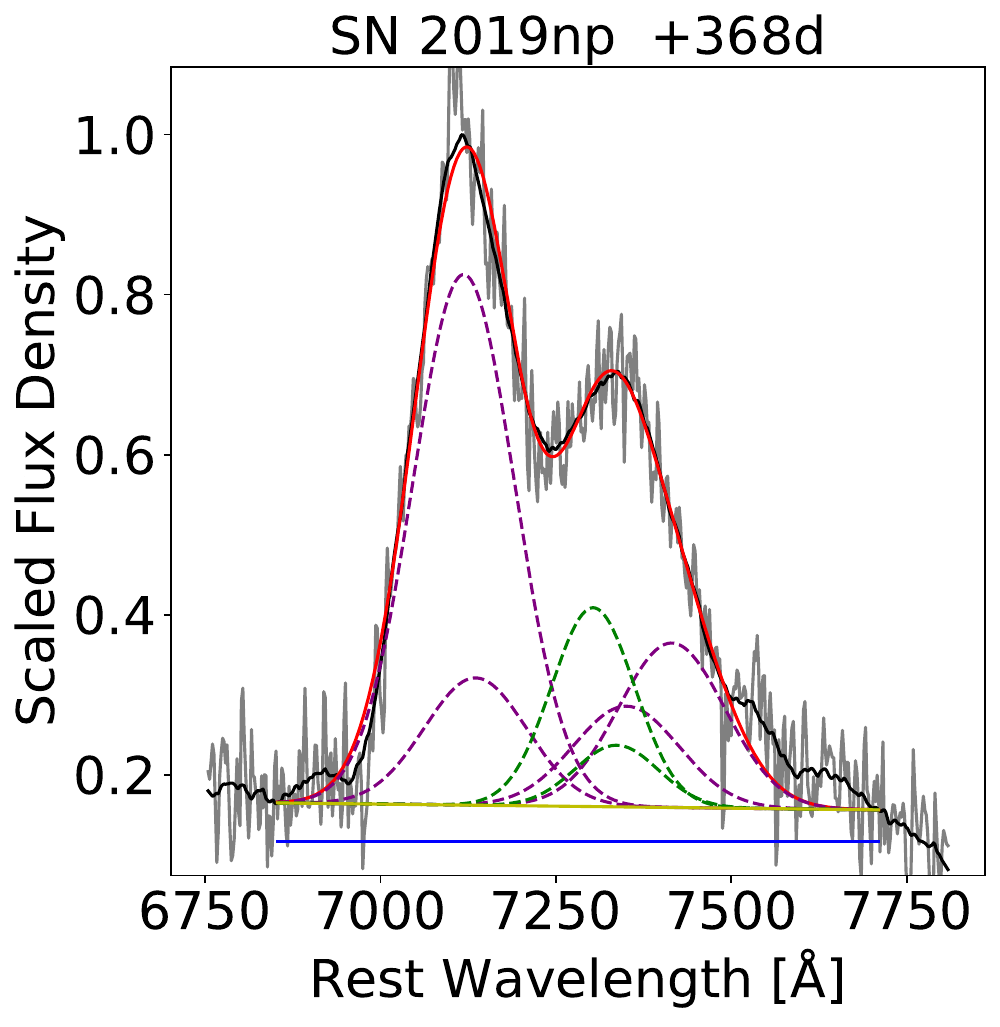}
\includegraphics[width=0.5\columnwidth]{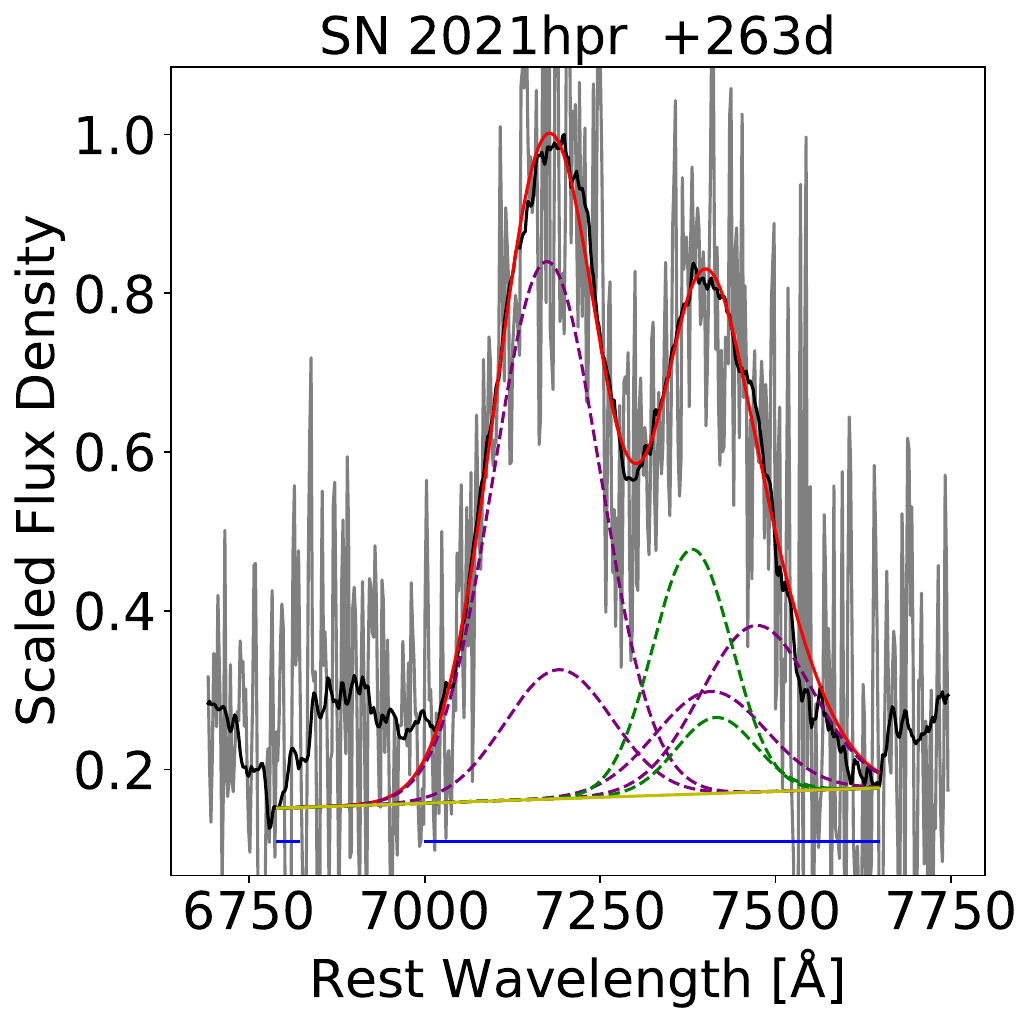}
\includegraphics[width=0.5\columnwidth]{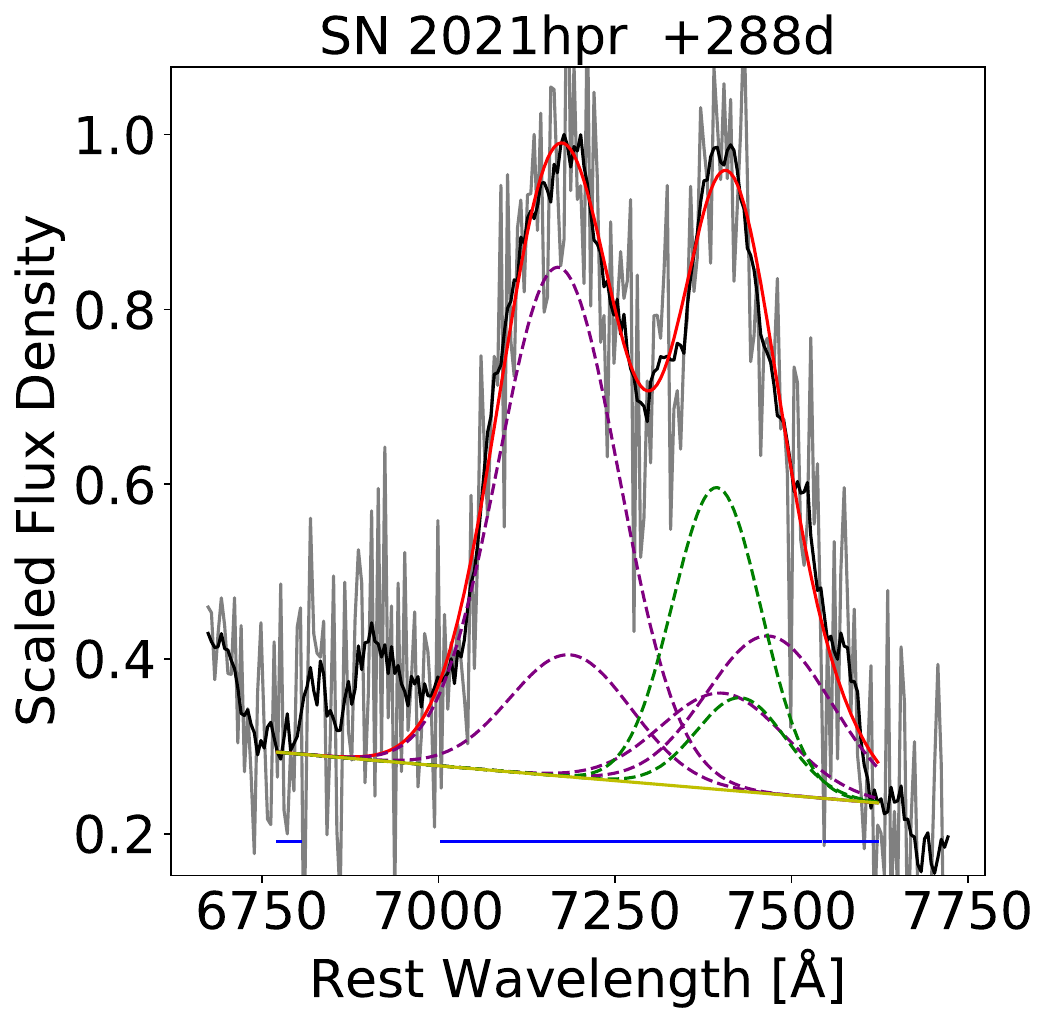}
\includegraphics[width=0.5\columnwidth]{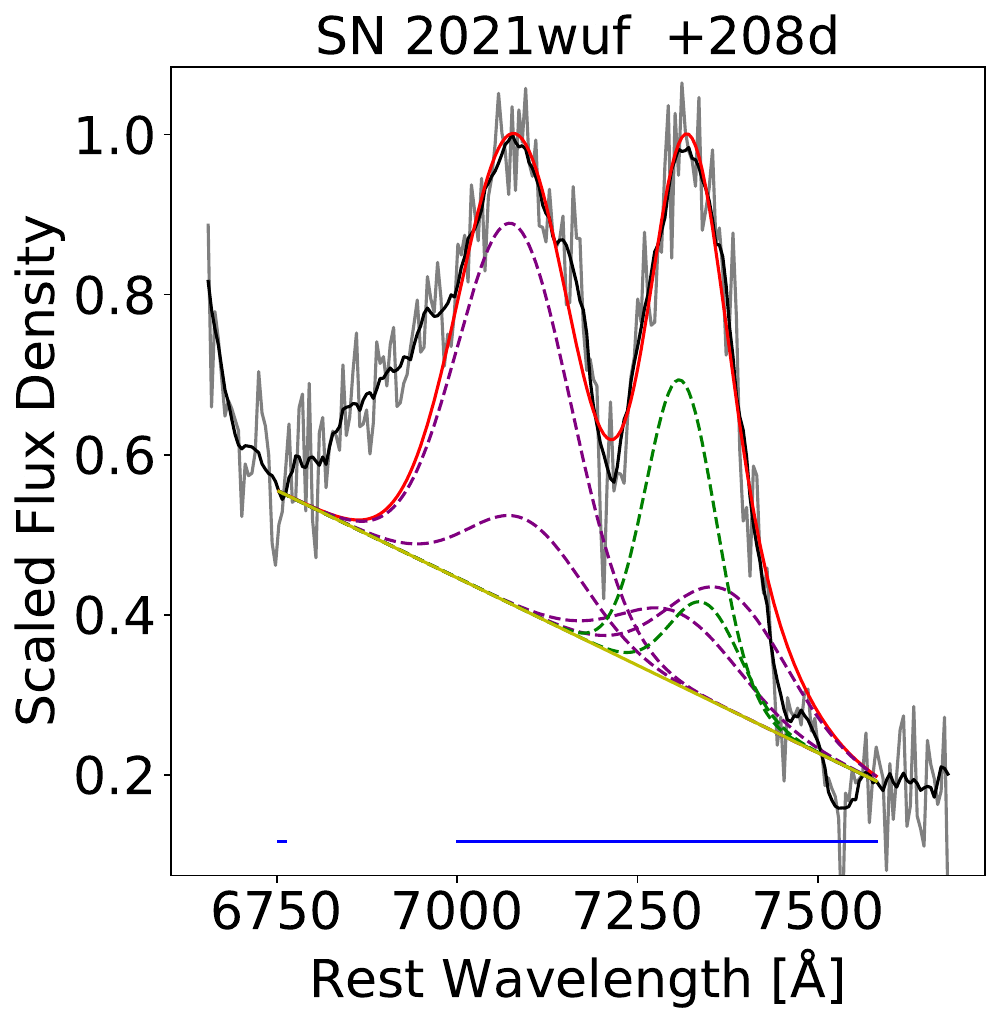}
\includegraphics[width=0.5\columnwidth]{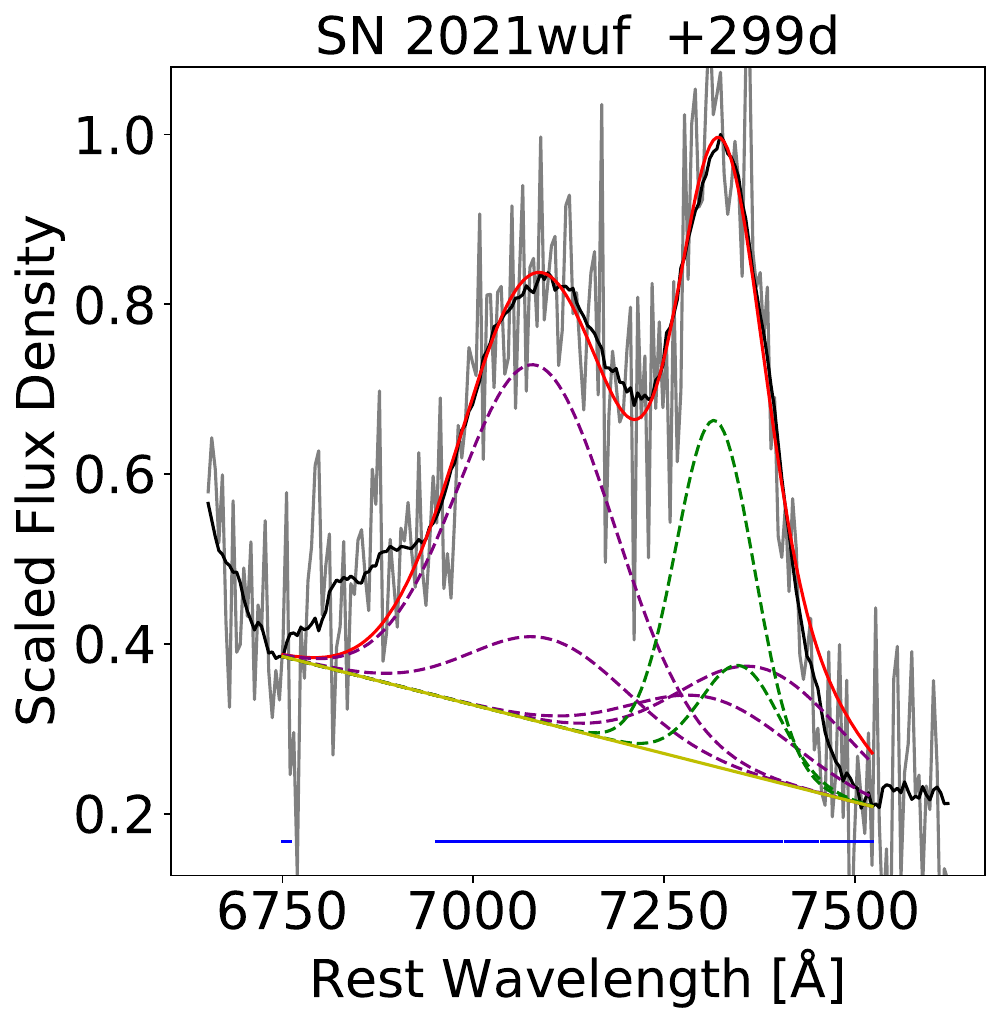}
\includegraphics[width=0.5\columnwidth]{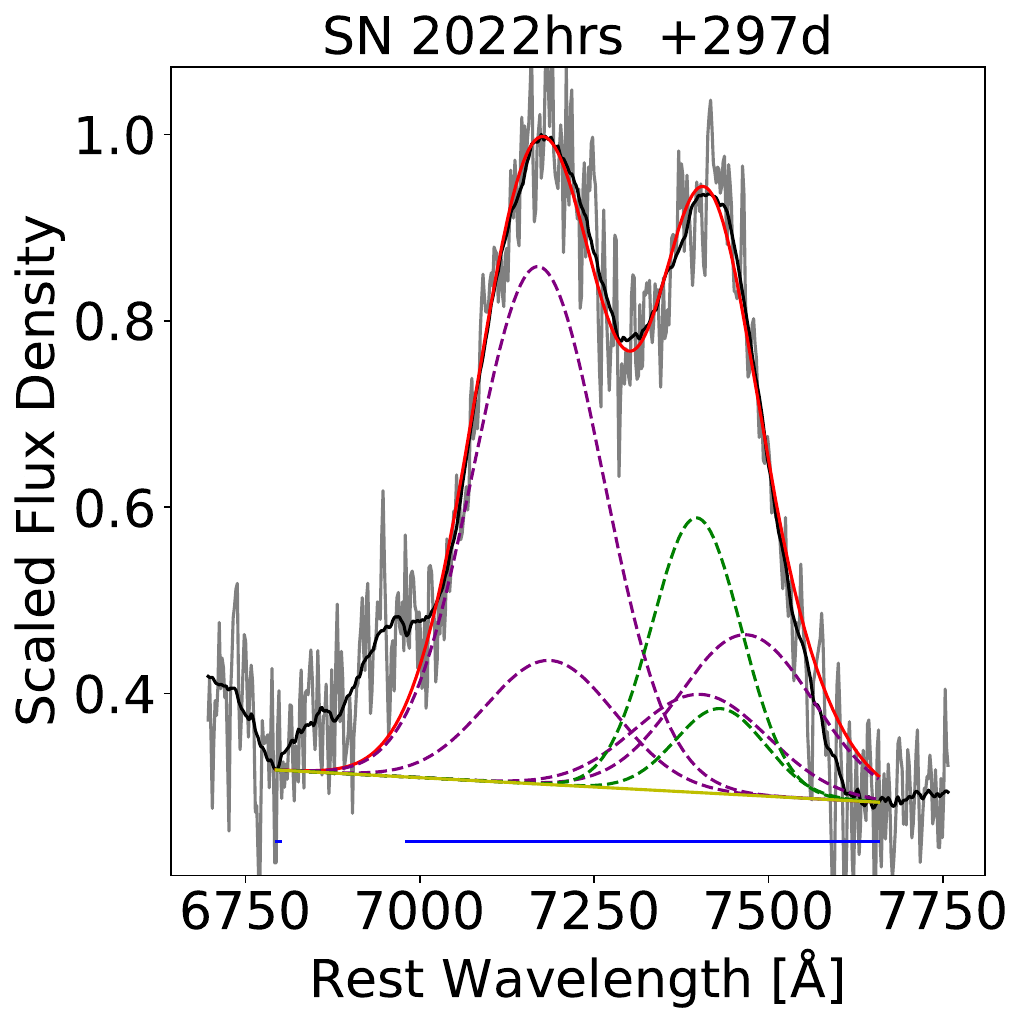}
\includegraphics[width=0.5\columnwidth]{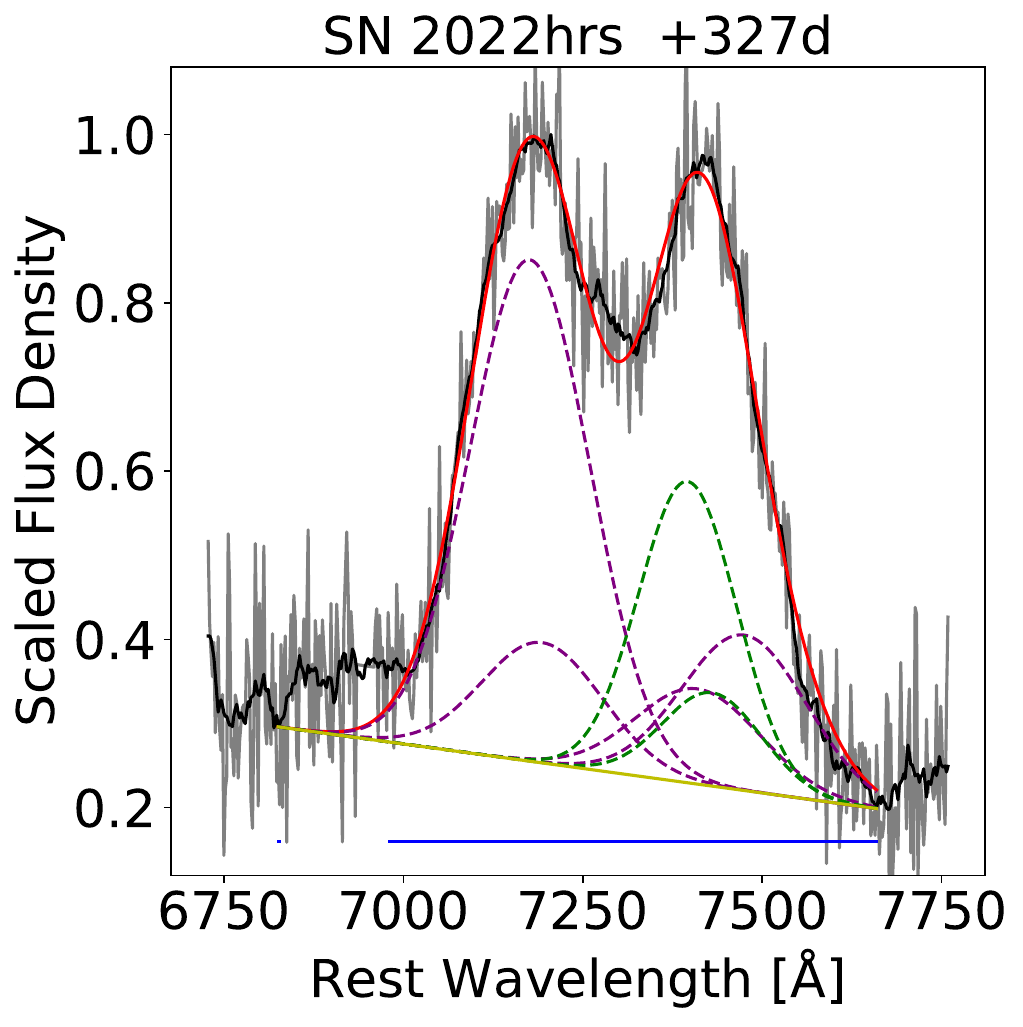}
\includegraphics[width=0.5\columnwidth]{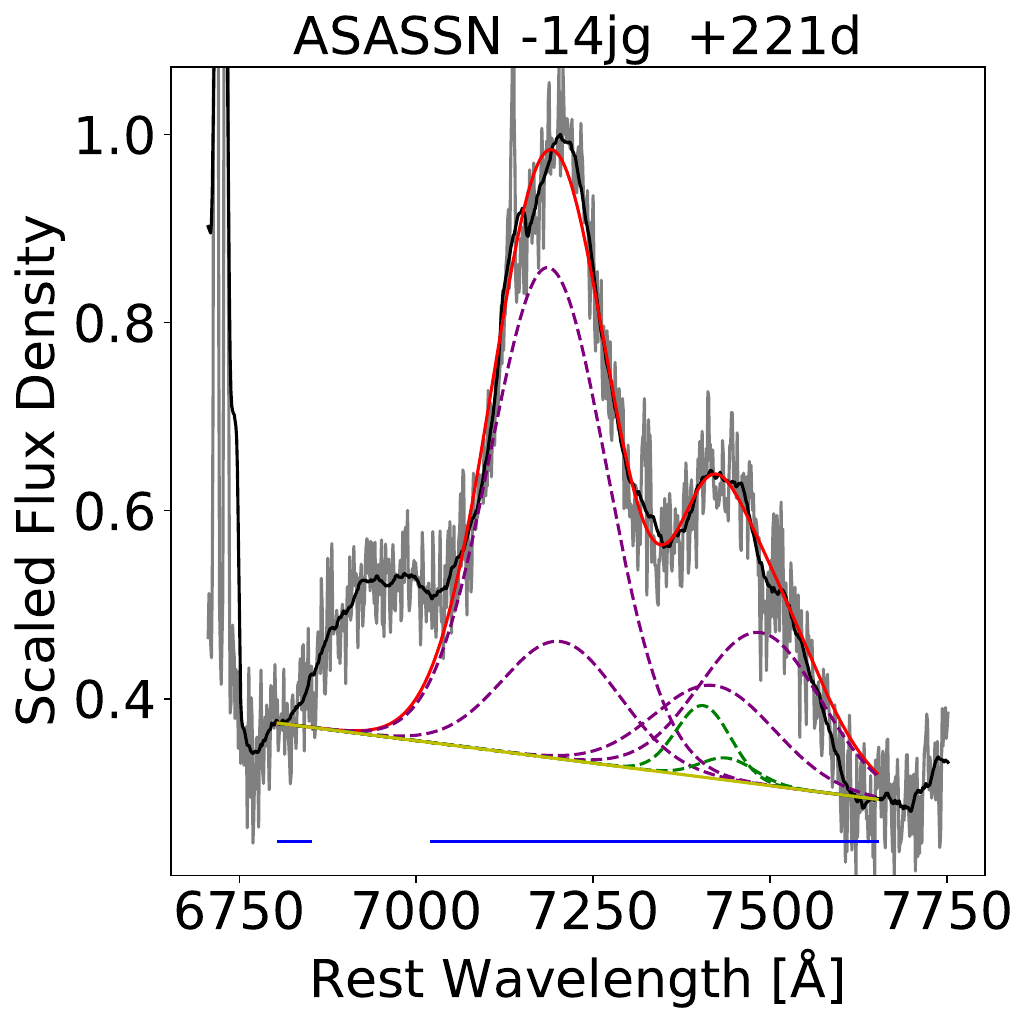}
\end{figure*}

\begin{figure*}
\centering
\includegraphics[width=0.5\columnwidth]{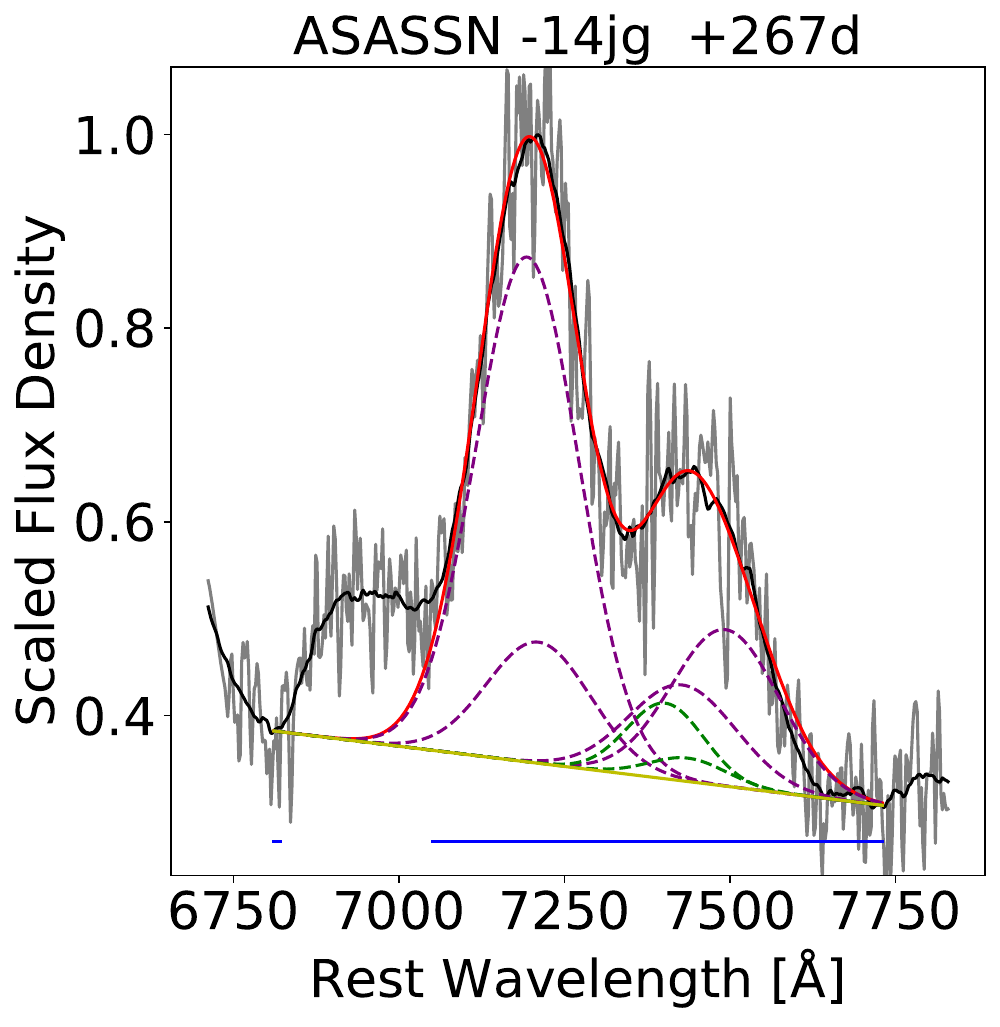}
\includegraphics[width=0.5\columnwidth]{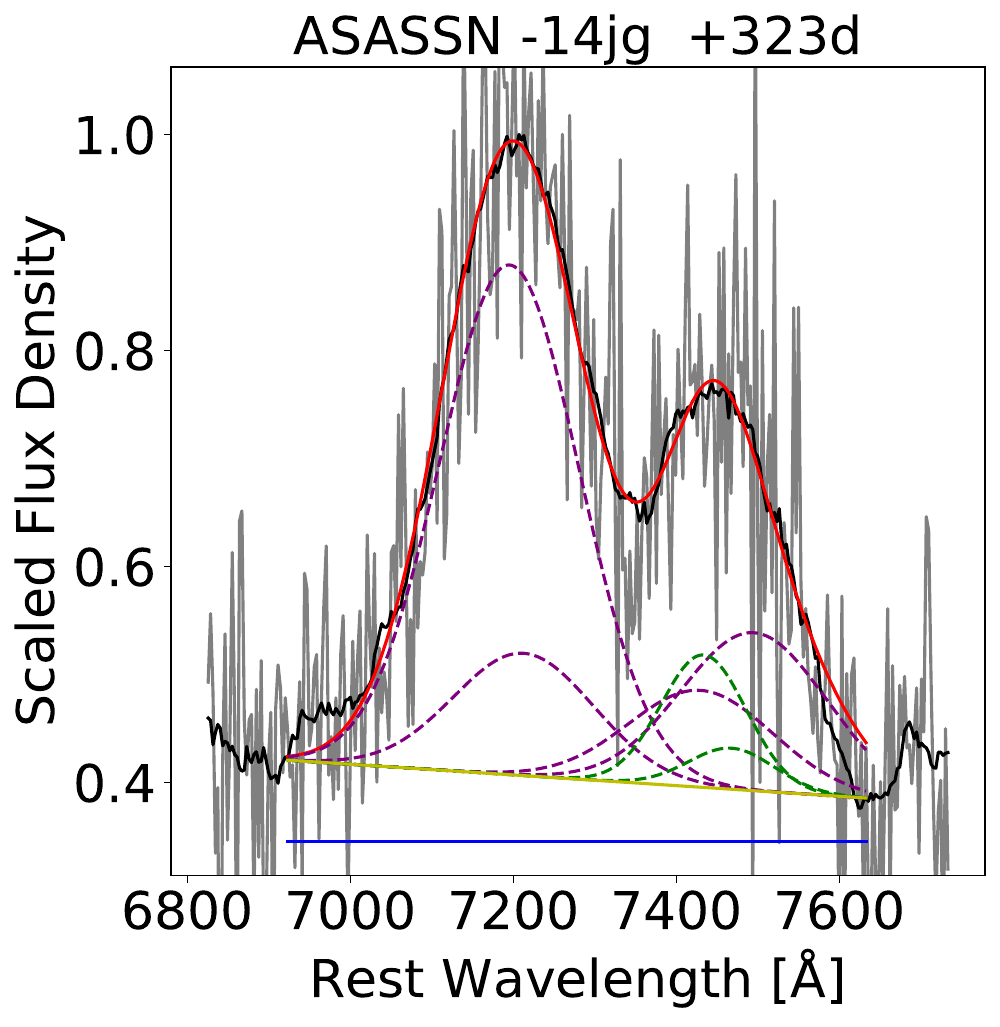}
\caption{Best fits to the 7300~{\AA} region dominated by [Fe~II] and [Ni~II] features. The reddening-corrected spectra are shown in grey while the smoothed spectra are shown in black. The overall fits are shown in red, the [Fe~II] features are in purple dashed lines, and the [Ni~II] features are in green dashed lines. The [Ca~II] features for SN~1986G and SN~2003gs are in pink. The fitting regions are indicated with blue lines,  and the pseudocontinuum is in yellow.}
    \label{fig:all_fits}
\end{figure*}


\bsp	
\label{lastpage}
\end{document}